**Shorter but not more catalytic: an active-site-resolved model of how tau reshapes amyloid-β secondary nucleation**

Andrey V. Kuznetsov

Department of Mechanical and Aerospace Engineering, North Carolina State University,

Raleigh, NC 27695-7910, USA; email: avkuznet@ncsu.edu

**Abstract**

Tau cross-seeding creates a mechanistic puzzle. Tau fibrils accelerate amyloid-β42 (Aβ42) nucleation and produce a larger population of shorter Aβ fibrils, yet homotypic secondary nucleation occurs at rare catalytic sites rather than uniformly over fibril mass. Does making more, shorter fibrils make the Aβ population more catalytic? A coupled tau–Aβ model is extended to track fibril number, fibril mass, total active secondary-nucleation sites, and the fraction of sites exposed after plaque consolidation. The model recovers the preceding plaque-surface-limited formulation when active sites form in direct proportion to fibril mass. At matched fibril mass, shortening alone has little effect: the departure of active-site density from mass proportionality scales inversely with mean fibril size, so the extra fibrils matter only if each birth event introduces about as many sites as elongation adds over a mean-length fibril. At equal time, the sign of the homotypic response is controlled mainly by a Damköhler number comparing monomer consumption by elongation with monomer clearance. Tau coupling increases homotypic secondary nucleation while clearance controls free monomer, but suppresses it once elongation becomes the dominant monomer sink. In a composite scenario, these regimes occur in sequence. Across all cases examined, including 800 Latin-hypercube samples and simulations with mean fibril lengths of 0.3–30 μm, direct tau catalysis increases accumulated oligomer-equivalent exposure. The model predicts that shorter tau-seeded fibrils are not intrinsically more catalytic unless their sites are more numerous per unit mass, more active, or more exposed; matched-mass site counting and secondary-nucleation measurements can distinguish these alternatives.

**Keywords:** Alzheimer's disease; amyloid-β42; tau; heterotypic nucleation; active secondary-nucleation sites; secondary nucleation; fibril growth; oligomer kinetics; mathematical modeling

## 1. Introduction

The molecular consequences of amyloid-β (Aβ) aggregation cannot be inferred from plaque mass alone. Once Aβ fibrils have formed, their surfaces can catalyze secondary nucleation and generate additional soluble oligomeric intermediates [1]. At the same time, many Aβ42 oligomers return to monomer rather than becoming growth-competent fibrils [2]. The biologically relevant burden therefore emerges from a competition among nucleation, elongation, oligomer conversion, dissociation, and clearance. This distinction between how much fibrillar protein is present and how catalytically active that protein is provides the starting point for the present study.

A recent tau–Aβ cross-seeding experiment made that distinction especially important. PHF- and CTE-fold tau fibrils accelerate heterotypic Aβ42 primary nucleation, with CTE tau acting as the stronger catalyst, and the resulting Aβ42 fibrils are shorter [3]. CTE seeds also narrow the distribution of fibril morphologies, although no distinct Aβ fibril structure was resolved. The kinetic explanation is straightforward: faster nucleation creates more growing fibrils, so each fibril receives a smaller share of the free monomer available for elongation. Tau can therefore reshape the Aβ fibril population even when it does not simply increase fibril mass. What the experiment does not establish is whether this altered population is also more effective at catalyzing subsequent Aβ secondary nucleation.

Our preceding coupled tau–Aβ model examined the long-term consequence of this cross-seeding mechanism [4]. Tau aggregates were represented in a neuronal soma and Aβ aggregation in the surrounding extracellular/interstitial region, with an effective fraction of somatic tau fibrillar material providing an additional route to Aβ nucleation. Because fibril number and fibril mass were tracked separately, the model reproduced the qualitative shift toward more numerous, shorter Aβ fibrils while total fibrillar mass remained nearly unchanged. It also predicted a substantial increase in soluble oligomer-equivalent exposure.

That model, however, left one mechanistic question unresolved. Homotypic Aβ secondary-nucleation activity was assumed to scale with the geometrically accessible fraction of fibril mass. Under that assumption, two fibril populations with the same accessible mass are equally catalytic even if one contains more active sites than the other. The model could therefore describe the redistribution of Aβ mass among fibrils, but not whether the growth history of those fibrils changes their catalytic character.

Recent active-site measurements provide a reason to relax that assumption. One such study [5] found that Aβ secondary nucleation is associated predominantly with rare structural growth defects rather than with a uniformly active fibril surface. For one Aβ42 preparation, the authors estimated approximately one active site per 143 monomers; for Aβ40, changing the growth conditions reduced the measured site frequency from approximately one per 99 monomers to one per 834 monomers and reduced secondary-nucleation activity. These numbers are preparation-specific and should not be transferred directly to brain plaques or tau-seeded Aβ42, but they establish the key physical point: equal fibril masses need not have equal catalytic activity.

Taken together, the cross-seeding and active-site studies pose a specific unanswered question. Tau makes Aβ42 fibrils more numerous and shorter [3], while rare growth-associated sites can control secondary nucleation [5]; neither study measured the active-site density of tau-seeded Aβ42 fibrils. More fibrils at a fixed mass could therefore carry more sites, the same number of sites, or fewer sites. If site formation is tied mainly to elongation, redistributing a fixed mass among more fibrils need not increase the total site burden. If each fibril-birth event contributes additional catalytic sites, the opposite can occur. Fibril length itself is therefore not the mechanistic variable of interest; the relevant quantities are site formation during fibril birth and growth, the fraction of sites that remains exposed, and the free monomer available to use those sites.

The present study develops that mechanistic chain explicitly. The coupled tau–Aβ model is extended by introducing the concentration of catalytically active Aβ secondary-nucleation sites as an independent state variable and by allowing only the geometrically exposed subset of those sites to catalyze homotypic secondary nucleation. The analysis asks, in sequence, four questions: when does producing more fibrils change the active-site density at matched fibril mass; when can a gain in accessible sites overcome tau-induced monomer depletion; what controls the sign of the homotypic response at a given time; and which conclusions survive when growth history, masking, oligomer dissociation, soluble-protein clearance, fragmentation, fibril turnover, and alternative accessibility laws are introduced. No assumption is made that shorter fibrils are intrinsically more catalytic.

Relative to the preceding model [4], the new ingredients are an active-site balance, a secondary-nucleation coefficient defined per accessible active site, a growth-history-dependent site yield, and the application of plaque-surface shielding to active sites rather than directly to fibril mass. The paper is organized as follows. Section 2 formulates the model; Section 3 derives exact and quasi-steady results that identify what must be true for shorter fibrils to become more catalytic; Section 4 tests those conditions numerically and reveals a switch between clearance-controlled and elongation-controlled monomer regimes; and Section 5 relates that switch to experiment and identifies the measurements that would decide whether tau-seeded Aβ fibrils are intrinsically different catalysts.

## 2. Mathematical model

## 2.1. Coupled geometry and tau subsystem

The model retains the two-compartment geometry of the preceding formulation: tau aggregation is represented within a neuronal soma of volume $V_{tau}$, whereas Aβ aggregation occurs in the surrounding effective extracellular/interstitial volume $V_{A\beta}$. The purpose of this separation is not merely geometric. It forces the tau-to-Aβ coupling to be expressed as an explicitly rescaled, effective exposure of extracellular Aβ to intracellularly generated tau fibrillar material.

The two control volumes therefore play different roles. Tau concentrations are defined per unit somatic volume, whereas all Aβ concentrations are defined per unit extracellular/interstitial volume. Before somatic tau fibril mass can enter an Aβ nucleation rate, it is multiplied by the soma-to-interstitium volume ratio. Fig. 1 summarizes this geometry and, more importantly, the flow of causality in the model: tau fibrils drive heterotypic Aβ nucleation; Aβ fibril birth and elongation create catalytic sites; plaque consolidation controls which sites remain exposed; and the exposed sites drive homotypic secondary nucleation.

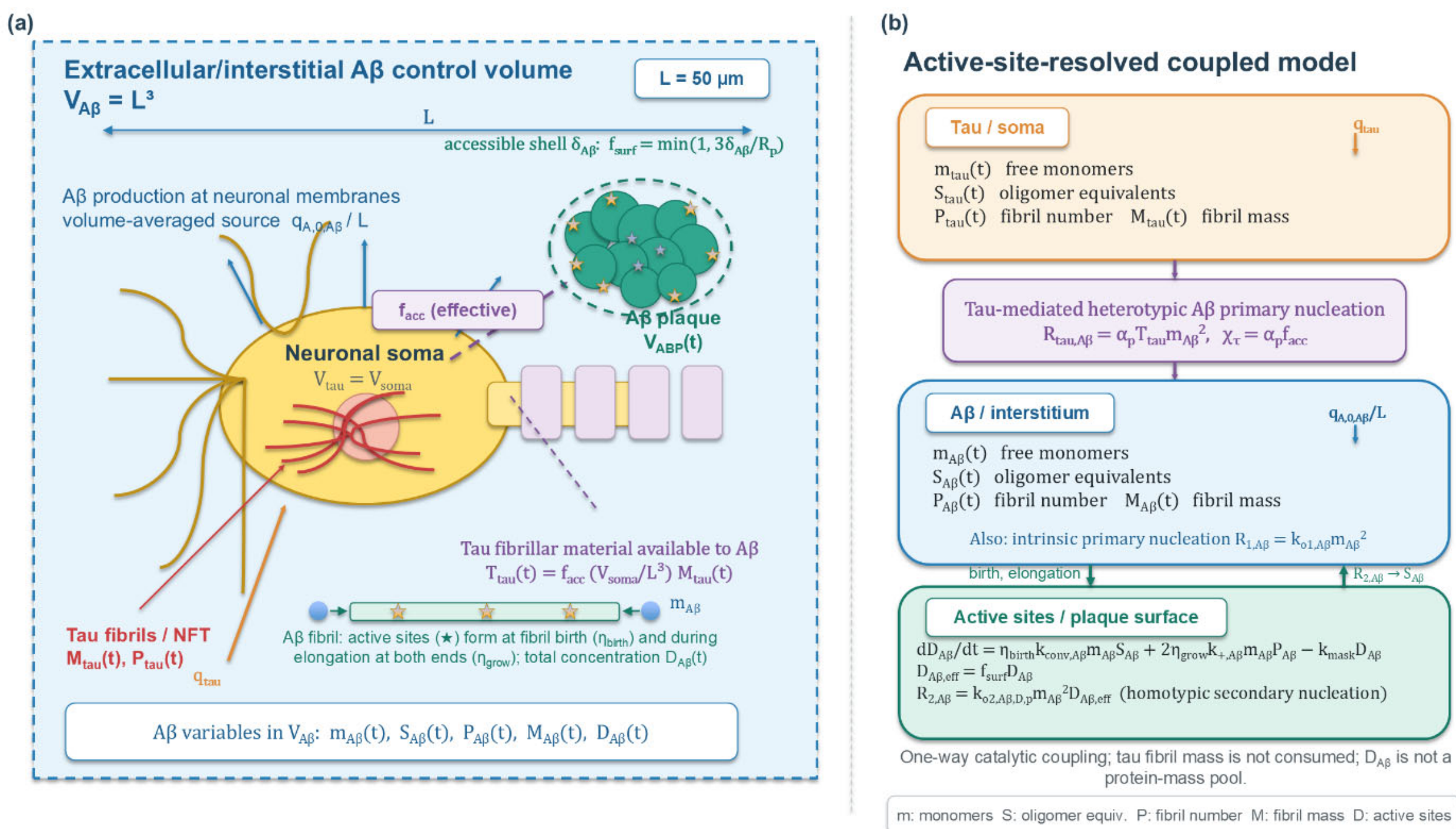

**Fig. 1.** Definition sketch of the active-site-resolved coupled tau–Aβ model. (a) Geometry. Tau aggregates inside a representative neuronal soma of volume $V_{tau} = V_{soma}$; Aβ is produced at neuronal membranes and aggregates in the surrounding extracellular/interstitial control volume $V_{A\beta} = L^3$. An effective fraction $f_{acc}$ of the tau fibrillar material, representing material released, exposed, or otherwise available to extracellular Aβ rather than penetration of Aβ into the soma, catalyzes heterotypic Aβ nucleation with the effective coupling coefficient $\chi_\tau = \alpha_p f_{acc}$. Active secondary-

nucleation sites (stars) are created when an Aβ fibril is born (yield $\eta_{birth}$) and as monomers add to both fibril ends (yield $\eta_{grow}$). In a compact plaque, only the sites within a thin outer shell of thickness $\delta_{A\beta}$ remain exposed (gold stars); interior sites are buried (gray stars). (b) Model structure. The tau subsystem [Eqs. (S32)–(S35)] drives heterotypic Aβ primary nucleation [Eqs. (1) and (3)]; fibril birth and elongation in the Aβ subsystem [Eqs. (9)–(12)] generate active sites [Eq. (13)]; the accessible sites $D_{A\beta,eff} = f_{surf} D_{A\beta}$ [Eqs. (15)–(18)] set the homotypic secondary-nucleation rate $R_{2,A\beta}$ [Eq. (4)], which feeds the Aβ oligomer pool.

The tau subsystem follows free monomer $m_{tau}$, soluble on-pathway oligomer equivalents $S_{tau}$, fibril number $P_{tau}$, and fibril mass $M_{tau}$ (Table S5). Its kinetics include primary and fibril-surface secondary nucleation, oligomer dissociation and conversion, elongation, fragmentation, and protein turnover. Because the present paper asks what happens downstream of tau-catalyzed Aβ nucleation, these tau equations are not modified.

The coupling is therefore one-way: tau changes Aβ kinetics, whereas Aβ does not feed back on tau. The four tau balances are reproduced in Section S3 of Supplementary Materials [Eqs. (S32)–(S35)] for completeness. Simulations begin at the onset of the modeled AD pathology with zero initial tau concentrations, and tau influences the Aβ subsystem only through the accessible tau fibrillar-material concentration defined below.

### 2.2. Aβ kinetics with explicit catalytic sites

#### 2.2.1. How tau enters the Aβ compartment

Tau fibrillar material that Aβ can reach is represented by the effective concentration

$$T_{tau}(t) = f_{acc} \frac{V_{soma}}{L^3} M_{tau}(t) \tag{1}$$

Eq. (1) is a product of three factors. The first, the accessibility fraction $f_{acc}$, selects the part of the tau fibrillar material that is available to interstitial Aβ. The second, the volume ratio $V_{soma}/L^3$, rescales a concentration defined in the soma to one defined in the Aβ control volume. The third, $M_{tau}(t)$, is the total tau fibril-mass concentration in the soma. Multiplying them yields $T_{tau}(t)$, the concentration of tau fibrillar material available to catalyze heterotypic Aβ nucleation.

The accessibility fraction $f_{acc}$ is an effective, phenomenological quantity. Tau fibrils form inside neurons, whereas the Aβ considered here is extracellular, so $f_{acc}$ does not describe penetration of Aβ into an intact soma. It lumps together routes by which tau fibrillar material becomes available to extracellular Aβ, including tau-aggregate release and fibril exposure through membrane damage, as well as the extracellular neurofibrillary tangles documented in the AD hippocampus [6]. The parameter expresses the resulting available material as a fraction of the somatic tau fibril burden. Tau release, membrane disruption, extracellular transport, and uptake are not modeled explicitly (Section 5).

The heterotypic coefficient depends on the tau polymorph $p$ and is written $\alpha_p$; PHF and CTE tau fibrils are assigned different values, reflecting the different catalytic activities measured for the two folds [3].

Because the coupling is one-way, $\alpha_p$ and $f_{acc}$ enter the Aβ equations only through their product. It is therefore convenient to define the effective heterotypic coupling coefficient $\chi_\tau = \alpha_p f_{acc}$, so that $R_{tau,A\beta} = \chi_\tau(V_{soma}/L^3)M_{tau}m_{A\beta}^2$ in Eq. (3). In the present model, $\chi_\tau$ is the only parameter through which tau accessibility and tau polymorph affect Aβ: the PHF and CTE folds are experimental mappings onto $\chi_\tau$ ($\chi_\tau = 1.86\times10^{-9}$ and $4.45\times10^{-9}$ $\mu M^{-2}\ s^{-1}$ for $f_{acc} = 1$), not distinct mechanisms. Results are therefore reported for PHF and CTE coupling at given values of $f_{acc}$, which are simply points on the $\chi_\tau$ axis; the corresponding values of $\chi_\tau$ are given in the figure captions and tables, where the uncoupled case is denoted $\chi_\tau = 0$.

**2.2.2. Why fibril mass and catalytic activity must be separated**

The Aβ subsystem tracks five variables: the free-monomer concentration $m_{A\beta}$, soluble on-pathway oligomer-equivalent concentration $S_{A\beta}$, fibril-number concentration $P_{A\beta}$, fibril-mass concentration $M_{A\beta}$, and concentration of active secondary-nucleation sites $D_{A\beta}$. All five quantities are expressed in μM, with $D_{A\beta}$ interpreted as micromoles of active sites per liter.

The added state variable $D_{A\beta}$ is the molar concentration of sites capable of catalyzing Aβ secondary nucleation. It is a kinetic descriptor, not an additional protein-mass pool. Experiments indicate that these sites are associated predominantly with rare structural growth defects [5], but the model deliberately tracks catalytic sites rather than assigning each site to a unique microscopic defect. The derived quantity $D_{A\beta,eff}$ represents the subset of sites that remains accessible to interstitial Aβ after fibrils consolidate into a plaque.

This separation creates the central distinction of the paper. Fibril mass tells us how much Aβ is incorporated into fibrils; active-site concentration tells us how much catalytic capacity those fibrils contain; and plaque accessibility tells us how much of that capacity can actually contact soluble monomer. Homotypic secondary nucleation is therefore driven by accessible active sites rather than by total fibril mass.

### 2.2.3. Three nucleation routes and fibril growth

The three Aβ nucleation pathways, namely intrinsic primary nucleation, heterotypic nucleation on tau, and homotypic secondary nucleation on Aβ fibril surfaces, proceed at the rates

$$R_{1,A\beta} = k_{o1,A\beta} m_{A\beta}^2 \quad (2)$$

$$R_{tau,A\beta} = \alpha_p T_{tau} m_{A\beta}^2 \quad (3)$$

$$R_{2,A\beta} = k_{o2,A\beta,D,p} m_{A\beta}^2 D_{A\beta,eff} \quad (4)$$

Eqs. (2)–(4) define the three distinct Aβ nucleation rates. In $R_{1,A\beta}$, $k_{o1,A\beta}$ is the intrinsic primary-nucleation coefficient and $m_{A\beta}^2$ supplies the effective second-order dependence on free Aβ monomer. In $R_{tau,A\beta}$, $\alpha_p$ is the tau-polymorph-dependent catalytic coefficient, $T_{tau}$ is the accessible tau fibrillar-material concentration, and $m_{A\beta}^2$ is the monomer factor. In $R_{2,A\beta}$, $k_{o2,A\beta,D,p}$ is the rate coefficient per accessible Aβ active site, $m_{A\beta}^2$ is the monomer factor, and $D_{A\beta,eff}$ is the accessible active-site concentration.

The exponent 2 on the monomer concentration is an effective reaction order; it does not mean that each nucleation event consumes exactly two monomers. As in the coarse-grained description used previously, a newly formed on-pathway oligomer is counted as one activated-monomer equivalent in the mass balance.

For reference, the remaining kinetic contributions entering the balances are

$$\text{Conversion rate} = k_{\mathrm{conv},A\beta} m_{A\beta} S_{A\beta} \quad (5)$$

$$\text{Elongation rate} = 2k_{+,A\beta} m_{A\beta} P_{A\beta} \quad (6)$$

$$\text{Fragmentation rate} = k_{-,A\beta} M_{A\beta} \quad (7)$$

$$\text{Dissociation rate} = \left(k_{d1,A\beta} + k_{d2,A\beta} M_{A\beta}\right) S_{A\beta} \quad (8)$$

Eqs. (5)–(8) define the four kinetic contributions used below. The conversion term $k_{conv,A\beta} m_{A\beta} S_{A\beta}$ describes an oligomer combining with a monomer to form an elongation-competent fibril. The elongation term $2k_{+,A\beta} m_{A\beta} P_{A\beta}$ describes monomer addition at both ends of all Aβ fibrils; the factor 2 counts the two ends. The fragmentation term $k_{-,A\beta} M_{A\beta}$ creates additional fibrillar species from existing fibrillar mass. The dissociation term $\left(k_{d1,A\beta} + k_{d2,A\beta} M_{A\beta}\right) S_{A\beta}$ combines spontaneous oligomer dissociation with fibril-catalyzed dissociation.

The dissociation rate is retained in its original mass-dependent form. The present model does not assume that the sites responsible for secondary nucleation are necessarily the same sites responsible for oligomer dissociation. Earlier work [7] demonstrated that fibril surfaces can catalyze Aβ oligomer dissociation, but the relationship between the structural sites responsible for oligomer formation and dissociation remains to be established.

**2.2.4. Monomer, oligomer, fibril, and active-site balances**

The free-Aβ-monomer balance is

$$\frac{dm_{A\beta}}{dt} = -R_{1,A\beta} - R_{2,A\beta} - \alpha_p T_{tau} m_{A\beta}^2 + k_{d1,A\beta} S_{A\beta} + k_{d2,A\beta} S_{A\beta} M_{A\beta}$$

$$-2k_{+,A\beta} m_{A\beta} P_{A\beta} - k_{conv,A\beta} m_{A\beta} S_{A\beta} - m_{A\beta} \frac{\ln(2)}{T_{1/2,m,A\beta}} + \frac{q_{A,0,A\beta}}{L} \quad (9)$$

Free Aβ monomer in Eq. (9) is depleted by the three nucleation channels, namely intrinsic primary nucleation, homotypic secondary nucleation at accessible sites, and heterotypic nucleation on tau (first three terms), and is restored when oligomers dissociate spontaneously or on fibril surfaces (fourth and fifth terms). Further losses arise from monomer addition at the two tips of every fibril, which accounts for the factor 2 in the sixth term, from the single monomer consumed each time an oligomer is converted into an elongation-competent fibril (seventh term), and from effective first-order monomer clearance/turnover with rate constant $\ln(2)/T_{1/2,m,A\beta}$. Monomer production enters through the volume-averaged source $q_{A,0,A\beta}/L$ (last term).

Soluble on-pathway Aβ oligomers obey

$$\frac{dS_{A\beta}}{dt} = R_{1,A\beta} + R_{2,A\beta} + \alpha_p T_{tau} m_{A\beta}^2 - k_{d1,A\beta} S_{A\beta} - k_{d2,A\beta} S_{A\beta} M_{A\beta}$$

$$-k_{conv,A\beta} m_{A\beta} S_{A\beta} - S_{A\beta} \frac{\ln(2)}{T_{1/2,S,A\beta}} \quad (10)$$

Each nucleation event that removes a monomer in Eq. (9) adds an oligomer in Eq. (10), so the first three terms of Eq. (10) are the nucleation terms of Eq. (9) with the sign reversed. Oligomers are lost by spontaneous and fibril-catalyzed dissociation (fourth and fifth terms), by conversion into elongation-competent fibrils through monomer addition (sixth term), and by effective first-order oligomer clearance/turnover with rate constant $\ln(2)/T_{1/2,S,A\beta}$. The Finke–Watzky model provides a minimal coarse-grained A→B, A+B→2B description in which A is monomeric protein and B is an aggregated/product state [8]; related applications have also interpreted a conformationally activated or misfolded monomer as the nucleating state [9]. In the present model, assigning one modeled oligomer unit one monomer equivalent of mass is a bookkeeping convention, not a statement about the molecular mass of a physical oligomer.

Consequently, $S_{A\beta}$ is not the molar concentration of physical oligomer particles but a soluble oligomer-equivalent concentration, expressed in activated-monomer equivalents: it measures the Aβ held in soluble on-pathway oligomeric intermediates, not the number of oligomers. The same qualification applies to the accumulated exposure $\Xi_{A\beta}$ defined in Eq. (20).

The Aβ fibril-number concentration obeys

$$\frac{dP_{A\beta}}{dt} = k_{conv,A\beta} m_{A\beta} S_{A\beta} + k_{-,A\beta} M_{A\beta} \tag{11}$$

Eq. (11) describes the two processes that increase the number of Aβ fibrils. Conversion of an oligomer into an elongation-competent fibril by monomer addition proceeds at a rate set by the coefficient $k_{conv,A\beta}$ and the reactant product $m_{A\beta} S_{A\beta}$. Breakage of existing fibrils is governed by the fragmentation coefficient $k_{-,A\beta}$ acting on the fibrillar mass $M_{A\beta}$. Elongation does not appear, since it lengthens fibrils without changing how many there are.

The fibril-mass concentration obeys

$$\frac{dM_{A\beta}}{dt} = 2k_{+,A\beta} m_{A\beta} P_{A\beta} + 2k_{conv,A\beta} m_{A\beta} S_{A\beta} \tag{12}$$

Aβ fibril mass in Eq. (12) increases through elongation (first term), which depends on the elongation coefficient $k_{+,A\beta}$, the free-monomer concentration $m_{A\beta}$, and the fibril-number concentration $P_{A\beta}$, with a factor 2 for the two growing tips. Mass is also added whenever a monomer and an oligomer combine into a new fibril (second term), at a rate governed by $k_{conv,A\beta}$ and $m_{A\beta} S_{A\beta}$; here the factor 2 arises because both the monomer and the oligomer, itself one activated-monomer equivalent, become part of the fibril. Turnover of fibrils and plaques is not included in Eqs. (11)–(13); its effect is examined in Section 4.5 by adding first-order removal of $P_{A\beta}$, $M_{A\beta}$, and $D_{A\beta}$ with a common half-life $T_{1/2,F}$.

The new equation describing the concentration of active sites is

$$\frac{dD_{A\beta}}{dt} = \eta_{birth} k_{conv,A\beta} m_{A\beta} S_{A\beta} + 2\eta_{grow} k_{+,A\beta} m_{A\beta} P_{A\beta} - k_{mask} D_{A\beta} \tag{13}$$

The first term on the right-hand side of Eq. (13) creates active sites when a new Aβ fibril is born: $\eta_{birth}$ is the mean number of sites introduced per conversion event and $k_{conv,A\beta} m_{A\beta} S_{A\beta}$ is the fibril-birth rate. The second term creates sites during elongation: $\eta_{grow}$ is the mean site yield per incorporated monomer, $k_{+,A\beta} m_{A\beta} P_{A\beta}$ is the one-end elongation rate, and the factor 2 counts the two fibril ends. The final term removes catalytic activity at the first-order masking or deactivation rate $k_{mask}D_{A\beta}$ without removing fibril mass.

Eq. (13) allows the model to ask whether two fibril populations of equal mass can have different catalytic histories. Active sites can enter in two ways: when a new fibril becomes growth competent and as monomers are subsequently incorporated during elongation. The corresponding yields, $\eta_{birth}$ and $\eta_{grow}$, are new effective parameters and have not been measured for PHF- or CTE-tau-seeded Aβ42 fibrils.

Fragmentation is kept conceptually separate. In the baseline model it increases fibril number without automatically creating new active sites, because a newly exposed fibril end need not be a secondary-nucleation site. Section 4.5 tests the alternative possibility by adding $\eta_{frag}$ fragmentation-generated sites per fragmentation event ($\eta_{frag} = 0$ in all other simulations). This distinction is important: a process can make fibrils more numerous without necessarily making the fibril population more catalytic.

**2.3. How fibril growth history can change active-site formation**

Recent measurements [5] suggest that growth conditions can alter the number of catalytic sites incorporated into a fibril. The reference simulations therefore begin with a constant elongation-site yield, but the sensitivity analysis also allows that yield to depend on the monomer environment experienced during growth.

$\eta_{grow}$ is constant in the reference simulations; in the growth-history study of Section S2.1, its dependence on monomer supersaturation is represented by

$$\eta_{grow}(u) = \eta_{min} + \frac{(\eta_{max}-\eta_{min})u^h}{K_u^h+u^h}, \qquad u = max[0, \frac{m_{A\beta}-m_{eq,A\beta}}{m_{eq,A\beta}}] \qquad (14)$$

In Eq. (14), $\eta_{min}$ is the baseline elongation-site yield, $\eta_{max} - \eta_{min}$ the largest possible increment, $K_u$ the characteristic supersaturation, and $h$ the Hill exponent; $u$ is the fractional excess of free Aβ monomer over its equilibrium concentration $m_{eq,A\beta}$ and is zero at or below equilibrium. Physically, $u$ measures the driving force for elongation: far above equilibrium, monomers add rapidly and are more likely to be incorporated in imperfect, kinetically trapped conformations that can act as catalytic sites, whereas growth close to equilibrium is slower and allows more ordered packing. Eq. (14) therefore assigns higher site yields to fibril segments grown at high supersaturation.

Eq. (14) is intentionally a constitutive hypothesis rather than an experimentally established law for tau-seeded Aβ42. Its purpose is to ask whether a memory of growth conditions could change the active-site yield enough to alter the conclusions. The possible link between the modeled sites and kinetically trapped structural defects is motivated by recent active-site experiments [5]. A quantitatively predictive treatment close to equilibrium would also require reversible elongation, which is outside the present model.

**2.4. Observables and numerical protocol**

The total active-site density, plaque-surface accessibility factor, accessible active-site concentration, and equivalent plaque radius are defined as

$$\sigma_{D,A\beta}(t) = \frac{D_{A\beta}(t)}{M_{A\beta}(t)}, \qquad M_{A\beta}(t) > 0 \tag{15}$$

$$f_{surf}(t) = min[1, \frac{3\delta_{A\beta}}{R_p(t)}] \tag{16}$$

$$D_{A\beta,\text{eff}}(t) = f_{surf}(t) D_{A\beta}(t) \tag{17}$$

$$R_p(t) = \left[\frac{3V_{ABP}(t)}{4\pi}\right]^{1/3} \tag{18}$$

Eqs. (15)–(18) translate the modeled active-site burden into the catalytic capacity that is actually available to soluble Aβ. Eq. (15) gives the active-site density per fibril-incorporated monomer, whereas Eq. (17) gives the accessible-site concentration after geometric shielding is applied. The shielding itself is defined by Eq. (16). We represent the compact deposit as an equivalent sphere of radius $R_p$ [Eq. (18)] and assume that active sites are distributed throughout its volume, but that only a thin outer shell of thickness $\delta_{A\beta}$—approximately one fibril diameter (Table 1; Section S1.6)—is exposed to the surrounding interstitial fluid. Sites deeper in the deposit are treated as buried and therefore do not catalyze secondary nucleation. For $\delta_{A\beta} \ll R_p$, the exposed shell occupies $[R_p^3 - (R_p - \delta_{A\beta})^3]/R_p^3 \approx 3\delta_{A\beta}/R_p$ of the total volume, which gives the large-deposit limit in Eq. (16). The minimum with unity makes the same expression physically consistent for very small deposits: when $R_p < 3\delta_{A\beta}$ (0.03 μm for the reference shell thickness), essentially all active sites remain exposed. As the deposit grows, however, $R_p \propto M_{A\beta}^{1/3}$ and $f_{surf} \propto M_{A\beta}^{-1/3}$; thus, in the mass-proportional limit, the accessible active-site burden grows only as $M_{A\beta}^{2/3}$ rather than linearly with fibril mass. This geometric shielding becomes strong quickly in the reference simulation: $f_{surf}$ falls below 0.04 within four months and reaches $8.4\times10^{-3}$ at 33 years (Section 4.2 and Fig. 3a). Eq. (16) is therefore a closure: an assumed relation that represents unresolved physics—here, the packing of fibrils in a plaque—through a variable that the model does resolve, namely fibril mass; alternative accessibility closures are examined in Section S2.4. Eqs. (16) and (18) retain the geometry of the preceding model, but the shielding factor now acts on the active-site concentration rather than directly on fibril mass. The protein-equivalent compact-deposit volume is obtained from the fibrillar Aβ mass as $V_{ABP}(t) = 2L^3\, MW_{A\beta}\, M_{A\beta}(t)/(a_{21}\, \rho_{A\beta})$ [4] [Eq. (53) therein], with $MW_{A\beta}$ and $\rho_{A\beta}$ (a conventional protein-material density rather than a measured bulk-plaque density) given in Table S1; the factor 2 arises because the plaque is assumed to lie between two neighboring neurons that both contribute Aβ monomers.

The average fibril size, in monomer units, is

$$\bar{n}_F(t) = \frac{M_{A\beta}(t)}{P_{A\beta}(t)}, \qquad P_{A\beta}(t) > 0 \tag{19}$$

In Eq. (19), $\bar{n}_F$ is the average number of monomer units per fibril; its conversion into a physical mean fibril length requires an independently established relationship between monomer number and fibril length for the relevant Aβ polymorph.

Accumulated soluble Aβ oligomer-equivalent exposure is defined by

$$\Xi_{A\beta}(t) = \int_0^t S_{A\beta}\,(t')\,dt' \tag{20}$$

$\Xi_{A\beta}$ is a model diagnostic with units of μM s, not a directly measurable quantity. Ratios of $\Xi_{A\beta}$ between simulations, the main quantities reported in Section 4, do not depend on the mapping between oligomer equivalents and physical particles, provided that the mean oligomer size and detectability do not depend on tau coupling; absolute values of $\Xi_{A\beta}$ should not be compared directly with measured oligomer concentrations.

The equations are solved with the MATLAB stiff solver ode15s at relative and absolute tolerances of $10^{-10}$, reusing the tau-subsystem code of the preceding model [4]. The accumulated exposure in Eq. (20) is evaluated by cumulative trapezoidal quadrature on $2\times10^4$ uniformly spaced output times, ten times the minimum of $2\times10^3$ output times required for convergence when oligomer dissociation is included, as described previously [4]. The accessibility factor of Eqs. (15)–(18) is applied only to homotypic secondary nucleation, not to the $D_{A\beta}$ balance, elongation, dissociation, or protein-mass balance. No Michaelis–Menten monomer-saturation term is introduced because the modeled free-monomer concentration is far below the reported Aβ42 secondary-nucleation saturation range. The MATLAB scripts that generate all numerical results are described in Section S5 of the Supplementary Materials; each 33-year simulation takes less than one second.

The reference numerical protocol uses $\chi_\tau = 0$, $1.86\times10^{-10}$, and $1.86\times10^{-9}$ μM$^{-2}$ s$^{-1}$ ($f_{acc} = 0$, 0.1, and 1 with the PHF heterotypic coefficient), $\eta_{birth} = 2\sigma_0$, $\eta_{grow} = \sigma_0$, $k_{mask} = 0$, $k_{-,A\beta} = 0$, and both oligomer dissociation and soluble-protein clearance switched off. This is the mass-proportional reference case. The same calculations are then repeated with the CTE coefficient $\alpha_{CTE} = 4.45\times10^{-9}$ μM$^{-2}$ s$^{-1}$ ($\chi_\tau = 4.45\times10^{-10}$ and $4.45\times10^{-9}$ μM$^{-2}$ s$^{-1}$) while all other parameters are held fixed. Eq. (14) is used only in the growth-history study.

For simulations using zero-Aβ-aggregate initial conditions,

$$m_{A\beta}(0) = S_{A\beta}(0) = P_{A\beta}(0) = M_{A\beta}(0) = D_{A\beta}(0) = 0 \tag{21}$$

The dependent and derived variables are summarized in Table S5, and the principal parameters of the active-site extension and of the tau–Aβ coupling in Table 1; the complete parameter set, with sources and derivations, is given in Table S1 and Section S1 of the Supplementary Materials. These sources comprise plaque-spacing and Aβ-production estimates for the geometry and source terms ([10]; [11]; [12]; [13]), a conventional protein density ([14]), measurements of tau sorting and turnover ([15]; [16]; [17]; [18]; [19]), tau aggregation kinetics ([20]; [21]; [22]; [23]; [24]), oligomer-exchange time scales ([25]; [26]), Aβ42 aggregation and clearance kinetics ([27]; [28]; [29]; [30]), and the clinical timeline that sets the simulated interval ([31]; [32]; [33]; [34]). Inherited parameters retain the conventions of the preceding model [4]; new active-site-related coefficients are distinguished from experimentally measured quantities.

Table 1. Principal parameters of the active-site extension and of the tau–Aβ coupling. The last column gives the values used in the reference case and the ranges examined in the sensitivity studies. Inherited kinetic parameters, sources, and derivations are given in Table S1 and Section S1 of the Supplementary Materials.

| Symbol | Definition | Units | Value/range | Value used / status |
|---|---|---|---|---|
| **Tau–Aβ coupling and plaque accessibility** | | | | |
| $f_{acc}$ | Effective fraction of tau fibril mass available to extracellular Aβ (phenomenological) | — | 0–1 | 0, 0.1, 1; extend as needed |
| $\alpha_p$ | Tau-fold-dependent Aβ heterotypic nucleation coefficient (PHF/CTE) | $\mu M^{-2}\ s^{-1}$ | PHF $1.86\times10^{-9}$; CTE $4.45\times10^{-9}$ | Fold-specific; inherited |
| $\chi_\tau$ | Effective heterotypic coupling coefficient, $\alpha_p f_{acc}$ | $\mu M^{-2}\ s^{-1}$ | 0–$4.45\times10^{-9}$ | 0; $1.86\times10^{-10}$; $4.45\times10^{-10}$; $1.86\times10^{-9}$; $4.45\times10^{-9}$ |
| $\delta_{A\beta}$ | Thickness of the catalytically accessible surface layer of an Aβ plaque | μm | 0.010 | 0.010; 0.0025–0.1 (Section 4.5) |
| **New active-site-resolved secondary-nucleation parameters** | | | | |
| $\sigma_{ref} = D_{A\beta,ref}/M_{A\beta,ref}$ | Measured reference density of active Aβ42 sites, $\sigma_{ref}$ | sites/monomer | ≈1/143 ≈ $6.99\times10^{-3}$ | Reference only; not tau-seeded value |
| $k_{o2,A\beta,M}^{ref}$ | Mass-based Aβ secondary-nucleation coefficient for comparison | $\mu M^{-2}\ s^{-1}$ | $1.00\times10^{-8}$ | Calibration comparison, not used in the new equations |
| $k_{o2,A\beta,D,p}$ | Secondary-nucleation coefficient per accessible active site in $R_{2,A\beta} = k_{o2,A\beta,D,p} m_{A\beta}^2 D_{A\beta,eff}$ | $\mu M^{-2}\ s^{-1}$ | Unmeasured; conditional ≈$1.43\times10^{-6}$ | Fit to matched experiments; no universal value |
| $\sigma_0$ | Reference active-site density used in limiting case | sites/monomer | $\sigma_{ref}$ as illustrative choice | 1/143 only for matched reference case |
| $\eta_{birth}$ | Mean number of active sites introduced per new fibril during conversion | sites/fibril | Unmeasured | $2\sigma_0$ in mass-proportional limit; 0–$2\times10^6\sigma_0$ (Section 4.3) |
| $\eta_{grow}$ | Mean number of active sites introduced per monomer incorporated by elongation | sites/monomer | Unmeasured | $\sigma_0$ in mass-proportional limit; varies with $u$ via Eq. (14) (Section S2.1) |
| $k_{mask}$ | Intrinsic loss or masking rate of catalytic-site activity without fibril-mass loss; separate from geometric shielding | $s^{-1}$ | Unmeasured | 0 reference; $10^{-10}$–$10^{-7}$ (Section S2.2) |

## 3. Analytical results: what must be true for shorter fibrils to be more catalytic?

Before assigning values to the new active-site parameters, the equations themselves answer several parts of the central question. The results below establish a sequence of increasingly restrictive tests: first, when the new model collapses exactly to the old mass-based model; second, when more fibrils at the same mass can change the number of active sites; third, whether any site gain can overcome the quadratic loss caused by monomer depletion; and finally, how monomer clearance changes the answer.

### 3.1. Benchmark: recovery of the mass-proportional model

Consider first the limit in which sites are introduced at a constant frequency per incorporated Aβ monomer, with no intrinsic masking or fragmentation-generated sites, and assume that

$$\eta_{\mathrm{grow}} = \sigma_0, \qquad \eta_{\mathrm{birth}} = 2\sigma_0, \qquad D_{A\beta}(0) = \sigma_0 M_{A\beta}(0) \tag{22}$$

In Eq. (22), the site yield per fibril-birth event is $2\sigma_0$ because each conversion event adds two monomer equivalents to fibril mass, and the initial site density is also $\sigma_0$.

Under these conditions, Eqs. (12) and (13) give $D_{A\beta}(t) = \sigma_0 M_{A\beta}(t)$ at all times.

To verify this, multiply Eq. (12) by $\sigma_0$ and subtract the result from Eq. (13) with $k_{mask} = 0$. This gives $d(D_{A\beta} - \sigma_0 M_{A\beta})/dt = (\eta_{birth} - 2\sigma_0)k_{conv,A\beta}m_{A\beta}S_{A\beta} + 2(\eta_{grow} - \sigma_0)k_{+,A\beta}m_{A\beta}P_{A\beta}$, which vanishes identically under Eq. (22). Hence $D_{A\beta} - \sigma_0 M_{A\beta}$ retains its initial value of zero. Fragmentation does not affect this result because it changes $P_{A\beta}$ but neither $M_{A\beta}$ nor $D_{A\beta}$ when it creates no active sites ($\eta_{frag} = 0$; Section 4.5).

Consequently, after the geometric accessibility mapping in Eqs. (15)–(18), the secondary-nucleation rate becomes

$$R_{2,A\beta} = k_{o2,A\beta,D,p} m_{A\beta}^2 D_{A\beta,eff} = \left(k_{o2,A\beta,D,p}\sigma_0\right) m_{A\beta}^2 M_{A\beta} f_{surf} \tag{23}$$

where $f_{surf}$ is the exposed plaque fraction of Eq. (16). This limiting case is more than a consistency check. It defines the null hypothesis for the rest of the paper: if sites are introduced uniformly per incorporated monomer, the active-site variable contains no information beyond fibril mass. Any genuinely new effect of fibril shortening must therefore arise from a departure from mass-proportional site formation, from a change in site accessibility or activity, or from a change in the free monomer that reaches those sites.

### 3.2. Matched-mass criterion: when do more fibrils carry more active sites?

A central result follows by considering constant $\eta_{birth}$ and $\eta_{grow}$, no active-site masking ($k_{mask} = 0$), and no fragmentation ($k_{-,A\beta} = 0$, as in the reference case). Integrating Eq. (13) and using the fibrillar mass and number balances yields

$$D_{A\beta}(t) - D_{A\beta}(0) = \eta_{grow}\left[M_{A\beta}(t) - M_{A\beta}(0)\right] + \left(\eta_{birth} - 2\eta_{grow}\right)\left[P_{A\beta}(t) - P_{A\beta}(0)\right] \tag{24}$$

Eq. (24) separates the history of a fibril population into two contributions. Elongation assigns a baseline number of sites to the accumulated fibril mass; fibril birth then adds or subtracts a correction according to whether a birth event contributes more or fewer sites than would be expected from the two monomer equivalents incorporated at conversion.

This immediately gives three regimes. If $\eta_{birth} = 2\eta_{grow}$, catalytic-site concentration is determined by fibril mass alone, so redistributing a fixed mass among more fibrils changes nothing. If $\eta_{birth} > 2\eta_{grow}$, fibril birth carries an excess site yield and more fibrils can be more catalytic at the same mass. If $\eta_{birth} < 2\eta_{grow}$, creating more fibrils can actually lower the total site burden. The experimentally observed shortening of tau-seeded Aβ fibrils therefore has no unique implication for secondary nucleation until the relative site yields of fibril birth and elongation are known.

For two systems with identical initial conditions and the same constant site-incorporation coefficients, evaluated at the same fibrillar mass and with $k_{mask} = k_{-,A\beta} = 0$, Eq. (24) gives

$$\Delta D_{A\beta} = \left(\eta_{birth} - 2\eta_{grow}\right)\Delta P_{A\beta} \tag{25}$$

The same result can be written in a more intuitive form by dividing by fibril mass. For zero initial Aβ aggregates, the active-site density equals the elongation-site yield plus a birth-related correction that is inversely proportional to the mean fibril size. As fibrils become long, the correction becomes small. Consequently, fibril shortening changes the site density appreciably only when each birth event introduces a site burden comparable to that accumulated during elongation over an entire mean-length fibril.

This scaling also explains why an apparently large per-birth enhancement may still have little effect in a tissue-scale model with long fibrils. Extremely large birth yields are not ruled out experimentally, because $\eta_{birth}$ has not been measured, but they would shift the interpretation from a small nascent nucleus carrying a few fixed sites toward a lineage effect in which fibrils nucleated on tau retain a different site yield throughout their growth. Section S2.6 tests that alternative explicitly.

### 3.3. More sites are not enough: competition with monomer depletion

Even if tau-mediated nucleation were to increase the number of active sites, secondary nucleation could still fall. Tau coupling creates more growing fibrils and therefore consumes free Aβ monomer more rapidly, while plaque consolidation can hide sites from the surrounding solution. Because the homotypic secondary-nucleation rate is linear in accessible sites but quadratic in free monomer, a modest loss of monomer can outweigh a substantial gain in catalytic surface.

For tau-coupled and tau-uncoupled systems with the same secondary-nucleation coefficient per accessible active site,

$$\frac{R_{2,A\beta}^{(tau)}}{R_{2,A\beta}^{(0)}} = \left(\frac{m_{A\beta}^{(tau)}}{m_{A\beta}^{(0)}}\right)^2 \frac{D_{A\beta,eff}^{(tau)}}{D_{A\beta,eff}^{(0)}} \tag{26}$$

Consequently, tau-mediated nucleation increases the instantaneous homotypic Aβ secondary-nucleation rate only if

$$\frac{D_{A\beta,eff}^{(tau)}}{D_{A\beta,eff}^{(0)}} > \left(\frac{m_{A\beta}^{(0)}}{m_{A\beta}^{(tau)}}\right)^2 \tag{27}$$

Eq. (27) is the corresponding decision rule: tau raises homotypic secondary nucleation only when the gain in exposed catalytic sites is larger than the squared penalty associated with the drop in free monomer. This inequality turns the qualitative question 'are there more sites?' into the mechanistic question 'are there enough additional exposed sites to compensate for monomer depletion?'

At matched fibril mass, the geometric exposure factor is identical in the two systems because it depends only on mass. In the mass-proportional limit, the total active-site concentrations are also identical. Under those conditions tau cannot increase homotypic secondary nucleation by creating more fibrils; it can do so only if the free monomer rises, or if tau-seeded fibrils differ in site density, per-site activity, or accessibility.

This restriction applies only to the homotypic Aβ pathway. Tau also provides a direct heterotypic nucleation route. Total oligomer production can therefore increase even while homotypic secondary nucleation is being suppressed, a distinction that becomes central in the numerical results.

### 3.4. The monomer-control switch: a Damköhler-number criterion

The remaining question is what sets the free monomer. The numerical behavior can be anticipated by a quasi-steady reduction of the monomer balance. In the reference case, free Aβ relaxes much faster than fibril number and fibril mass change, so the monomer can be viewed as responding almost instantaneously to two competing sinks: elongation at fibril ends and soluble-protein clearance at rate $\lambda_{m,A\beta} = \ln(2)/T_{1/2,m,A\beta}$. Balancing these sinks against the source gives the quasi-steady free monomer

$$m_{A\beta,qs} = \frac{q_{A,0,A\beta}}{L\left(2k_{+,A\beta}P_{A\beta}+\lambda_{m,A\beta}\right)} \tag{28}$$

The ratio of the two loss rates in the denominator defines a Damköhler number,

$$Da = \frac{2k_{+,A\beta}P_{A\beta}}{\lambda_{m,A\beta}} \tag{29}$$

When *Da* is much larger than one, elongation is the dominant sink and adding fibrils lowers the free monomer approximately in inverse proportion to fibril number. When *Da* is much smaller than one, clearance sets the monomer concentration and the fibril population has little influence on it. *Da* therefore predicts whether tau-created fibrils mainly consume monomer or mainly add catalytic mass. Substituting Eq. (28) into Eq. (26) gives, for tau-coupled and uncoupled systems,

$$\frac{R_{2,A\beta}^{(tau)}}{R_{2,A\beta}^{(0)}} = \left[\frac{2k_{+,A\beta}P_{A\beta}^{(0)}+\lambda_{m,A\beta}}{2k_{+,A\beta}P_{A\beta}^{(tau)}+\lambda_{m,A\beta}}\right]^2 \frac{D_{A\beta,eff}^{(tau)}}{D_{A\beta,eff}^{(0)}} \tag{30}$$

Eq. (30) converts that physical picture into a bound. At matched fibril mass, mass-proportional site formation gives the same accessible-site concentration with and without tau. Tau creates more fibrils, so in the elongation-controlled regime the lower monomer necessarily suppresses homotypic secondary nucleation; in the clearance-controlled regime the matched-mass rate is unchanged. A higher matched-mass rate therefore requires either a strong departure from mass-proportional site formation (Section 3.2) or an intrinsic difference in the tau-seeded fibrils themselves: more sites per mass, greater per-site activity, or greater accessibility.

At equal time the situation is different because tau can also change how much fibril mass has accumulated. In the elongation-controlled limit, nearly all supplied Aβ is eventually deposited and the masses converge, so monomer depletion wins. In the clearance-controlled limit, free monomer is set by clearance and barely depends on fibril number, and tau accelerates the accumulation of fibril mass; the extra mass then carries extra accessible sites and raises homotypic secondary nucleation. The same framework therefore predicts opposite signs in different monomer-control regimes.

The supplementary analysis gives the corresponding quasi-steady oligomer relation, the protein-mass constraint used to verify the computations, and the identification of the per-site secondary-nucleation coefficient from matched measurements (Section S4). These results provide the analytical benchmarks against which the numerical simulations are tested next.

## 4. Numerical results: testing the mechanism

The simulations now follow the logic established analytically. We first verify the mass-proportional limit, then ask what tau coupling does to fibril number, mass, and monomer. We next increase the site yield at fibril birth to determine how large it must be before shorter fibrils become more catalytic. Finally, we relax the reference assumptions one by one and introduce soluble-protein clearance to determine when the sign of the homotypic response reverses. The calculations are mechanistic sensitivity analyses with tissue-scale effective coefficients; they test how the mechanisms combine, not the magnitudes expected in a specific experiment.

Table 2 provides the map for this section. Three results recur across the scenarios. First, at matched fibril mass, tau coupling does not significantly increase the accessible-site concentration unless fibril birth is an exceptionally strong site source, accessibility depends explicitly on fibril length, or tau-nucleated fibrils have different site properties. Second, at equal time, the sign of the homotypic secondary-nucleation response follows the monomer-control regime: it is negative when elongation dominates and positive when clearance dominates, with a crossover near *Da* of order unity. Third, accumulated oligomer-equivalent exposure rises in every scenario examined because tau adds a direct heterotypic nucleation pathway. Sections 4.2–4.4 establish, in the reference case, why tau does not increase the accessible-site concentration at matched fibril mass yet still increases oligomer-equivalent exposure; Section 4.5 tests the robustness of those conclusions, and Section 4.6 shows how the monomer-control regime determines the sign of the homotypic secondary-nucleation response.

**Table 2.** Qualitative summary of the effects of tau coupling at $t$ = 33 years: ↑ increase, ↓ decrease, = unchanged to within $10^{-4}$, ≈ changed by less than 1%, relative to the uncoupled case ($\chi_\tau = 0$) of the same scenario. Numbers are ranges of the corresponding ratios for the coupling cases of Table S2 ($\chi_\tau = 1.86\times10^{-10}$ to $4.45\times10^{-9}$ μM$^{-2}$ s$^{-1}$) or, for parameter sweeps, for the cases given in the indicated section. *Da* is the Damköhler number of Eq. (29) for the uncoupled system at 33 years.

| Scenario (Section) | Da | $M_{A\beta}$, equal time | $D_{A\beta,eff}$, matched mass | $R_{2,A\beta}$, equal time | $R_{2,A\beta}$, matched mass | $\Xi_{A\beta}$ (33 yr) |
|---|---|---|---|---|---|---|
| Reference, mass-proportional sites (4.2–4.4) | ∞ | = | = | ↓ 0.16–0.71 | ↓ | ↑ 1.4–6.3 |
| Birth-dominated site formation, $\eta_{birth}/(2\eta_{grow}) \le 10^5$ (4.3, 4.4) | ∞ | = | ↑ only for $\eta_{birth}/(2\eta_{grow}) \gtrsim 10^3$; +19% at $10^5$ | ↓ | ↓ | ↑ |
| Growth-history law, Eq. (14), parameter sweep (S2.1) | ∞ | = | ↓ 0.60–1.00 | ↓ 0.12–0.38 | ↓ | ↑ 2.6–8.6 |
| Masking, $k_{mask} = 10^{-8}$ s$^{-1}$ (S2.2) | ∞ | = | = | ↓ 0.06–0.41 | ↓ | ↑ 2.5–16 |
| Oligomer dissociation (S2.3) | ∞ | = | = | ↓ 0.27–0.80 | ↓ | ↑ 1.2–3.7 |
| Soluble-protein clearance (S2.3) | ≪ 1 | ↑ 26–610 | = | ↑ 26–268 | = | ↑ 39–915 |
| Fragmentation, $\bar{n}_F = 55$ (S2.3) | ∞ | = | = | = | = | ↑ 1.06–1.96 |
| Plaque-accessibility closures (4.5) | ∞ | = | = | ↓ 0.07–0.996 | ↓ | ↑ 1.005–14.9 |
| Fibril removal, $T_{1/2,F}$ = 1–100 yr (4.5) | ∞ | = | = | ↓ 0.04–0.69 | ↓ | ↑ 1.4–16.6 |
| Fragmentation-generated sites (4.5) | ∞ | = | = | = | = | ↑ 1.02–1.96 |
| Mean fibril length 0.3–30 μm (4.5) | ∞ | = | = | ≈ or ↓ 0.68–1.00 | as equal time | ↑ 1.9–7.7 |
| Length-dependent accessibility, $\gamma$ = 0.5–3 (4.5) | ∞; 0.023 | = | ↑ (= with fragmentation) | as matched mass when *Da* = ∞ | ↓ for γ < 2, ↑ for γ > 2 when Da = ∞; ↑ for γ > 0 when Da ≪ 1 | ↑ |

| Scenario (Section) | Da | $M_{A\beta}$, equal time | $D_{A\beta,eff}$, matched mass | $R_{2,A\beta}$, equal time | $R_{2,A\beta}$, matched mass | $\Xi_{A\beta}$ (33 yr) |
|---|---|---|---|---|---|---|
| Length-dependent fragmentation, $\beta_{frag} = -1$ to 2 (4.5) | ∞ | = | ≈ (≤ 0.7%) | ↓ 0.20–1.00 | as equal time | ↑ 1.5–6.5 |
| Latin-hypercube study A, 400 samples (4.5) | ∞ | = | ≈ (≤ 0.16%) | ↓ 0.09–0.999 | ↓ | ↑ 1.001–11.6 |
| Regime grid and Latin-hypercube study B (4.6) | < 1 to ∞ | ↑ for $Da < 1$ | =, except at negligible mass | ↓ for $Da > 1$; ↑ for $Da < 0.1$; mixed for $0.1 < Da < 1$ | ≤ 1, except at negligible mass | ↑ 1.002–$3.5\times10^3$ |
| Combined dissociation and clearance (4.6) | 0.023 | ↑ 1.7–5.4 | = | ↑ 1.38–2.61 | ↓ 0.92–0.99 | ↑ 1.6–5.1 |
| Composite with 1-μm fibrils (4.6) | rises to 70 | ↑; 1.1–1.4 at 33 yr | = | ↑ before $Da \approx 1$; ↓ 0.39–0.73 at 33 yr | = | ↑ 12–185 |
| Tau-nucleated fibrils, site-abundance factor r, activity factor $r_k$ (S2.6) | ∞ | = | $1 + (r - 1)\varphi_\tau$ | ↓ unless $rr_k \gtrsim 3$–10 | as equal time | ↑ 3.6–5.9 |

### 4.1. Verification: the active-site model reproduces the expected limits

The one-way coupling makes the tau subsystem identical in all simulations and to that of the preceding model [4]. Three additional tests were satisfied in every run. In the mass-proportional limit, the active-site density remains equal to the prescribed reference density and the Aβ solution coincides with the preceding model. With soluble-protein clearance removed, total Aβ mass follows the imposed source to within a relative error below $10^{-14}$ (round-off level). Finally, the accumulated oligomer-equivalent exposure is converged with respect to the output grid (Section S5 and Fig. S2). These checks establish that the effects discussed below arise from the new mechanisms rather than from numerical drift.

### 4.2. Tau makes more, shorter fibrils but not more active sites in the reference case

The reference simulations compare uncoupled Aβ aggregation with PHF- and CTE-mediated coupling while keeping active-site formation exactly proportional to fibril mass. Because the tau-fold coefficient and the accessibility fraction enter only through their product, CTE coupling is equivalent to PHF coupling at a proportionally larger effective coupling strength. Thus the two tau folds provide experimentally grounded points on the same coupling axis unless additional fold-specific site properties are introduced.

The first effect of tau is the one expected from the cross-seeding experiment: more fibrils compete for the same monomer pool. By 33 years, increasing tau coupling lowers free Aβ monomer and raises fibril number, with the strongest response for CTE coupling. Yet fibril mass remains 228.8 μM in all five reference simulations to within 0.001%. Tau therefore redistributes essentially the same Aβ mass among a larger number of fibrils rather than increasing the final fibril burden (Fig. 2).

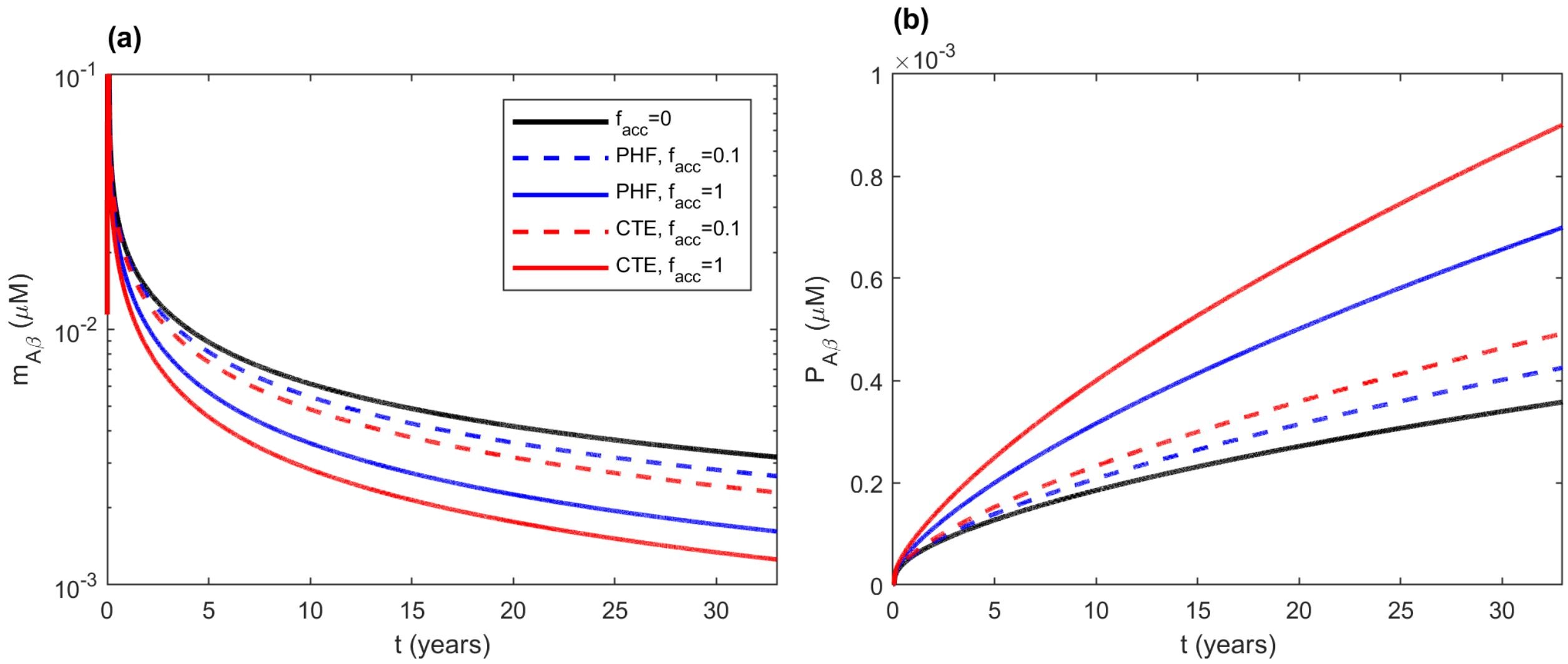

**Fig. 2.** Reference case ($\eta_{birth} = 2\sigma_0$, $\eta_{grow} = \sigma_0$, $k_{mask} = 0$, no oligomer dissociation or soluble-protein clearance) for $f_{acc} = 0$ (black), PHF coupling (blue), and CTE coupling (red) at $f_{acc} = 0.1$ (dashed) and 1 (solid), that is, $\chi_\tau = 1.86\times10^{-10}$ and $1.86\times10^{-9}$ (PHF) and $4.45\times10^{-10}$ and $4.45\times10^{-9}$ μM$^{-2}$ s$^{-1}$ (CTE). (a) Free Aβ monomer concentration $m_{A\beta}$. (b) Aβ fibril-number concentration $P_{A\beta}$.

That redistribution shortens the fibrils in the same direction reported experimentally [3], but it does not create additional catalytic capacity in the mass-proportional model. Mean fibril size falls from $6.4\times10^5$ monomer units without tau coupling to $3.3\times10^5$ for PHF and $2.5\times10^5$ for CTE coupling at full accessibility, while total and exposed active-site concentrations remain identical because they are fixed by fibril mass. The absolute fibril lengths are unrealistically large under the reference tissue-scale parameters (about 150 μm for two protofilaments with a 0.47-nm cross-β layer spacing [35]; Section S2.4); only the direction of shortening is compared with experiment, and Section 4.5 repeats the analysis for mean lengths of 0.3–30 μm. Plaque consolidation, rather than fibril shortening, is what strongly limits accessibility: fewer than 1% of sites remain exposed at 33 years (Fig. 3).

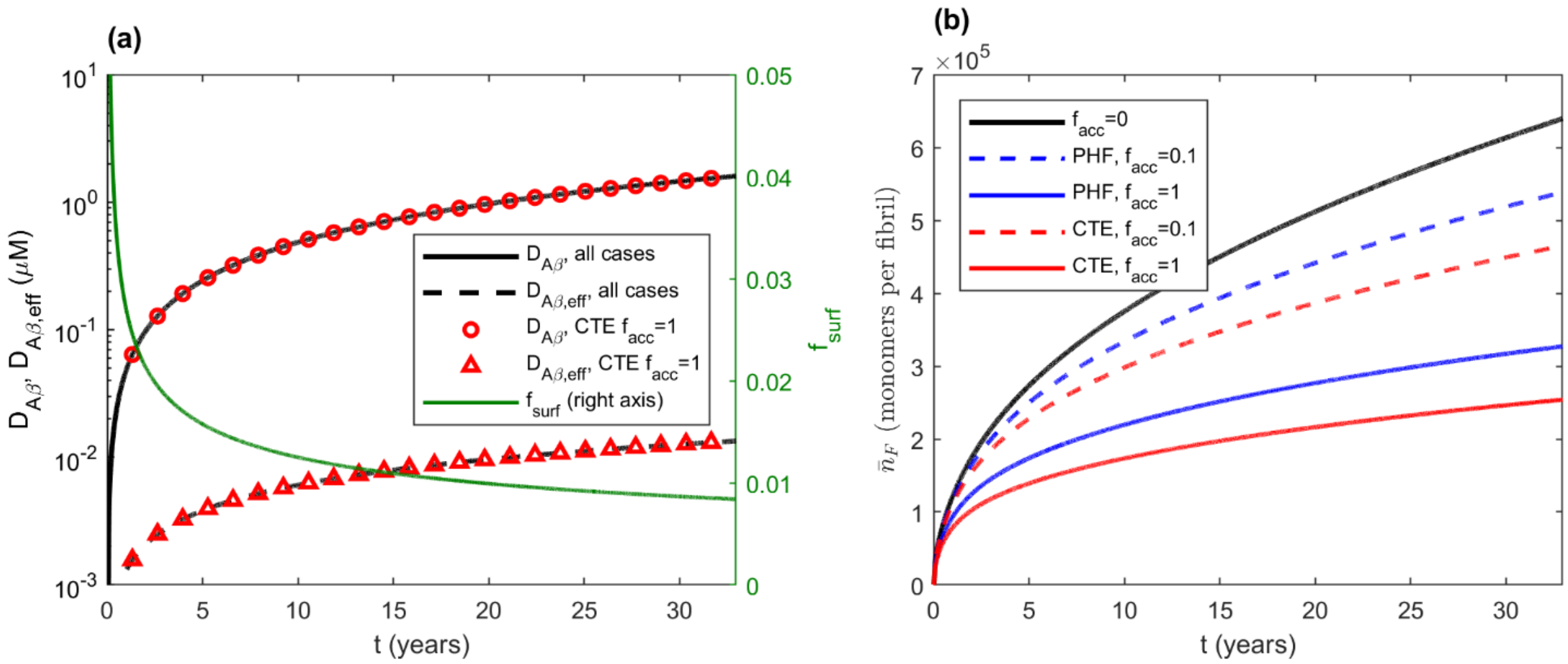

**Fig. 3.** Active sites and fibril size in the reference case. (a) Total active-site concentration $D_{A\beta}$ (solid black), accessible active-site concentration $D_{A\beta,eff}$ (dashed black), both on the left axis, and exposed fraction $f_{surf}$ (green, right axis). The curves for all five cases of Fig. 2 coincide; symbols show the CTE, $f_{acc} = 1$ case. (b) Mean fibril size $\bar{n}_F = M_{A\beta}/P_{A\beta}$ [Eq. (19)]; line styles as in Fig. 2.

### 4.3. How strong must fibril-birth site formation be to matter?

To move away from the null hypothesis, the elongation-site yield is held fixed while the site yield assigned to each fibril-birth event is varied over many orders of magnitude. Because fibril mass is still effectively matched between coupled and uncoupled systems, this experiment isolates the only question that matters here: can the additional fibrils created by tau carry enough extra sites to change catalytic-site density?

The answer is quantitatively restrictive. The simulations fall exactly on the analytical relation of Section 3.2, and active-site density changes by less than 1% until the birth yield exceeds the mass-proportional value by roughly three orders of magnitude. Differences of order 10% between tau-coupled and uncoupled systems require several hundred to several thousand sites per fibril-birth event. Thus, for the long fibrils generated by the reference tissue-scale model, simply making more fibrils is not enough; each new fibril would have to carry an unusually large catalytic legacy before matched-mass site density changes appreciably.

Oligomer exposure behaves differently. Accumulated oligomer-equivalent exposure responds strongly to tau accessibility but remains nearly insensitive to the birth-site yield over the range in which active-site density is unchanged. In the mass-proportional case, exposure rises 1.41- and 3.82-fold for PHF coupling and 1.89- and 6.34-fold for CTE coupling at the two accessibility levels. The key inference is therefore already visible: tau can raise oligomer exposure substantially even when the Aβ fibrils it creates are not more catalytically active.

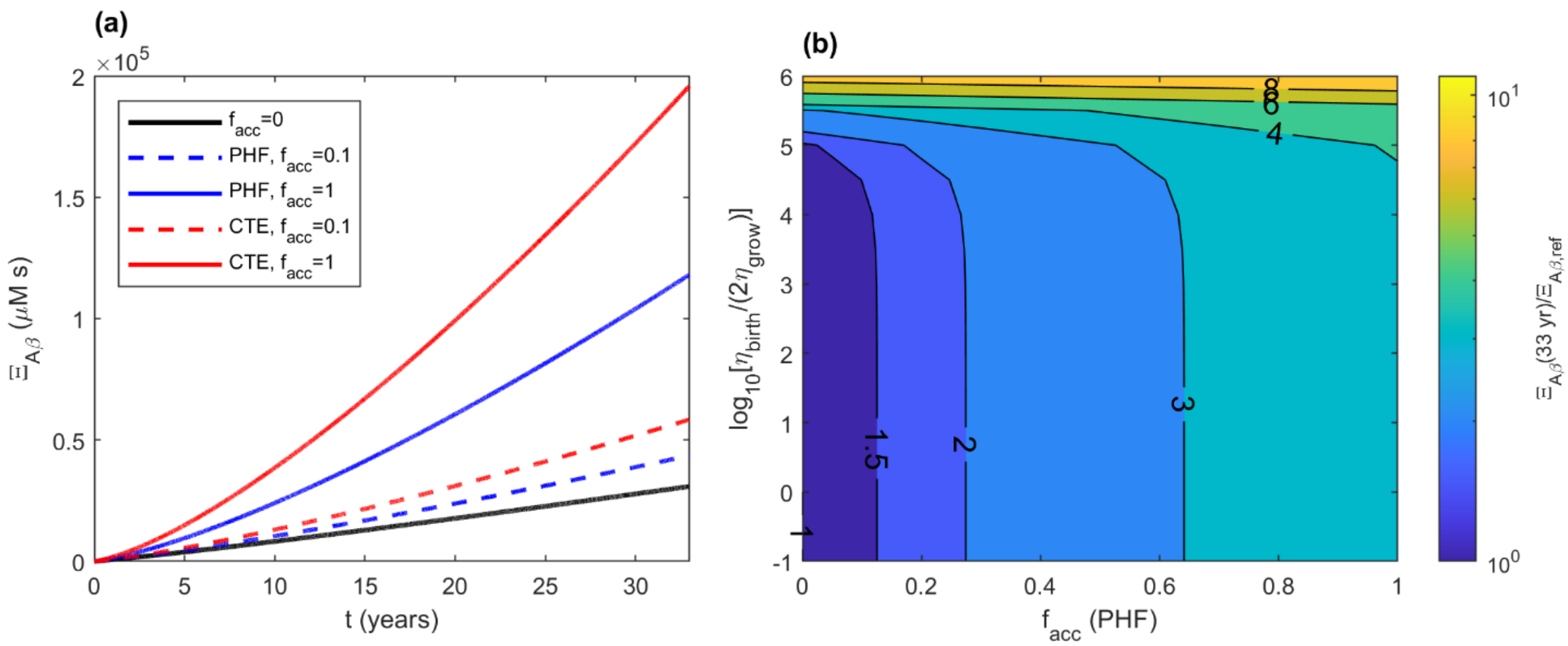

**Fig. 4.** Accumulated soluble Aβ oligomer-equivalent exposure. (a) $\mathit{\Xi}_{A\beta}(t)$ in the reference case; line styles as in Fig. 2. (b) $\mathit{\Xi}_{A\beta}$(33 yr), normalized by its value for $f_{acc} = 0$ in the mass-proportional case ($\mathit{\Xi}_{A\beta,ref} = 3.09\times10^4$ μM s), as a function of the accessibility fraction $f_{acc}$ for PHF coupling ($\chi_\tau = \alpha_{PHF} f_{acc}$) and of the fibril-birth site yield $\eta_{birth}/(2\eta_{grow})$, with $\eta_{grow} = \sigma_0$.

### 4.4. Why more fibrils can produce less homotypic secondary nucleation

The next calculation separates the two factors in Eq. (26): the change in exposed active sites and the squared change in free monomer. This directly tests whether any catalytic-site gain is large enough to compensate for the stronger monomer consumption caused by the larger fibril population.

In the reference case the exposed-site ratio is exactly one, so the entire homotypic response comes from monomer depletion. At 33 years, the secondary-nucleation rate falls to 71% and 26% of control for PHF coupling and to 53% and 16% for CTE coupling across the two accessibility levels. Even an extreme birth-site yield that raises exposed sites by 19% cannot overcome the squared monomer penalty. Total oligomer formation nevertheless rises because the heterotypic tau pathway becomes dominant, accounting for up to 90–95% of oligomer formation at full coupling. Tau therefore increases oligomer exposure while simultaneously suppressing the homotypic Aβ feedback loop; this is the central mechanistic separation of the model (Fig. 5).

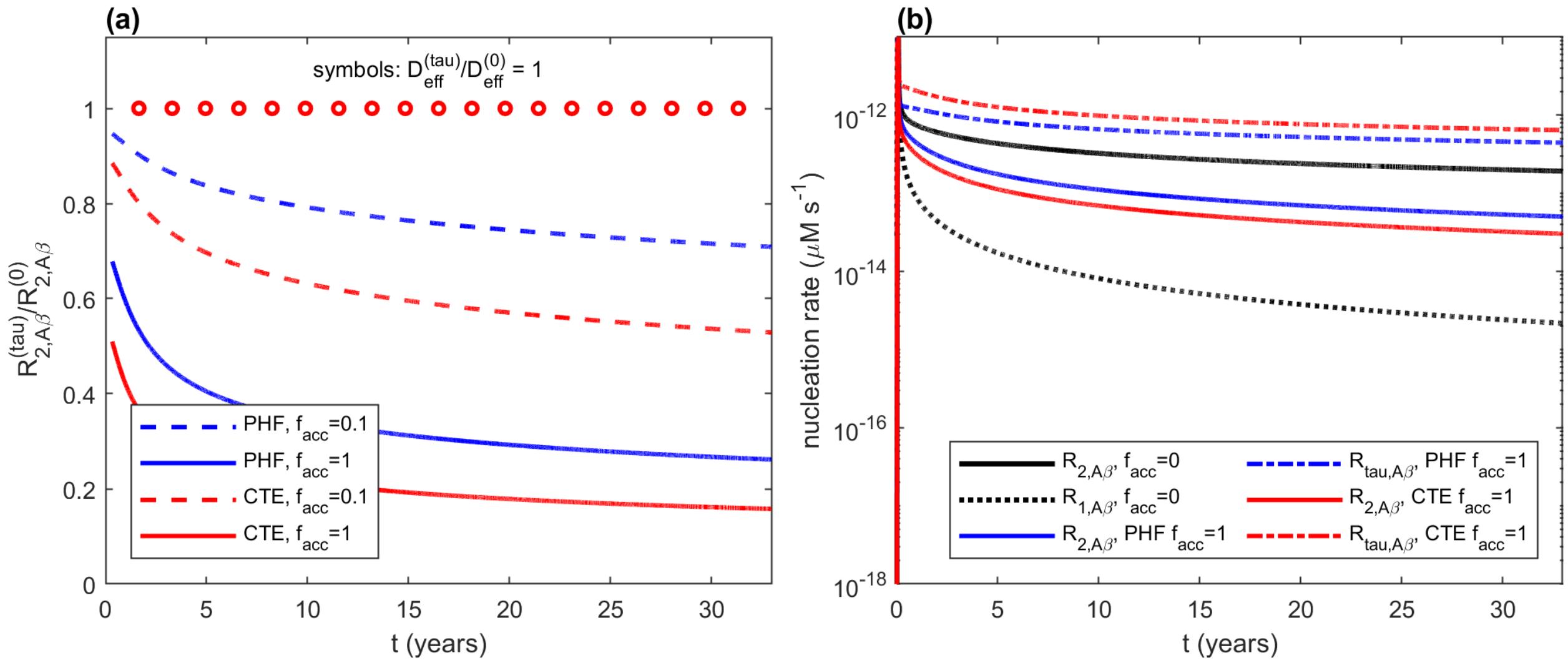

**Fig. 5.** Competition between accessible-site gain and monomer depletion [Eqs. (26) and (27)] in the reference case. (a) Homotypic secondary-nucleation rate in tau-coupled systems relative to $f_{acc}$ = 0 (lines; styles as in Fig. 2) and accessible-site ratio $D_{A\beta,eff}^{(tau)}/D_{A\beta,eff}^{(0)}$ (symbols, identically 1); the first four months are omitted. (b) Homotypic secondary-nucleation rate $R_{2,A\beta}$, intrinsic primary-nucleation rate $R_{1,A\beta}$, and heterotypic rate $R_{tau,A\beta}$ for $f_{acc}$ = 0 and for $f_{acc}$ = 1 with PHF and CTE coupling.

### 4.5. Robustness: what can overturn the reference conclusion?

The reference case is deliberately simple, so each assumption was relaxed in turn. The tests include growth-history-dependent site formation, masking, oligomer dissociation, soluble-protein clearance, fragmentation, alternative plaque-accessibility laws, fibril removal, fragmentation-generated sites, realistic fibril lengths, morphology-dependent accessibility and fragmentation, broad parameter uncertainty, and a distinct site phenotype for fibrils nucleated on tau. Table 2 summarizes the outcomes and Section S2 gives the full calculations.

Growth history, masking, dissociation, clearance, and fragmentation. A supersaturation-dependent elongation-site yield can lower the exposed-site concentration under tau coupling, and masking reduces site density in both coupled and uncoupled systems. Oligomer dissociation changes the absolute exposure by more than two orders of magnitude but leaves the tau-induced exposure ratios qualitatively similar. Soluble-protein clearance is the important exception: once clearance controls the monomer, tau accelerates the accumulation of fibril mass and can raise homotypic secondary nucleation at equal time. Fragmentation at the literature-based unliganded rate does not change the main conclusion.

Accessibility, turnover, realistic fibril lengths, and parameter uncertainty. Across a 40-fold range of plaque-shell thickness, alternative porous and no-burial closures, slow fibril removal, fragmentation-generated sites, and fragmentation rates giving mean fibril lengths of 0.3–30 μm, tau still fails to create a meaningful matched-mass increase in exposed active sites under mass-based accessibility. In 400 Latin-hypercube samples (a stratified random design that varies all sampled parameters simultaneously; study A in Section S2.4) without soluble-protein clearance, tau coupling lowers homotypic secondary nucleation and raises oligomer-equivalent exposure in every sample. The effect sizes vary widely, but the direction of the effect does not.

Morphology-dependent accessibility identifies the clearest route to a different result. If shorter fibrils expose more catalytic sites per unit mass, tau-induced shortening can raise homotypic secondary nucleation at matched mass. The required strength of this length effect depends on monomer control: any positive dependence is sufficient when clearance controls the monomer, whereas a much stronger dependence ($\gamma > 2$ in Eq. (S30)) is required when elongation controls it. By contrast, making the fragmentation rate depend on fibril length changes matched-mass active-site exposure by less than 1% in the cases tested.

A second route is a genuinely different catalytic phenotype for fibrils nucleated on tau. When such fibrils have both a different active-site abundance and a different activity per site, the homotypic response depends on the product of those two changes. In the reference regime that product must be several-fold larger, approximately 3–10 times or more, before tau-seeded fibrils overcome monomer depletion. This result makes site counting and kinetic activity complementary measurements rather than substitutes.

The robustness analysis therefore separates structural conclusions from parameter-dependent magnitudes. The structural result is that shorter fibrils do not become more catalytic merely because they are shorter: a matched-mass increase requires an additional mechanism that changes site yield, site activity, or site exposure. The numerical reductions in homotypic secondary nucleation and fold increases in oligomer exposure depend strongly on the chosen effective coefficients and should be read as illustrations, not predictions of experimental magnitude. This distinction is important because the qualitative matched-mass result follows from the model structure itself whenever all fibrils share the same site yields and accessibility depends only on fibril mass.

### 4.6. The controlling switch is monomer balance: a Damköhler-number transition

The quasi-steady analysis accurately reproduces the full simulations and provides the organizing variable missing from the earlier sections. After the initial transient, the monomer relation of Eq. (28) and the secondary-nucleation ratio of Eq. (30) agree closely with the numerical solutions across the reference, dissociation, clearance, fragmentation, regime-grid, and Latin-hypercube calculations (Fig. S16). The complicated parameter dependence can therefore be interpreted through one physical competition: elongation versus clearance.

When *Da* is large, elongation controls the monomer. Tau adds fibril ends, free monomer falls, and homotypic secondary nucleation is suppressed even though oligomer exposure rises through the heterotypic pathway. When *Da* is small, clearance rather than elongation sets the free monomer concentration, so additional fibrils barely change it. Tau can then accelerate fibril accumulation, increase the exposed-site burden at a given time, and raise homotypic secondary nucleation. A grid of source and clearance conditions shows the crossover near *Da* of order unity (Fig. S17), while matched-mass secondary nucleation remains at or below the uncoupled value as predicted analytically.

The second 400-sample Latin-hypercube study (study B; details in Section S2.4) confirms that this monomer-control balance dominates the uncertainty. Tau increases accumulated oligomer-equivalent exposure in every sample. Equal-time homotypic secondary nucleation is lower in every sample with $Da > 1$ and higher in most samples with $Da < 1$, with the exceptions concentrated near the transition. Clearance strength is the leading correlate of the tau effect, ahead of any active-site parameter.

Combining oligomer dissociation with soluble-protein clearance makes the same point in a physiologically richer setting. Here *Da* is well below one, so tau raises fibril mass, exposed active sites, and equal-time homotypic secondary nucleation while leaving active-site density unchanged. At matched fibril mass, exposed sites are again identical and the homotypic rate is slightly lower. Tau therefore increases the homotypic pathway at a given time because there is more fibrillar material, not because the fibrils themselves are more catalytic.

In the composite simulation, the two regimes occur in sequence (Fig. S18). With fragmentation adjusted to give a mean fibril length of about 1 μm, *Da* rises from far below one early in deposition to above one as fibrils accumulate. Tau coupling brings this transition forward because it creates fibrils sooner. Accordingly, tau strongly enhances homotypic secondary nucleation during the clearance-controlled phase but suppresses it later, after the expanding fibril population becomes the dominant monomer sink. The precise transition times depend on the effective tissue-scale parameters and are not predictions of disease chronology; the sequence of regimes is the mechanistic prediction.

**5. Discussion**

The question posed in the Introduction has a qualified answer: tau can make Aβ fibrils shorter without making them intrinsically more catalytic. The reason is that fibril mass, catalytic-site abundance, and site exposure are distinct physical quantities. That separation is supported by evidence that secondary nucleation is associated with rare structural features [5], that secondary nucleation and elongation occur at different reactive sites [36], and that Aβ42 self-replication occurs on small isolated fibril sites [37]. It is also consistent with interventions that suppress secondary nucleation without removing fibril mass, including BRICHOS [38] and aducanumab [39]. The active-site variable introduced here formalizes this distinction: a fibril population can change its mass, its number, its catalytic-site density, or the fraction of sites exposed to solution, and these changes need not move together.

This distinction resolves the apparent paradox created by tau cross-seeding. Tau-seeded Aβ42 fibrils are experimentally shorter [3], but shorter fibrils are not automatically more active fibrils. The model treats the resulting filaments as homomolecular Aβ fibrils, consistent with the observation that Aβ40 and Aβ42 can cross-react during primary nucleation while assembling into separate homomolecular fibrils [40]. What matters after nucleation is therefore not the mere existence of more fibrils, but whether their formation history leaves a different density, activity, or exposure of secondary-nucleation sites.

Under the inherited tissue-scale parameters, fibrils become long enough that the contribution of a single birth event is diluted over a large fibril mass. Unless fibril birth introduces hundreds to thousands of sites, tau-induced shortening leaves active-site density essentially unchanged at matched mass. Tau nonetheless raises oligomer-equivalent exposure, reproducing the qualitative conclusion of the preceding model [4], but the present formulation identifies the mechanism more sharply: the increase is driven by direct heterotypic catalysis and is partly offset, rather than amplified, by the homotypic Aβ secondary-nucleation pathway when elongation controls the monomer. The key new result is therefore not simply that tau changes aggregation, but that the same tau perturbation can strengthen one nucleation pathway while weakening another.

The Damköhler number turns this competition into a stage-dependent hypothesis. In a closed aggregation assay, in which monomer is not cleared, the system is elongation controlled and the matched-mass prediction can be tested directly. In tissue, however, *Da* should evolve as the fibril population grows. Early in deposition, when few fibril ends are present, production and clearance can control free monomer; tau then accelerates both fibril accumulation and homotypic secondary nucleation. This situation may be most relevant when tau pathology precedes substantial amyloid deposition [41] and [42]. The ability of pre-existing aggregates to induce or accelerate cerebral Aβ deposition is demonstrated by seeding experiments in APP-transgenic mice [43] and by heterologous amyloid cross-seeding [44]. As fibrils accumulate, they become an increasingly strong monomer sink, and tau-induced addition of fibril ends then lowers the free monomer and suppresses homotypic secondary nucleation. The decline of soluble or CSF Aβ42 during amyloid deposition [45], reduced Aβ clearance in late-onset AD [46], the fall of interstitial Aβ42 as plaques accumulate in mice [47], the inverse association between cortical amyloid load and CSF Aβ42 [48], and altered soluble Aβ42 turnover in presenilin mutation carriers [49] are qualitatively compatible with such a shift. These observations do not validate the model or locate the transition in human disease, but they make the predicted change of regime experimentally meaningful.

The largest uncertainty is therefore experimental, not mathematical. Active-site densities have been measured for specific *in vitro* fibril preparations, including Aβ40 as well as Aβ42 [5], whereas the tau cross-seeding experiments established changes in Aβ42 fibril length and morphology but did not measure secondary-nucleation-site density [3]. The model consequently does not assign a measured active-site phenotype to PHF- or CTE-seeded Aβ42. In the present parameterization, the two tau folds differ only in their heterotypic nucleation coefficient, so a fold-specific effect that cannot be reproduced by rescaling the effective coupling would point directly to differences in site formation, site accessibility, or per-site catalytic activity.

A second uncertainty concerns plaque geometry. The model compresses all catalytic sites into one concentration and represents burial by a spherical thin-shell factor, whereas real plaques are porous, irregular, and protein-rich. The closure should therefore be read as a minimal mapping from total to exposed catalytic capacity, not as a literal plaque morphology. Its qualitative premise, that the plaque periphery is the region most available to soluble Aβ, is consistent with the halo of oligomeric Aβ around plaques [50] and with protofibrillar Aβ42 hotspots at plaque surfaces not covered by microglia [51]. The sensitivity analysis shows that geometry strongly affects the magnitude of exposed-site burden, while the matched-mass logic survives unless accessibility itself depends on fibril morphology.

The tau–Aβ contact is also represented phenomenologically. The accessibility fraction combines tau release, membrane disruption, extracellular transport, and related routes into one effective parameter, so the model cannot predict how coupling evolves as neurons die or tau pathology spreads. A spatially resolved model would make that coupling time dependent and would distinguish intracellular, extracellular, and transported tau species. The present formulation instead asks a narrower question: given an effective amount of tau fibrillar catalyst available to Aβ, what follows for the catalytic properties of the Aβ population that forms?

The absolute numerical scales require similar caution. The reference tissue-scale source and aggregation coefficients generate very long fibrils when fragmentation and fibril turnover are absent, which is why birth-generated sites matter only at extreme yields. The sensitivity analysis shows that the central matched-mass result persists when fragmentation is increased enough to produce mean lengths of 0.3–30 μm, but the *in vitro* cross-seeding and active-site experiments operate at micromolar monomer concentrations and hour-scale kinetics that are not reproduced by the decades-long tissue parameterization. The numerical effect sizes should therefore be interpreted as mechanistic sensitivity results, not as predictions of a particular experiment or patient trajectory.

The model also keeps oligomer dissociation separate from secondary-nucleation activity. Fibril surfaces can catalyze Aβ oligomer dissociation [7], but it is not known whether the same microscopic sites that generate oligomers also accelerate their loss. Coupling both processes to the same active-site variable would therefore impose an untested mechanistic identity. Retaining separate terms makes the model less compact but avoids building that assumption into the answer.

The decisive experiments are now clear. First, active-site density should be measured for control, PHF-seeded, and CTE-seeded Aβ42 fibrils at matched fibril mass, ideally using a site-counting assay such as BRICHOS binding [5] together with secondary-nucleation kinetics on the same preparations; this separates site abundance from activity per site. Second, time-resolved site counts during nucleation-dominated and elongation-dominated intervals could distinguish the birth and growth site yields. Third, aging fibrils at approximately constant mass would constrain site masking. Finally, measurements of soluble-Aβ clearance, fibril-end concentration, and elongation rate in tissue or animal models would determine *Da* and therefore identify which monomer-control regime applies. These experiments would test the model at the points where its conclusions can change, rather than only comparing overall aggregation curves.

Table 3 collects these predictions in experimentally actionable form. Its purpose is to separate statements that follow from the model structure from statements that require new measurements of tau-seeded Aβ fibrils.

**Table 3.** Testable predictions of the active-site-resolved model.

| Prediction | Condition in the model | Measurement needed |
| --- | --- | --- |
| At matched fibril mass, tau-seeded and control Aβ42 fibrils have the same active-site density unless birth-generated sites are important. | Equal site yields for all fibrils; departure $(\eta_{birth}/\eta_{grow} - 2)/\bar{n}_F$ [Eq. (24)]. | Active-site counts (e.g., BRICHOS binding) per monomer, with fibril number or length distribution, for control, PHF-seeded, and CTE-seeded fibrils at equal mass. |
| Tau coupling does not raise the homotypic secondary-nucleation rate at matched fibril mass. | Approximately mass-proportional sites (negligible birth sites), accessibility a function of mass only, quasi-steady monomer [Eq. (30)]. | Seeded secondary-nucleation kinetics with equal masses of fibrils grown with and without tau seeds, with tau seeds removed or otherwise controlled for. |
| The sign of the equal-time change in homotypic secondary nucleation is controlled mainly by *Da*. | $Da \gg 1$: decrease; $Da \ll 1$: increase [Eqs. (29) and (30)]; near $Da \approx 1$ the accessible-site ratio also matters. | Fibril-number concentration, elongation rate, and soluble-Aβ clearance rate in the system of interest. |
| PHF and CTE coupling differ only by a rescaling of $f_{acc}$. | Polymorph-independent $\eta_{birth}$, $\eta_{grow}$, and $k_{o2,A\beta,D,p}$ (Section 4.2). | Aβ42 kinetics with PHF and CTE seeds at seed concentrations in the ratio $\alpha_{CTE}/\alpha_{PHF} \approx 2.4$ (CTE lower). |
| A supersaturation-dependent site yield lowers the site density under tau coupling. | Eq. (14) (Section S2.1). | Site density of fibrils grown at different, controlled monomer supersaturations. |
| Tau-nucleated fibrils raise homotypic secondary nucleation at matched mass | Two-lineage model (Section S2.6). | Site density of fibrils grown on tau seeds, separated from homotypically nucleated |

| Prediction | Condition in the model | Measurement needed |
|---|---|---|
| only if the product of their site abundance and activity per site is several-fold higher. | | fibrils, together with seeded secondary-nucleation kinetics of the same preparations to separate $r$ from $r_k$. |
| Tau coupling raises homotypic secondary nucleation early in deposition and lowers it after *Da* exceeds unity. | Clearance-controlled free monomer before deposition; *Da* increasing with fibril number (composite scenario, Section 4.6). | Interstitial or CSF Aβ42 relative to pre-deposition levels, with fibril burden and secondary-nucleation markers, in regions with and without tau pathology at different amyloid stages. |
| Shorter fibrils raise homotypic secondary nucleation at matched mass only if their deposits expose more sites per unit mass. | Length-dependent accessibility, Eq. (S30): any $\gamma > 0$ when clearance controls the monomer, $\gamma > 2$ when elongation controls it (Section 4.5). | Exposed (accessible) site density in compacted deposits of tau-seeded and control fibrils of equal mass, with fibril-length distributions. |

## 6. Conclusions

This study was motivated by a simple experimental observation with a nontrivial consequence: tau cross-seeding produces more, shorter Aβ fibrils, but fibril length alone does not tell us whether those fibrils are better catalysts of Aβ secondary nucleation. By resolving fibril number, fibril mass, catalytic-site concentration, and plaque-scale site exposure separately, the model shows why. At matched fibril mass, making more fibrils increases catalytic capacity only when fibril birth contributes a site burden large enough to compete with the sites accumulated during elongation, or when tau seeding changes site yield, activity, or exposure. Shorter fibrils are therefore not intrinsically more catalytic in this model.

The second result is dynamic. Tau coupling creates more fibrils and thereby changes the balance that sets free Aβ monomer. When elongation is the dominant monomer sink, the extra fibril ends deplete monomer and suppress homotypic secondary nucleation even though direct tau-catalyzed nucleation raises oligomer-equivalent exposure. When clearance controls the monomer, tau accelerates fibril accumulation and can increase homotypic secondary nucleation at the same time. A Damköhler number comparing elongation with clearance captures this switch, and a composite simulation shows the two regimes occurring successively as deposition progresses.

The strongest experimental test is therefore a matched-mass comparison of control and tau-seeded Aβ42 fibrils that measures active-site density, site accessibility, and activity per site in addition to fibril length. Such measurements would determine whether tau merely changes how Aβ mass is partitioned among fibrils or also changes the catalytic phenotype of the fibrils themselves. Until those data are available, the model supports a restrained conclusion: tau can strongly increase oligomer-generating pressure without requiring tau-seeded Aβ fibrils to be intrinsically more active catalysts, and the sign of the homotypic response depends on how the surrounding system controls free monomer.

Data accessibility. This article has no additional data.

Authors' contributions. AVK is the sole author of this paper.

Competing interests. The author declares no competing interests.

Funding statement. The author acknowledges the support provided by the National Science Foundation (grant DMS-2451660) and the Alexander von Humboldt Foundation through the Humboldt Research Award.

Ethics Statement. None.

**Shorter but not more catalytic: an active-site-resolved model of how tau reshapes amyloid-β secondary nucleation**

Andrey V. Kuznetsov

Department of Mechanical and Aerospace Engineering, North Carolina State University,

Raleigh, NC 27695-7910, USA; email: avkuznet@ncsu.edu

## Supplementary Materials

Supplementary Materials for: Shorter but not more catalytic: an active-site-resolved model of how tau reshapes amyloid-β secondary nucleation. The supplement is organized around five questions that support the main paper: where the parameter values come from and how strongly each is constrained (Section S1); which conclusions survive changes in kinetic, geometric, and morphological assumptions (Section S2); which equations govern the upstream tau subsystem (Section S3); which analytical and conservation relations explain and verify the simulations (Section S4); and how the calculations can be reproduced (Section S5). Sections S6 and S7 collect the supplementary figures and tables. Unless stated otherwise, parameters are those of the reference case in Table S1.

### S1. Parameter foundations: what is measured, inherited, and assumed

The purpose of this section is not simply to list parameter values, but to make their evidentiary status explicit. Each note identifies whether a quantity is taken directly from experiment, inherited from the preceding tissue-scale model, estimated by combining measurements obtained under different conditions, or introduced as a phenomenological sensitivity parameter. The notes follow the row order of Table S1 so that every value can be traced to its rationale; equations in this section are numbered (S1)–(S29).

**Geometry, source terms, and coupling parameters (S1.1–S1.8). These entries define the two control volumes, the effective tau and Aβ supply rates, heterotypic tau–Aβ coupling, and plaque-surface accessibility.**

**S1.1** $V_{tau}$. The diameter of the soma, $D_{soma}$, is assumed to be 20 μm. Treating the soma as a sphere gives $V_{tau} = V_{soma} = \pi D_{soma}^3/6 = 4.19\times10^3$ μm$^3$.

**S1.2** $L$. The spacing between neighboring plaques in Fig. 1(a) of [10] is approximately 100 μm, and $L$ is set to approximately half that value, 50 μm. Because $q_{A,0,A\beta}$ is scaled in proportion to $L$ (Section S1.4), the volumetric Aβ production rate $q_{A,0,A\beta}/L$ does not depend on this choice.

**S1.3** $q_{tau}$. Tau is synthesized predominantly in the neuronal soma and is normally transported along the axon. In AD, tau is missorted into, and accumulates within, the somatodendritic compartment [15]; in neurons exposed to Aβ, this missorting follows failure of the tau diffusion barrier at the axon initial segment [16]. With supply-limited growth, the tau fibril mass in the soma is $M_{tau} = q_{tau} a_{21} t / V_{soma}$, so that the tangle volume fraction at the end of the simulation is $q_{tau} MW_{tau} t_{end} / (\rho_{tau} V_{soma})$. Reference [6] does not report a neurofibrillary-tangle volume fraction of 0.081. Rather, it reports that extracellular neurofibrillary tangles account for only 2.2–17.2% (mean 8.1%) of the CA1 pyramidal-neuron loss observed in AD. Thus, 0.081 cannot be interpreted as a measured tangle volume fraction. To preserve the tau-source calibration used in the preceding model and MATLAB code, the present calculations retain 0.081 as an inherited phenomenological terminal tangle-volume fraction. With this assumption, $q_{tau} = 0.081 \rho_{tau} V_{soma} / (MW_{tau} t_{end}) = 1.10\times10^{-23}$ mol s$^{-1}$, with $\rho_{tau} = 1.35\times10^{-12}$ g μm$^{-3}$ (Section S1.8), $V_{soma} = 4.19\times10^{3}$ μm$^{3}$ (Section S1.1), and $t_{end} = 1.04\times10^{9}$ s (Section S1.29). The resulting tau supply rate should therefore be regarded as an inherited calibration parameter rather than as a quantity directly measured in the neuropathological study.

The value of $MW_{tau}$ varies among the six major human CNS isoforms, for which molecular weights of 36.7–45.9 kDa are listed in [15]. The model uses $4\times10^{4}$ g mol$^{-1}$ (40 kDa), the value implemented in the MATLAB code.

**S1.4** $q_{A,0,A\beta}$. According to [11], Aβ42 monomers are produced in the human brain at a total rate of approximately

$$\dot{N}_{A\beta} = 0.33 \times 10^{-6}\ \mu\text{mol s}^{-1} \quad \text{(S1)}$$

This estimate corresponds to approximately $2\times 10^{11}$ Aβ42 molecules produced per second in the human brain, the lower end of the 2–4 molecules per neuron per second for about $10^{11}$ neurons given in that study, which regards this rate as a conservative lower bound. Dividing this value by the estimated brain volume,

$$V_{brain} = 1.5 \times 10^{15}\ \mu\text{m}^3 \quad \text{(S2)}$$

gives the average volumetric production rate

$$\dot{Q}_{A\beta} = \frac{\dot{N}_{A\beta}}{V_{brain}} = 2.2 \times 10^{-22}\ \mu\text{mol}\ \mu\text{m}^{-3}\ \text{s}^{-1} \quad \text{(S3)}$$

Since $1\ \text{L} = 10^{15}\ \mu\text{m}^3$, this is equivalent to the concentration source

$$Q_{A\beta} = 2.2 \times 10^{-7}\ \mu\text{M s}^{-1} \quad \text{(S4)}$$

For the local cubic control volume of side length $L = 50$ μm, $V_{A\beta} = L^3 = 1.25 \times 10^5$ μm$^3$, the corresponding total Aβ42 production rate within the control volume is

$$\dot{N}_{A\beta,\mathrm{CV}} = \dot{Q}_{A\beta} L^3 = 2.75 \times 10^{-17} \text{ μmol s}^{-1} \tag{S5}$$

These values are the same as those used in the earlier formulation [13].

Aβ is generated and released at neuronal membranes distributed throughout the local tissue [12], rather than entering the control volume through a single physical face. Accordingly, $Q_{A\beta}$ is interpreted as a volume-averaged membrane source. More generally, if $J_{A\beta}$ denotes the Aβ production/release flux per unit neuronal membrane area and $A_m$ is the total neuronal membrane area contained within the control volume, then

$$Q_{A\beta} = \frac{A_m}{V_{A\beta}} J_{A\beta} \tag{S6}$$

To retain the source-term notation used in the present model and in earlier work, the equivalent parameter $q_{A,0,A\beta}$, with units of μM μm s$^{-1}$, is defined such that

$$\frac{q_{A,0,A\beta}}{L} = Q_{A\beta} \tag{S7}$$

Therefore,

$$q_{A,0,A\beta} = Q_{A\beta} L = (2.2 \times 10^{-7} \text{ μM s}^{-1})(50\ \mu m) = 1.1 \times 10^{-5} \text{ μM μm s}^{-1} \tag{S8}$$

Thus, $q_{A,0,A\beta} = 1.1 \times 10^{-5}$ μM μm s$^{-1}$ is adopted as the baseline value. The term $q_{A,0,A\beta}/L$ in the governing equation should not be interpreted as transport through one face of the control volume; it represents the spatially averaged production and release of Aβ42 from neuronal membranes distributed throughout the local extracellular/interstitial tissue.

One conceptual point is worth emphasizing: $Q_{A\beta} = 2.2 \times 10^{-7}$ μM s$^{-1}$ is the physical volumetric source; $q_{A,0,A\beta}$ is merely its equivalent representation in the existing $q/L$ formulation. Consequently, if $L$ is changed, $q_{A,0,A\beta}$ should change proportionally so that $q_{A,0,A\beta}/L = Q_{A\beta}$ remains fixed. This avoids reintroducing an unintended dependence of the actual Aβ production rate on control-volume size.

**S1.5** $\alpha_p$. In the present model, the tau-catalyzed nucleation rate is written as

$$R_{tau,A\beta} = \alpha_p T_{tau} m_{A\beta}^2 \tag{S9}$$

where $T_{tau}$, given by Eq. (1), is the effective concentration of tau fibrillar material accessible to Aβ and $m_{A\beta}$ is the Aβ monomer concentration. Thus, $\alpha_p$ determines the increase in the Aβ primary-nucleation rate produced by a unit concentration of accessible tau fibrils.

The coefficient $\alpha_p$ is called fold-dependent because different structural forms of tau fibrils have different catalytic activities. PHF and CTE tau fibrils expose different surface structures, and a recent cross-seeding study [3] found that CTE fibrils catalyze Aβ42 nucleation more strongly than PHF fibrils. Therefore,

$$\alpha_{CTE} > \alpha_{PHF} \tag{S10}$$

Physically, $\alpha_p$ is an effective heterotypic nucleation coefficient: it incorporates the ability of a particular tau fibril fold to bind Aβ monomers and promote their conversion into new Aβ nuclei. It is not a tau consumption rate, because tau acts catalytically and is not consumed by the reaction.

The fold-dependent heterotypic coefficient $\alpha_p$ was calibrated directly from [3], Supplementary Table 1B. In [3], $k_n$ denotes the fitted effective Aβ42 primary-nucleation rate constant obtained for a given concentration of preformed tau fibril seeds. Thus, $k_n$ is not an additional parameter of the present model; rather, its dependence on tau seed concentration is used here to quantify the tau-induced increase in Aβ primary nucleation. Linear regression of the fitted $k_n$ values against tau seed concentration over the range 0–1 $\mu M$ gives slopes of $6.686\ \mathrm{M}^{-1}\mathrm{h}^{-1}$ per μM of PHF seeds and $16.010\ \mathrm{M}^{-1}\mathrm{h}^{-1}$ per μM of CTE seeds. Because these slopes represent the increase in the effective Aβ primary-nucleation coefficient per unit tau fibril concentration, they correspond directly to the heterotypic coefficients $\alpha_{PHF}$ and $\alpha_{CTE}$ in the present model. Converting $\mathrm{M}^{-1}$ to $10^{-6}\ \mu\mathrm{M}^{-1}$ and $\mathrm{h}^{-1}$ to $\mathrm{s}^{-1}$ gives

$$\alpha_{PHF} = 1.86 \times 10^{-9}\ \mu\mathrm{M}^{-2}\mathrm{s}^{-1} \tag{S11}$$

and

$$\alpha_{CTE} = 4.45 \times 10^{-9}\ \mu\mathrm{M}^{-2}\mathrm{s}^{-1} \tag{S12}$$

These coefficients correspond to the fully accessible tau seed concentrations used in the *in vitro* experiments of [3]. In the tissue-scale model, only a fraction of intracellular tau fibril mass is assumed to be accessible to extracellular/interstitial Aβ; this compartmental accessibility is represented separately by $f_{acc}$.

**S1.6** $\delta_{A\beta}$. Only the active sites in a thin surface layer of a compact plaque are taken to be reachable by free Aβ monomer. The thickness of this layer is set to 0.010 μm, approximately the diameter of one Aβ42 fibril, so that about one layer of fibrils at the plaque surface is exposed. For a spherical deposit of radius $R_p$, the thin-shell approximation then gives $f_{surf} = min(1, 3\delta_{A\beta}/R_p)$ [Eq. (16)], the same closure and shell thickness as in Eq. (11) of the preceding model. The value has not been measured; Section 4.5 varies it from 0.0025 to 0.1 μm and tests two alternative closures.

**S1.7** $MW_{A\beta}$. $MW_{A\beta}$ corresponds to the molecular weight of Aβ42, as reported in [11]. The value used is $4.51\times10^{3}$ g $mol^{-1}$.

**S1.8** $\rho_{A\beta}$. Protein density is commonly approximated as 1.35 g/$cm^3$, independent of the specific protein in question [14]. Accordingly, $\rho_{A\beta}$ = 1.35 g $cm^{-3}$ = $1.35\times10^{-12}$ g $\mu m^{-3}$; the same value is used for the tau density $\rho_{tau}$ in Section S1.3.

**Tau aggregation kinetics (S1.9–S1.19). These estimates combine tau-specific measurements with inherited tissue-scale calibrations. Where different tau constructs, inducers, or kinetic models must be combined, the limitation is stated explicitly rather than treating the resulting value as a direct measurement.**

**S1.9** $n_{c,tau}$. An effective primary-nucleation order $n_{c,tau} = 1$ is used here, following the global kinetic fit of [24]; those authors note that the primary-nucleation rate is ill determined in their data and could be bounded only from above, so this order is the best-fitting rather than a tightly constrained value. Kundel et al. [23] obtained $n_c = 2$ for heparin-induced full-length wild-type tau but $n_c = 1$ for the P301S variant under otherwise identical conditions, indicating that the apparent primary-nucleation order is sensitive to sequence as well as to construct and solution conditions; their study is used here principally to constrain elongation and fragmentation.

**S1.10** $k_{o1,tau}$. The value 0.01 $h^{-1}$ = $2.78\times10^{-6}$ $s^{-1}$ is retained from the preceding model for continuity, but it should not be attributed numerically to [21]. In that study, aggregation of 2 μM full-length 2N4R tau induced by 75 μM arachidonic acid (ARA) was followed by light scattering and fitted with the Finke–Watzky two-step scheme on time courses plotted in minutes over 250 min. Figure 6A gives the wild-type (WT) $k_1$ value as approximately 0.011 on that plotted time scale; because the paper does not print a unit for $k_1$, it cannot be equated with the 0.01 $h^{-1}$ adopted here. If the fitted time unit is minutes, the rounded value 0.01 $min^{-1}$ corresponds to $1.67\times10^{-4}$ $s^{-1}$, exactly 60 times the adopted $2.78\times10^{-6}$ $s^{-1}$ (= 0.01 $h^{-1}$), because 1 h = 60 min; the approximate WT value 0.011 $min^{-1}$ read from Fig. 6A corresponds to $1.83\times10^{-4}$ $s^{-1}$, about 66 times the adopted value. The authors also state explicitly that $k_1$ and $k_2$ are used to compare nucleation and elongation rates qualitatively among WT and FTDP-17 mutants. Thus, Combs and Gamblin [21] support the use of a first-order Finke–Watzky nucleation step under those specific ARA-induced conditions, but not the numerical value adopted here. The quantitative constraint retained here is instead the upper limit $k_{+,tau}k_{n,tau} < 10^{-2}$ $M^{-1}$ $s^{-2}$ reported by [24], combined with $k_{+,tau} = 4.2\times10^{-3}$ $\mu M^{-1}$ $s^{-1}$ from [23], which gives $k_{n,tau} < 2.38\times10^{-6}$ $s^{-1}$. The inherited value $2.78\times10^{-6}$ $s^{-1}$ is approximately 17% above this bound and is therefore treated as an order-of-magnitude tissue-scale estimate used for sensitivity analysis rather than as a directly measured constant. This corrects the numerical attribution in note f of the preceding model [4]; the adopted value itself is retained here to preserve consistency with that model.

**S1.11** $k_{o2,tau}$. The value is estimated from the ratio $k_{2,tau}/k_{n,tau} > 6\times10^{13}$ $M^{-2}$ = 60 $\mu M^{-2}$ inferred from the fitted rate-constant products reported by [24] for $tau_{304-380,C322S}$ at pH 8.0: Table 1 gives $k_+k_2 = 6\times10^{11}$ $M^{-3}$ $s^{-2}$ and $k_+k_n < 10^{-2}$ $M^{-1}$ $s^{-2}$ (upper limit only; best fit $n_{c,tau} = 1$), with $n_2 = 2$ and $\sqrt{K_{2,tau}}$= 30 nM, combined with $k_{n,tau} = k_{o1,tau} = 2.78\times10^{-6}$ $s^{-1}$ (Section S1.10), giving $k_{o2,tau} > 1.67\times10^{-4}$ $\mu M^{-2}$ $s^{-1}$. As in note g of the preceding model [4], the numerical derivation is correct, but this lower bound scales linearly with the adopted $k_{o1,tau}$ and inherits the uncertainty discussed in Section S1.10. If the approximate WT $k_1$ read from [21] were instead interpreted on that experiment's minute time scale and substituted into the ratio from [24], then 0.01 $min^{-1}$ would give $k_{o2,tau} > 1.00\times10^{-2}$ $\mu M^{-2}$ $s^{-1}$, exactly 60 times the adopted lower bound; using the approximate WT value 0.011 $min^{-1}$ read from Fig. 6A would give $k_{o2,tau} >$ $1.10\times10^{-2}$ $\mu M^{-2}$ $s^{-1}$, about 66 times larger. That cross-condition substitution is not used because the studies [21] and [24] employed different tau constructs, aggregation inducers, and kinetic models. Because Eq. (S17) provides only a lower bound, and because Rodriguez Camargo et al. [24] used a multistep secondary-nucleation mechanism that saturates with $\sqrt{K_{2,tau}} \approx 30$ nM, the value obtained here should be interpreted as a low-monomer-limit, cross-construct estimate rather than as a uniquely determined *in vivo* constant. The present unsaturated term is therefore a coarse-grained approximation whose magnitude was checked directly. In the regime the model occupies, $m_{tau} \approx 9\times10^{-5}$ μM, the saturation factor $1 + m_{tau}^2/K_{2,tau}$ is $1 + 9\times10^{-6}$, so the unsaturated term is the correct limit, and saturation changes the tau secondary-nucleation rate by less than $10^{-5}$ in relative terms. The approximation is therefore immaterial in Eqs. (S32) and (S33). The same factor is important in Eq. (S20), where it is included: at the free cytosolic tau concentration of 100 nM adopted there, it equals 12.1.

The ratio $k_{2,tau}/k_{n,tau}$ of Eq. (S17) (Section S1.13), which gives the lower bound on $k_{o2,tau}$ above, also constrains, through Eq. (S14), the fibril-catalyzed dissociation constant $k_{d2,tau}$, so that $n_{c,tau}$, $n_{2,tau}$, $k_{o2,tau}$, and $k_{d2,tau}$ combine information from multiple experimental sources and should be regarded as approximate cross-construct estimates. Importantly, $k_{d2,tau}$ depends through Eq. (S14) on the ratio $k_{2,tau}/k_{n,tau}$, not on either absolute constant separately. Therefore, a common rescaling of $k_{n,tau}$ (= $k_{o1,tau}$) and $k_{2,tau}$ (= $k_{o2,tau}$), such as the 60-fold minute-versus-hour reinterpretation discussed above, leaves the $k_{d2,tau}$ estimate unchanged.

**S1.12** $k_{d1,tau}$. A consistent finding in oligomer-explicit aggregation studies is that a substantial fraction of oligomers dissociates back to monomer before conversion into growth-competent fibrils; at representative intracellular tau concentrations, the relevant kinetic comparison is $k_{conv,tau} m_{tau} < k_{d1,tau}$ [2], [22].

NMR relaxation and EPR studies of amyloid oligomers report monomer-exchange and oligomer-dissociation times ranging from milliseconds (monomer exchange with large oligomers) to a few hours (oligomer dissociation). Two independent measurements on other amyloid proteins illustrate this range: monomer release from large, NMR-invisible Aβ(1–40) "dark-state" oligomers occurs on a millisecond scale ($k_{d1}$ = 73 $s^{-1}$ [25]), whereas α-synuclein oligomer dissociation occurs on a time scale of several hours or less ($k_{d1} > 5.6\times10^{-5}$ $s^{-1}$, i.e., > 0.2 $h^{-1}$ [26]). Neither measurement concerns tau directly; for tau, an order-of-magnitude estimate within this range is adopted:

$$k_{d1,tau} \approx 10^{-4}\text{–}10^{-3}\ s^{-1} \tag{S13}$$

For wild-type K18 tau, Shammas et al. [22] directly fitted an oligomer dissociation rate of 0.61 $h^{-1}$ = $1.69\times10^{-4}$ $s^{-1}$. This tau-specific value lies within the adopted order-of-magnitude range and supports the value $1.69\times10^{-4}$ $s^{-1}$ used in the dissociation scenario (Table S1).

**S1.13** $k_{d2,tau}$. Direct measurements of fibril-surface-mediated dissociation of tau oligomers are not presently available. The corresponding rate constant $k_{d2,tau}$ was therefore estimated from the experimentally measured catalytic effect of tau fibril surfaces on oligomer formation. Dear et al. [7] showed that a fibril surface that catalyzes oligomer formation should also accelerate the reverse oligomer-dissociation process by approximately the same kinetic factor. In the notation used in that kinetic framework,

$$k_{d2,tau} \simeq \frac{2}{3}\frac{k_{2,tau}}{k_{n,tau}} m_{tau}^{n_{2,tau}-n_{c,tau}} k_{d1,tau} \tag{S14}$$

Here, $k_{n,tau}$ denotes the intrinsic primary-nucleation rate constant for tau aggregation, $k_{2,tau}$ denotes the fibril-surface-mediated secondary-nucleation rate constant, $n_{c,tau}$ is the reaction order with respect to tau monomer concentration for primary nucleation, and $n_{2,tau}$ is the corresponding reaction order for secondary nucleation. These quantities are introduced here only because they are the parameters used in [24] to characterize the relative catalytic strength of primary and fibril-surface-mediated tau nucleation. They are not additional parameters of the present model and do not enter its governing equations independently; they are used only to obtain a literature-based estimate of $k_{d2,tau}$.

Rodriguez Camargo et al. [24] reported for the $tau_{304-380,C322S}$ fragment

$$k_{+,tau}k_{2,tau} = 6 \times 10^{11}\ \mathrm{M^{-3}s^{-2}} \tag{S15}$$

and

$$k_{+,tau}k_{n,tau} < 10^{-2}\ \mathrm{M^{-1}s^{-2}} \tag{S16}$$

with $n_{2,tau} = 2$ and $n_{c,tau} = 1$. Because the same elongation constant $k_{+,tau}$ appears in both fitted products, it cancels when their ratio is taken, giving

$$\frac{k_{2,tau}}{k_{n,tau}} > 6 \times 10^{13}\ \mathrm{M^{-2}} \tag{S17}$$

Substitution of $n_{2,tau} = 2$ and $n_{c,tau} = 1$ into the expression above gives

$$k_{d2,tau} \gtrsim \frac{2}{3}(6 \times 10^{13}\ M^{-2})m_{tau}k_{d1,tau} \tag{S18}$$

For wild-type K18 tau, Shammas et al. [22] reported a free-oligomer dissociation rate of approximately

$$k_{d1,tau} = 0.61\ h^{-1} = 1.69 \times 10^{-4}\ \mathrm{s^{-1}} \tag{S19}$$

The secondary-nucleation mechanism fitted by [24] is multistep and saturates. If the saturation constant of that mechanism is written as $K_{2,tau}$, which has units of μM$^2$ because the reaction order is $n_{2,tau} = 2$, the unsaturated ratio in Eq. (S18) must be divided by the factor $1 + m_{tau}^2/K_{2,tau}$; the concentration at which secondary nucleation is half saturated is $\sqrt{K_{2,tau}} \approx 30$ nM. The relevant monomer concentration is that of the free cytosolic tau available for aggregation, for which Kundel et al. [23] adopt 100 nM in human neurons, with the great majority of intracellular tau being bound to microtubules. Evaluating Eq. (S18) at $m_{tau} = 100$ nM $= 1\times10^{-7}$ M, where the saturation factor is 12.1, gives

$$\frac{2}{3}(6 \times 10^{13})(1 \times 10^{-7})/12.1 = 3.3 \times 10^{5}\ \mathrm{M^{-1}} = 0.33\ \mathrm{\mu M^{-1}} \tag{S20}$$

Therefore,

$$k_{d2,tau} \approx 0.33k_{d1,tau} = 5.6 \times 10^{-5}\mathrm{\mu M^{-1}s^{-1}} \tag{S21}$$

Accordingly,

$$k_{d2,tau} = 5.6 \times 10^{-5}\mathrm{\mu M^{-1}s^{-1}} \tag{S22}$$

is the value used in the dissociation scenario. Because the estimate combines kinetic measurements obtained for different tau repeat-domain constructs and because the primary-nucleation contribution in [24] is reported only as an upper bound, the resulting value should be regarded as an approximate, lower-bound estimate. The influence of $k_{d2,tau}$ on the model predictions is therefore included in the dissociation scenario (Section S2.3). The alternative minute-based interpretation of [21] discussed in Section S1.10 does not change $k_{d2,tau}$, because Eq. (S14) uses only the ratio $k_{2,tau}/k_{n,tau}$; it would rescale $k_{o1,tau}$ and $k_{o2,tau}$ together by the same factor.

**S1.14** $k_{+,tau}$. Kundel et al. [23] used single-aggregate (SAVE) TIRF imaging to track the length distribution of full-length tau fibrils over aggregation times of up to two months, at monomer concentrations of 2 and 10 µM *in vitro*. The resulting length-versus-time data were fitted to a three-step kinetic model comprising primary nucleation, saturated elongation, and fragmentation (their Eq. (2), which adds saturation of elongation to the nucleation-elongation-fragmentation framework of [20]). The fitted elongation rate constant for WT tau (Fig. 2b in [23]) is

$$k_{+,tau} = 4.2\times10^{-3}\ \mu\text{M}^{-1}\ \text{s}^{-1}\ \text{(equivalently } 4.2\times10^{3}\ \text{M}^{-1}\ \text{s}^{-1}\ \text{as reported)} \qquad \text{(S23)}$$

The elongation constant $k_{+,tau}$ was fitted by [23] within a saturated-elongation form, $2k_{+,tau}\, m_{tau}\, P_{tau}/(1 + m_{tau}/K_{E,tau})$, with $K_{E,tau}$ = 3.5 µM, the critical concentration above which elongation becomes independent of monomer concentration. This constant is unrelated to $K_{2,tau}$ of Section S1.13, which governs the saturation of secondary nucleation and has different units. The present model operates at $m_{tau} \approx 9\times10^{-5}$ µM, so the saturation factor is $1 + 3\times10^{-5}$ and the unsaturated expression used in Eqs. (S32) and (S35) is the appropriate limit; saturated and unsaturated integrations were verified to agree to five significant figures.

**S1.15** $k_{-,tau}$. The same fit gives the fragmentation rate constant

$$k_{-,tau} = 5.6\times10^{-11}\ \text{s}^{-1} \qquad \text{(S24)}$$

The measurements of [23] directly constrain fibril elongation and fragmentation and therefore support the fragmentation term in Eq. (S34). They do not establish fragmentation as the predominant source of soluble oligomers. The isotope-labeling experiments of [24] instead show that the oligomeric intermediates are generated predominantly from soluble monomers by fibril-surface secondary nucleation, which is why that process appears in Eqs. (S32) and (S33).

**S1.16** $k_{conv,tau}$. Once a non-growth-competent tau oligomer forms, it may either dissociate back into monomers or undergo a structural conversion into a growth-competent fibrillar species. Shammas et al. [22] modeled this monomer-dependent oligomer-to-fibril conversion as a bimolecular process. The corresponding kinetic term has the same form as $k_{conv,tau} m_{tau} S_{tau}$ in the present model. For wild-type K18 tau, the conversion rate constant fitted in Supplementary Table 1 of [22] corresponds, in the present notation, to

$$k_{conv,tau} = (3.2 \pm 1.5) \times 10^{-2}\ \mu\text{M}^{-1}\text{h}^{-1} = (8.9 \pm 4.2) \times 10^{-6}\ \mu\text{M}^{-1}\text{s}^{-1} \qquad \text{(S25)}$$

The central value,

$k_{conv,tau} = 8.9 \times 10^{-6}\ \mu\text{M}^{-1}\text{s}^{-1}$ (S26)

is adopted as the baseline estimate in the present calculations.

**S1.17** $n_{min,tau}$. In Eq. (S34), fragmentation acts only on the fibril mass in excess of $n_{min,tau}P_{tau}$, so that no fragment shorter than $n_{min,tau}$ monomers is produced. Without this cutoff the mean tau fibril size would fall to about $2/(k_{-,tau}t)$, roughly 34 monomers (about 10 nm) after 33 years, far below the smallest aggregates (about 100 nm) resolved in the imaging experiments from which $k_{-,tau}$ was obtained [23]. The cutoff of $2\times10^3$ monomers corresponds to a filament length of about 590 nm. It is an assumed value; for any $n_{min,tau}$ above about $10^3$, fragmentation no longer contributes to $P_{tau}$, and $P_{tau}$, $M_{tau}/P_{tau}$, and $m_{tau}$ are unchanged.

**S1.18** $\lambda_{m,tau}$. $T_{1/2,m,tau} = 2.16\times10^5$ s (60 h) is adopted from [17]. The discussion of Fig. 1A in the Results section of that study explicitly states that the half-life of tau is approximately 60 h in HT22 cells; 60 h equals 216,000 s. This pulse-chase measurement characterizes cellular tau turnover rather than a free-monomer-specific half-life, so the value is used here as an effective clearance proxy. It is used only in the clearance scenario; the baseline sets $T_{1/2,m,tau} \to \infty$. The results are insensitive to it because $m_{tau}$ remains near $10^{-4}$ μM, so that monomer degradation removes negligible tau mass.

**S1.19** $\lambda_{S,tau}$. A direct experimental measurement of the intracellular degradation half-life of free tau oligomers is not currently available. The closest quantitative *in vivo* estimate is provided by measurements of soluble tau turnover in a mouse model of tauopathy. Yamada et al. [18] reported a half-life of approximately 9.7 days for soluble tau, whereas insoluble tau turned over considerably more slowly, with a half-life of approximately 34.2 days. Because the variable $S_{tau}$ in the present model represents a soluble, nonfibrillar tau oligomer pool, the soluble-tau turnover time provides a more appropriate experimental proxy than the insoluble-tau value. Accordingly,

$T_{1/2,S,tau} = 9.7\ \text{days} = 8.38 \times 10^5\ \text{s}$ (S27)

is used in the clearance scenario. This estimate is consistent in order of magnitude with measurements in human iPSC-derived neurons, for which a tau half-life of $6.74 \pm 0.45$ days has been reported [19]. Because neither study isolates tau oligomers specifically, $T_{1/2,S,tau}$ should be regarded as an experimentally constrained proxy for the turnover of the soluble tau pool rather than as a directly measured oligomer-specific half-life.

**Aβ aggregation kinetics and simulation horizon (S1.20–S1.29). These notes establish the Aβ nucleation, elongation, conversion, dissociation, clearance, and secondary-nucleation parameters**

**used in the numerical studies and distinguish baseline values from quantities explored mainly by sensitivity analysis.**

**S1.20** $k_{o1,A\beta}$. For intrinsic Aβ primary nucleation, Mosconi et al. [3] report fitted values for the tau-free control rate constant corresponding to $k_{o1,A\beta}$ in the present notation. The values are $0.78\ \mathrm{M^{-1}h^{-1}}$ for the PHF seed series and $0.60\ \mathrm{M^{-1}h^{-1}}$ for the CTE seed series, equivalent to $2.17 \times 10^{-10}$ and $1.67 \times 10^{-10}\ \mathrm{\mu M^{-1}s^{-1}}$, respectively. Dear et al. [29] give the corresponding intrinsic primary-nucleation constant as $1.1 \times 10^{-6}\ \mathrm{\mu M^{-1}h^{-1}} = 3.06 \times 10^{-10}\ \mathrm{\mu M^{-1}s^{-1}}$. Accordingly, $k_{o1,A\beta} = 2.17 \times 10^{-10}\ \mathrm{\mu M^{-1}s^{-1}}$, based on the PHF seed series of [3], is adopted as the baseline value, while the range $1.7\text{–}3.1 \times 10^{-10}\ \mathrm{\mu M^{-1}s^{-1}}$ spans the reported values (Table S1).

**S1.21** $k_{+,A\beta}$. For the Aβ fibril elongation rate constant $k_{+,A\beta}$, Supplementary Table 1 of Mosconi et al. [3] reports $k_{+,A\beta} = 3.5 \times 10^{8}\ \mathrm{M^{-1}h^{-1}}$, equivalent to $9.72 \times 10^{-2}\ \mathrm{\mu M^{-1}s^{-1}}$. This value is adopted as the baseline because it is consistent with the kinetic framework used in [3] to calibrate tau-mediated Aβ nucleation. Table S3 of Dear et al. [29] reports a larger Aβ elongation rate constant, $k_{+,A\beta} = 1.1 \times 10^{4}\ \mathrm{\mu M^{-1}h^{-1}} = 3.06\ \mathrm{\mu M^{-1}s^{-1}}$, which is listed for comparison. This comparison value is approximately 31.5 times larger than the baseline value, which indicates the uncertainty in $k_{+,A\beta}$.

**S1.22** $k_{conv,A\beta}$. Dear et al. [29] report for Aβ42 a pseudo-first-order oligomer-to-fibril conversion rate $k_c$ = $3.3\times10^{-2}\ \mathrm{h^{-1}}$ at an initial monomer concentration $m_0$ = 5 µM. With the present conversion term $k_{conv,A\beta} m_{A\beta} S_{A\beta}$ and $n_{conv}$ = 1, $k_{conv,A\beta} = k_c/m_0$ = $6.6\times10^{-3}\ \mathrm{\mu M^{-1}\ h^{-1}} = 1.83\times10^{-6}\ \mathrm{\mu M^{-1}\ s^{-1}}$. This concentration-normalized value is derived, not directly reported: the study gives the pseudo-first-order conversion rate at 5 µM, and the present model assumes first-order monomer dependence for Aβ42, although the underlying oligomer fit reported a conversion reaction order of about 2.7 [2]; at the sub-micromolar monomer concentrations of the model, the first-order form therefore gives faster conversion than an extrapolation with that order would.

**S1.23** $k_{-,A\beta}$. The Aβ fibril fragmentation rate constant, $k_{-,A\beta}$, was estimated from the kinetic measurements of Cook and Leyh [30], who fitted Aβ42 fibril-formation data using a nucleation–elongation–fragmentation model. For unliganded Aβ42, they reported a fragmentation rate constant of $(3.5 \pm 0.2) \times 10^{-11}\ \mathrm{s^{-1}}$. This literature value indicates that Aβ42 fibril fragmentation is a very slow process under quiescent conditions. In particular, the corresponding characteristic fragmentation time scale, $1/k_{-,A\beta}$, is approximately $9 \times 10^{2}$ years, which is far longer than the 33-year simulation period considered here. Fragmentation is therefore omitted in the reference case, with $k_{-,A\beta} = 0$; $3.5\times10^{-11}\ \mathrm{s^{-1}}$ is used in Section S2.3, and smaller values giving mean fibril lengths of 0.3–30 μm are used in Sections 4.5 and 4.6.

**S1.24** $k_{d1,A\beta}$. Dear et al. [29] report a combined Aβ42 oligomer dissociation rate of $k_d$ = 0.36 h$^{-1}$ = 1.00×10$^{-4}$ s$^{-1}$, where $k_d = k_{d1,A\beta} + k_{d2,A\beta} M_{A\beta}$. In the kinetic model of [29], oligomers are treated explicitly as a separate species. The data do not uniquely identify $k_{d1,A\beta}$ and $k_{d2,A\beta}$ separately. The present pair $k_{d1,A\beta}$ = 6.47×10$^{-7}$ s$^{-1}$ and $k_{d2,A\beta}$ = 1.99×10$^{-5}$ μM$^{-1}$ s$^{-1}$ is not, however, an arbitrary decomposition of that total, because a second and independent constraint fixes the ratio of the two constants. Eq. (2) of [7], evaluated with $n_2 = n_c = 2$ for Aβ42, gives $k_{d2,A\beta}/k_{d1,A\beta} = (2/3)(k_{o2,A\beta,Mref}/k_{o1,A\beta})$ = 30.7 μM$^{-1}$, with no dependence on monomer concentration. Imposing this ratio together with the fitted total at the reference fibril-mass concentration $M_{A\beta}$ = 5 μM, $k_d$ = 1.00×10$^{-4}$ s$^{-1}$ determines both constants uniquely, giving $k_{d1,A\beta}$ = 6.47×10$^{-7}$ s$^{-1}$ and $k_{d2,A\beta}$ = 1.99×10$^{-5}$ μM$^{-1}$ s$^{-1}$. The total dissociation rate itself remains an experimental input. Thus, the individual free and fibril-associated dissociation coefficients are derived from the reported total rate and the independent catalytic ratio; neither is directly reported.

**S1.25** $k_{d2,A\beta}$. The individual value of $k_{d2,A\beta}$ is not separately resolved by [29]. The value 1.99×10$^{-5}$ μM$^{-1}$ s$^{-1}$ (= 7.15×10$^{-2}$ μM$^{-1}$ h$^{-1}$) follows from combining the fitted total rate with the catalytic ratio $k_{d2,A\beta}/k_{d1,A\beta}$ = 30.7 μM$^{-1}$ of [7], Eq. (2), as set out in Section S1.24. Together with $k_{d1,A\beta}$ = 6.47×10$^{-7}$ s$^{-1}$, it reproduces the fitted total dissociation rate $k_d$ = 0.36 h$^{-1}$ at $M_{A\beta}$ = 5 μM exactly. Because the partition is fixed by an independent relation rather than chosen, a separate sensitivity analysis of it is not required; the remaining experimental uncertainty resides in the fitted total $k_d$.

**S1.26** $\lambda_{m,A\beta}$. The half-life of Aβ monomers, $T_{1/2,m,A\beta}$, is taken as 1.28 hours [27]. This is the Aβ turnover half-life assumed in the brain-slice diffusion model of that study, within the range of 0.7–3.8 h published for mice; it is used here to give an effective free-monomer clearance rate, ln(2)/(1.28 h) ≈ 1.50×10$^{-4}$ s$^{-1}$, in the clearance scenario only.

**S1.27** $\lambda_{S,A\beta}$. Direct measurements of the proteolytic half-life of free Aβ oligomers in brain tissue are not available. However, Takeda et al. [28] separated brain-interstitial-fluid Aβ by size-exclusion chromatography into high-molecular-weight (HMW; fractions 1–11), low-molecular-weight (LMW; fractions 14–19, described as LMW oligomers, which include dimers), and monomer (fractions 20–22) pools in awake APP/PS1 mice. After acute γ-secretase inhibition, the pooled 8-h LMW Aβ42 signal was 88.0% lower than baseline, leaving approximately 12% of the initial concentration. The authors stated that the *in vivo* half-life of interstitial-fluid Aβ oligomers could not be determined; therefore, treating the pooled 8-h change as a single first-order decay is an approximation. This gives

$$\lambda_{S,A\beta} = \frac{-\ln(0.12)}{8\ h} = 0.265\ \mathrm{h}^{-1} \tag{S28}$$

and hence

$$T_{1/2,S,A\beta} = \frac{\ln 2}{\lambda_{S,A\beta}} = 2.62\ h \approx 9.4 \times 10^3\ \mathrm{s} \tag{S29}$$

Accordingly, $T_{1/2,S,A\beta} = 2.6$ h is used in the clearance scenario as an effective first-order clearance proxy for low-molecular-weight Aβ42 oligomers, not as a half-life measured in that study. Because the post-inhibition value is a single 8-h collection, the 12% is a time average rather than an end-point concentration; interpreted as the time average of an exponential decay, it would give $T_{½} \approx 0.7$ h, so the adopted value should be regarded as an upper estimate of the oligomer half-life. Applying the same first-order approximation to the HMW decreases reported in the same experiment gives $T_{½} \approx 19$ h ($\lambda \approx 1.0\times10^{-5}\ \mathrm{s}^{-1}$) for HMW Aβ42 and $T_{½} \approx 12$ h ($\lambda \approx 1.6\times10^{-5}\ \mathrm{s}^{-1}$) for HMW Aβ40 (about 9 h and 5.5 h, respectively, under the time-average reading). Because $S_{A\beta}$ represents an oligomer-equivalent pool rather than a size-resolved species, $\lambda_{S,A\beta} \approx (1.0–3.5)\times10^{-5}\ \mathrm{s}^{-1}$ (spanning the end-point and time-average readings) is a useful candidate sensitivity range if larger oligomers contribute substantially; this slower range is not used in the present simulations.

**S1.28** $k_{o2,A\beta,Mref}$. For homotypic Aβ fibril-surface secondary nucleation, the rate constant denoted $k_{o2,A\beta}$, written $k_{o2,A\beta,Mref}$ in Table S1, is taken from [29]; that study reports $3.6 \times 10^{-5}\ \mu\mathrm{M}^{-2}\mathrm{h}^{-1} = 1.00 \times 10^{-8}\ \mu\mathrm{M}^{-2}\mathrm{s}^{-1}$; with $\sigma_{ref} = 1/143$ it gives the conditional per-site coefficient $k_{o2,A\beta,D,p}$ of Eq. (S41). Supplementary Table 1A of Mosconi et al. [3] gives an unsaturated Aβ42 self-aggregation value equivalent to $2.78 \times 10^{-9}\ \mu\mathrm{M}^{-2}\mathrm{s}^{-1}$, 3.6 times smaller. Both values are unsaturated, low-monomer estimates; the multistep saturated fits preferred in [3] determine only the product of $k_2$ and its saturation constant over the monomer range studied and therefore do not constrain $k_2$ at the low monomer concentrations of the model [4]. The value of [29] is adopted because it belongs to the same Aβ42 parameter set as $k_{conv,A\beta}$ and the total oligomer dissociation rate (it is the bulk-fit value with which those oligomer rates were fitted; Sections S1.22, S1.24), and it also fixes the ratio $k_{d2,A\beta}/k_{d1,A\beta}$ in Section S1.24, so that the inherited Aβ oligomer kinetics are mutually consistent.

**S1.29** $t_{end}$. The average age at onset of AD symptoms is about 75 years [33]. However, because brain accumulation of Aβ species and plaques is thought to precede clinical manifestations by up to 20–30 years, taken here as about 25 years [34], the onset of pathological changes, $t_{AD,onset}$, is estimated to occur around age 50.

The clinical phase is taken as 8 years, as in the preceding model [4]. Zanetti et al. [32] report a life expectancy after AD diagnosis of 3–10 years, with median survival decreasing with age at diagnosis from 8.3 years at 65 to 3.4 years at 90 [31], so 8 years lies toward the upper end for diagnosis at about 75–77 years, for which these medians suggest about 6 years. The total estimated disease timeline, $t_{end}$, from the onset of Aβ deposition to death, spans 33 years: 25 years of preclinical progression plus 8 years of symptomatic disease; the interval of about 2 years between symptom onset and diagnosis [33] is within the precision of this estimate. Because $t_{end}$ only sets the end of the simulated interval, a shorter clinical phase corresponds to evaluating the same time courses earlier, for example at 31 years for a clinical phase of 6 years.

Taken together, Section S1 deliberately separates quantities constrained by experiment from those needed to close the tissue-scale model. Section S2 then tests the assumptions that are least directly constrained, so that the conclusions can be separated from the particular baseline calibration.

## S2. Robustness analyses: which conclusions survive changes in the assumptions?

Section 4.5 of the main text summarizes the outcome; this section shows how that outcome is obtained. The tests proceed from local active-site physics to whole-system controls: growth-history-dependent site formation and masking (Sections S2.1–S2.2), oligomer dissociation, soluble-protein clearance, and fragmentation (Section S2.3), plaque geometry and broad parameter uncertainty (Section S2.4), explicit morphology dependence (Section S2.5), and finally the possibility that tau-nucleated fibrils have a distinct catalytic phenotype (Section S2.6). The common question is whether tau can increase homotypic Aβ secondary nucleation at matched fibril mass, rather than merely increasing total fibril burden at equal time.

### S2.1. Does growth history change active-site formation?

A further set of simulations makes $\eta_{grow}$ depend on monomer supersaturation. This tests whether enhanced tau-mediated nucleation reduces active-site formation by depleting the monomer pool, or whether changes in the number and growth histories of fibrils instead increase total active-site abundance. Eq. (14) also fixes the direction of the instantaneous effect. Because $\eta_{grow}$ is a non-decreasing function of $u$ when $\eta_{max} \geq \eta_{min}$, a tau-coupled system whose free-monomer concentration at a given time is lower than that of the uncoupled system necessarily has an equal or lower elongation-site yield at that time. Within this law, any increase in total active-site abundance caused by tau-mediated nucleation must therefore arise from fibril birth or from a shift of monomer incorporation toward earlier times of higher supersaturation, not from a higher yield at a given monomer concentration. Because $\eta_{grow}$ then varies in time, Eq. (24) no longer applies, and the full system must be integrated.

The parameters in Eq. (14) are treated as sensitivity parameters rather than assigned precise physiological values, because they have not been measured for tau-seeded Aβ42. Accordingly, Eq. (14) is used only in the growth-history analysis and is not part of the reference implementation. Its qualitative direction, however, is supported by the Aβ40 experiments of Hu et al. [5]. In those experiments, seeded fibrils grown close to equilibrium by gradually lowering the temperature from 60 to 37 °C contained about one active site per 834 monomers, compared with about one site per 99 monomers for fibrils grown at 37 °C. The lower site density under near-equilibrium growth is therefore consistent with the assumption in Eq. (14) that active-site yield increases with monomer supersaturation. However, because these measurements compare only two growth conditions for Aβ40, they do not establish the functional form or parameter values of Eq. (14), nor do they provide a quantitative law for tau-seeded Aβ42.

With the illustrative values $m_{eq,A\beta} = 10^{-3}$ μM, $\eta_{min} = \sigma_0/4$, $\eta_{max} = 2\sigma_0$, $K_u = 3$, and $h = 1$, the supersaturation $u$ falls from about 30 in the first months to 1.8, 1.5, and 0.6 at 33 years for $f_{acc} = 0$ and for PHF coupling at $f_{acc} = 0.1$ and 1 (Fig. S20a). The elongation yield follows $u$ (thin lines in Fig. S21a), and its elongation-flux-weighted mean over the 33 years equals $1.19\sigma_0$, $1.14\sigma_0$, and $0.93\sigma_0$ for the same three cases. Tau-mediated monomer depletion therefore lowers the active-site density: at 33 years, $\sigma_{D,A\beta}/\sigma_0 = 1.19$ for $f_{acc} = 0$, but 0.93 and 0.79 for PHF and CTE coupling at $f_{acc} = 1$ (Fig. S21a), so that $D_{A\beta,eff}$ is 22% (PHF) and 34% (CTE) lower than without tau coupling. The accessible-site ratio in Eq. (27) is then smaller than unity and reinforces the monomer penalty; $R_{2,A\beta}$ at 33 years is 25% (PHF) and 13% (CTE) of its uncoupled value. The tau-induced increase in $\Xi_{A\beta}$(33 yr) is correspondingly smaller than in the reference case: 3.15- instead of 3.82-fold for PHF, and 5.17- instead of 6.34-fold for CTE coupling at $f_{acc} = 1$ (Fig. S20b). Within this law, tau coupling decreases rather than increases active-site formation. Because the values of $m_{eq,A\beta}$, $K_u$, and $h$ are illustrative rather than calibrated, they were varied over $m_{eq,A\beta} = 10^{-4}$–$3\times10^{-2}$ μM, $K_u = 0.1$–30, and $h = 1$ and 2 (72 combinations; Fig. S6), and $\eta_{max}/\eta_{min}$ was varied from 1 to 128 with $\eta_{max} = 2\sigma_0$. For PHF coupling at $f_{acc} = 1$, $D_{A\beta,eff}$ at 33 years was lower than or equal to its uncoupled value in every combination (ratios 0.60–1.00; equal-time and matched-mass values coincide), $R_{2,A\beta}$ decreased to 0.12–0.38 of its uncoupled value, and $\Xi_{A\beta}$(33 yr) increased 2.6- to 8.6-fold. As $\eta_{max}/\eta_{min}$ increased from 1 (constant yield) to 2, 8, 32, and 128, the $D_{A\beta,eff}$ ratio decreased from 1.00 to 0.91, 0.78, 0.73, and 0.72. The reduction of $D_{A\beta,eff}$ vanishes in two limits in which the elongation yield is effectively constant: when $m_{eq,A\beta}$ is small and $u \gg K_u$, $\eta_{grow}$ stays close to $\eta_{max}$, and when $m_{eq,A\beta}$ exceeds the late free-monomer concentration, most of the fibril mass is deposited with $\eta_{grow}$ close to $\eta_{min}$. The reduction is largest, up to 40%, when $m_{eq,A\beta}$ lies between the late free-monomer concentrations of the coupled and uncoupled systems. The direction of the effect is thus fixed by the monotonic form of Eq. (14), whereas its magnitude depends on parameters that require the measurements listed in Section 5.

Interpretation. This test asks whether tau can change catalytic activity indirectly by changing the conditions under which Aβ fibrils grow. For the constitutive law examined here, tau-driven monomer depletion generally lowers rather than raises the elongation-site yield. Growth history therefore does not provide an automatic route from shorter fibrils to higher active-site density; the sign follows the assumed relation between monomer supersaturation and site formation.

### S2.2. Can masking erase catalytic sites without removing fibril mass?

The masking coefficient $k_{mask}$, which removes catalytic activity without removing fibril mass, is zero in the reference case. Positive values provide a direct test of whether the conclusions obtained from the matched-mass calculations require long-lived active sites. Hu et al. [5] argue that the very slow exchange of protein in fibril interiors suggests that structural growth defects persist in mature fibrils for long periods, which motivates examining small masking rates, but this argument does not establish that every catalytic site is permanent or justify setting $k_{mask}$ to zero universally. Because masking removes sites at a rate proportional to $D_{A\beta}$, Eq. (24) no longer holds when $k_{mask} > 0$: the site concentration then depends on when the sites were formed, not only on the net changes in fibril mass and number.

Masking lowers the site density on the time scale $1/k_{mask}$ (Fig. S21b): at 33 years, $\sigma_{D,A\beta}/\sigma_0$ equals 0.95, 0.62, 0.096, and 0.0096 for $k_{mask} = 10^{-10}$, $10^{-9}$, $10^{-8}$, and $10^{-7}$ s$^{-1}$. Because the fibril-mass histories are nearly identical, these values are the same for $f_{acc} = 0$ and 1 (symbols in Fig. S21b), and masking creates no matched-mass difference in the mass-proportional case. When birth-generated sites dominate ($\eta_{birth}/(2\eta_{grow}) = 10^5$), the relative increase in $D_{A\beta}$ caused by tau coupling at $f_{acc} = 1$ grows from 19% at $k_{mask} = 0$ to 38% at $k_{mask} = 10^{-7}$ s$^{-1}$ (Fig. S7b), because with masking $D_{A\beta}$ reflects recent rather than cumulative site formation; the sign of the difference does not change. Masking weakens the homotypic pathway: without tau coupling $\Xi_{A\beta}$(33 yr) falls to 0.13 of its reference value at $k_{mask} = 10^{-7}$ s$^{-1}$, whereas for $f_{acc} = 1$ it decreases by only 8%, because exposure is then dominated by the heterotypic channel (Fig. S7a). The relative effect of tau coupling on oligomer-equivalent exposure therefore increases with $k_{mask}$, from 3.8-fold to 26.5-fold, although the absolute exposure for $\chi_\tau = 1.86\times10^{-9}$ μM$^{-2}$ s$^{-1}$ changes little (from $1.18\times10^5$ to $1.09\times10^5$ μM s); the larger ratio reflects the smaller uncoupled exposure ($4.09\times10^3$ μM s at $k_{mask} = 10^{-7}$ s$^{-1}$).

Interpretation. Masking changes the absolute catalytic-site burden, but when it acts equally on tau-coupled and uncoupled fibrils, it does not by itself create a matched-mass difference. It is therefore important for the magnitude of secondary nucleation, not for the direction of the central tau-versus-control comparison.

**S2.3. What happens when oligomers dissociate, soluble proteins clear, or fibrils fragment?**

Oligomer dissociation and soluble-protein clearance are switched on separately using the scenario values in Table S1, after the active-site mechanisms have been characterized in the no-loss reference case. In these scenarios, fibril mass is no longer guaranteed to be nearly independent of $f_{acc}$, so equal-time comparisons need not be matched-mass comparisons. The comparisons of Sections 4.3 and 4.4 are therefore repeated at times when the compared systems have equal $M_{A\beta}$, obtained by interpolating the computed fibril-mass histories, and reported alongside the equal-time results. The clearance scenario removes soluble monomers and oligomers only; fibrils and plaques are not cleared in Eqs. (11)–(13), and slow fibril removal is examined separately in Section S2.4.

With oligomer dissociation, most oligomers revert to monomer before they are converted into fibrils, and $\Xi_{A\beta}$(33 yr) falls by more than two orders of magnitude, to 66 μM s for $f_{acc}$ = 0 (Fig. S22a). Free monomer remains high ($6.8\times10^{-2}$ μM at 33 years; Fig. S8a), fibril birth is slow, and $\bar{n}_F$ reaches $1.4\times10^{7}$. Fibril mass is again nearly independent of $f_{acc}$ (228.7–228.8 μM; Fig. S8b), so that the matched-mass times differ from 33 years by less than 0.03 years and equal-time and matched-mass comparisons coincide. The conclusions of the reference case are unchanged: $D_{A\beta,eff}$ is the same in all simulations, $R_{2,A\beta}$ is reduced by the squared monomer ratio to 80% (PHF, $f_{acc}$ = 0.1), 39% (PHF, $f_{acc}$ = 1), and 27% (CTE, $f_{acc}$ = 1) of its uncoupled value, and $\Xi_{A\beta}$(33 yr) increases by factors of 1.24, 2.53, and 3.71. The dissociation scenario also provides a more realistic description of oligomer kinetics. Without dissociation or clearance, oligomers are removed only by conversion, at the rate $k_{conv,A\beta}m_{A\beta}$, and their lifetime $1/\Lambda_{A\beta}$ [Eq. (S38)] reaches 5.5 years at 33 years in the uncoupled reference case (10.7 and 13.8 years for PHF and CTE coupling at $f_{acc}$ = 1), whereas with dissociation it is about 4 minutes, consistent with the kinetic measurements cited in Section 1, according to which most Aβ42 oligomers dissociate rather than convert into fibrils. The reference case therefore overstates the absolute exposure by more than two orders of magnitude, but the tau-induced ratios are similar (1.24–3.71 with dissociation and 1.41–6.34 in the reference case for the same coupling cases): the ratios, not the absolute exposures, are the robust outputs.

Soluble-protein clearance changes the response qualitatively (Fig. S22b). With an Aβ monomer half-life of 1.28 h, clearance fixes the free-monomer concentration at $q_{A,0,A\beta}T_{1/2,m,A\beta}/(L\ \ln(2)) \approx 1.46\times10^{-3}$ μM, independent of $f_{acc}$ (Fig. S8a), and aggregation becomes very slow: $M_{A\beta}$(33 yr) is only $2.6\times10^{-6}$ μM for $f_{acc}$ = 0. Fibrillar Aβ then originates mainly from tau-catalyzed nucleation, and $M_{A\beta}$(33 yr) increases to $6.9\times10^{-5}$, $6.6\times10^{-4}$, and $1.6\times10^{-3}$ μM for PHF coupling at $f_{acc}$ = 0.1 and 1 and CTE coupling at $f_{acc}$ = 1 (Fig. S8b). Because $m_{A\beta}$ is unchanged, tau coupling now raises $D_{A\beta,eff}$ by factors of 26, 150, and 268, and Eq. (27) is satisfied; $R_{2,A\beta}$ increases by the same factors, and $\Xi_{A\beta}$(33 yr) increases 39-, 383-, and 915-fold, from $6.55\times10^{-3}$ μM s to 0.257, 2.51, and 5.99 μM s. These large ratios arise mainly because the uncoupled exposure is extremely small: even the largest of these exposures is more than four orders of magnitude below the corresponding exposure without clearance (Table S2). The deposits remain small enough that $f_{surf}$ = 1 for $f_{acc}$ = 0, whereas $f_{surf}$ decreases to 0.59 (PHF) and 0.44 (CTE) for $f_{acc}$ = 1. At matched fibril mass, however, coupled and uncoupled systems have the same $D_{A\beta,eff}$ and $m_{A\beta}$, and hence the same $R_{2,A\beta}$; the PHF, $f_{acc}$ = 1 system reaches the final uncoupled fibril mass after 5.2 years, when its $\Xi_{A\beta}$ is already 9.5 times larger. Under clearance, the increase in homotypic secondary nucleation therefore reflects the larger fibril burden produced by tau coupling, not a change in site density. Section 4.6 shows that this behavior is characteristic of a clearance-controlled monomer pool ($Da \ll 1$) rather than an exception, and examines a scenario in which the Aβ source is raised so that substantial deposits form despite clearance.

Fibril fragmentation at the unliganded rate $k_{-,A\beta} = 3.5\times10^{-11}$ $s^{-1}$ increases $P_{A\beta}$(33 yr) by four orders of magnitude, to 4.2 μM, and reduces $\bar{n}_F$ to 55 (Fig. S9a). Free monomer is then consumed by elongation almost as fast as it is supplied ($m_{A\beta} = 2.7\times10^{-7}$ μM, independent of $f_{acc}$), and $\Xi_{A\beta}$(33 yr) decreases to $2.3\times10^{4}$ μM s for $f_{acc}$ = 0. Because fragmentation changes $P_{A\beta}$ but neither $M_{A\beta}$ nor $D_{A\beta}$, $\sigma_{D,A\beta} = \sigma_0$ still holds exactly, $R_{2,A\beta}$ is unaffected by tau coupling, and $\Xi_{A\beta}$ increases by only 5.5% ($f_{acc}$ = 0.1) and 47% ($f_{acc}$ = 1) (Fig. S9b).

Table S2 summarizes the reference, dissociation, and soluble-protein-clearance scenarios, together with the combined scenario of Section 4.6, for the same four coupling strengths; Table S3 gives the growth-history, masking, fragmentation, fragmentation-site, and fibril-removal scenarios. In every scenario in which elongation rather than clearance controls the free monomer (all except the soluble-protein-clearance scenario and the combined scenario of Section 4.6), tau coupling lowers the homotypic secondary-nucleation rate (or leaves it unchanged when fragmentation sets the fibril number) and raises oligomer-equivalent exposure through the heterotypic channel. The active-site density changes appreciably only when site formation is made dependent on growth history, and it then decreases.

Interpretation. These tests separate three effects that are easily conflated. Oligomer dissociation primarily changes oligomer lifetime, fragmentation primarily changes fibril number, and soluble-protein clearance changes how much Aβ remains available to accumulate. Of the three, clearance is the mechanism that can reverse the equal-time homotypic response because it moves the system between the monomer-control regimes identified in Section 3.4.

**S2.4. How sensitive are the conclusions to plaque geometry, fibril turnover, fibril length, and parameter uncertainty?**

**Plaque accessibility.** The thickness of the accessible shell, $\delta_{A\beta}$, was varied from 0.0025 to 0.1 μm, a 40-fold range around the reference value of 0.010 μm, and two alternative closures were tested (Fig. S10a): a power-law closure, $f_{surf} = min[1, (3\delta_{A\beta}/R_p)^{\beta}]$ with $\beta = 1/2$ and $\delta_{A\beta} = 0.010$ μm, which represents a porous or fractal deposit whose exposed fraction decreases more slowly with plaque size, and a dispersed-fibril limit, $f_{surf} = 1$, in which no sites are buried. Because $f_{surf}$ depends only on $M_{A\beta}$ in every closure, $D_{A\beta,eff}$ is again identical in coupled and uncoupled simulations at matched mass, and the sign of every effect is unchanged: tau coupling lowers $R_{2,A\beta}$ and raises $\Xi_{A\beta}$ for every $\delta_{A\beta}$ and closure (Fig. S10b and Table S4). The magnitudes, however, depend strongly on accessibility. As $f_{surf}$(33 yr) increases from $2.1\times10^{-3}$ ($\delta_{A\beta}$ = 0.0025 μm) to $8.4\times10^{-2}$ ($\delta_{A\beta}$ = 0.1 μm), the homotypic channel becomes more important, the heterotypic share of oligomer formation at 33 years for $\chi_\tau = 1.86\times10^{-9}$ μM$^{-2}$ s$^{-1}$ falls from 97% to 47%, $R_{2,A\beta}/R_{2,A\beta}^{(0)}$ rises from 0.12 to 0.72, and the tau-induced increase in $\Xi_{A\beta}$(33 yr) falls from 8.7- to 1.4-fold. The power-law closure gives nearly the same results as the shell closure with $\delta_{A\beta}$ = 0.1 μm, which has a similar exposed fraction. In the limit without burial, homotypic secondary nucleation dominates oligomer formation, the uncoupled exposure is 18 times larger than in the reference case ($5.5\times10^{5}$ μM s), and tau coupling at $\chi_\tau = 1.86\times10^{-9}$ μM$^{-2}$ s$^{-1}$ lowers $R_{2,A\beta}$ by only 4% and raises $\Xi_{A\beta}$ by only 4.5%. The plaque-accessibility closure therefore controls the absolute accessible-site burden and the size of the tau effect, but not its direction. The range of $\delta_{A\beta}$ also covers larger plaques: because $f_{surf} = 3\delta_{A\beta}/R_p$ and $R_p$ scales with the cube root of the deposited mass, collecting the Aβ from $N$ control volumes into one plaque is equivalent to dividing $\delta_{A\beta}$ by $N^{1/3}$. The case $\delta_{A\beta}$ = 0.0025 μm therefore also represents a plaque of radius about 14 μm, four times the reference radius of 3.6 μm and within the size range of dense-core plaques.

**Fibril turnover.** To represent slow removal of fibrillar Aβ, for example by phagocytosis, first-order loss terms $-(\ln 2/T_{1/2,F})P_{A\beta}$, $-(\ln 2/T_{1/2,F})M_{A\beta}$, and $-(\ln 2/T_{1/2,F})D_{A\beta}$ were added to Eqs. (11)–(13), with a common half-life $T_{1/2,F}$ between 1 and 100 years. Removing fibrils together with their sites leaves $\sigma_{D,A\beta} = \sigma_0$ in the mass-proportional case, and because the free-monomer and oligomer pools remain small, $M_{A\beta}$(33 yr) is still independent of $\chi_\tau$ (10.0 μM for $T_{1/2,F}$ = 1 year and 89.9 μM for 10 years). Tau coupling therefore again leaves $D_{A\beta,eff}$ unchanged at equal time and at matched mass. Faster turnover leaves fewer fibrils and more free monomer, so that the heterotypic channel gains relative importance: for $\chi_\tau = 1.86\times10^{-9}$ μM$^{-2}$ $s^{-1}$, $R_{2,A\beta}/R_{2,A\beta}^{(0)}$ decreases from 0.25 ($T_{1/2,F}$ = 100 years) to 0.064 (1 year), and $\Xi_{A\beta}/\Xi_{A\beta}^{(0)}$ increases from 3.9 to 9.6 (Fig. S11a). Fibril removal thus strengthens rather than weakens the conclusions of Sections 4.2–4.4.

**Fragmentation-generated sites.** With fragmentation at the unliganded rate $k_{-,A\beta} = 3.5\times10^{-11}$ s$^{-1}$, the optional term $\eta_{frag}k_{-,A\beta}M_{A\beta}$ was added to Eq. (13), with $\eta_{frag}$ between 0.1 and $10^3$ sites per fragmentation event. Fragmentation-generated sites quickly dominate the site balance: for $\eta_{frag} = 1$ they account for 72% of all sites formed in 33 years and $\sigma_{D,A\beta}/\sigma_0 = 3.6$, and for $\eta_{frag} = 10$ they account for 96% and $\sigma_{D,A\beta}/\sigma_0 = 27$ (Fig. S11b). Because the fragmentation rate $k_{-,A\beta}M_{A\beta}$ of Eq. (7) depends on fibril mass but not on fibril number, these sites are produced at the same rate in coupled and uncoupled simulations with equal $M_{A\beta}$, and $D_{A\beta,eff}/D_{A\beta,eff}^{(0)}$ remains equal to unity to within $10^{-4}$ for every $\eta_{frag}$. Because fragmentation also pins the free-monomer concentration, the homotypic rate is again unaffected by tau coupling, and $\Xi_{A\beta}$(33 yr) increases by factors of 1.20–1.47 for $\chi_\tau = 1.86\times10^{-9}$ μM$^{-2}$ s$^{-1}$ and 1.43–1.96 for $\chi_\tau = 4.45\times10^{-9}$ μM$^{-2}$ s$^{-1}$. The conclusion that tau-induced fibril shortening does not raise the catalytic-site burden therefore survives when fragmentation creates active sites. It relies on the mass-based fragmentation law of Eq. (7) and on a site yield per break that does not depend on tau coupling; a fragmentation rate that increased with fibril number at fixed mass would make fragmentation-generated sites depend on tau coupling.

**Fibril length.** The reference fibrils are about 150 μm long, and those of the fragmentation scenario are about 13 nm long (about 4.3 monomers per nanometer for two protofilaments with a 0.47-nm cross-β layer spacing [35]), so neither is consistent with the observed Aβ42 fibril lengths of a few micrometers or less. The fragmentation coefficient was therefore chosen so that the uncoupled mean fibril length at 33 years equals 0.3, 1, 3, 10, or 30 μm, which requires $k_{-,A\beta} = 1.5\times10^{-12}$, $4.5\times10^{-13}$, $1.5\times10^{-13}$, $4.4\times10^{-14}$, and $1.3\times10^{-14}$ s$^{-1}$, and the comparisons were repeated for PHF and CTE coupling at $f_{acc} = 1$ and $\eta_{birth}/(2\eta_{grow}) = 1$–$10^4$ (Fig. S13). With fibrils a few micrometers long or shorter, $P_{A\beta}$ is set mainly by fragmentation, which depends on $M_{A\beta}$ but not on $\chi_\tau$, so that tau coupling changes the fibril number and the free monomer little: for a mean length of 1 μm, $R_{2,A\beta}/R_{2,A\beta}^{(0)} = 0.999$ (PHF) and 0.998 (CTE), whereas $\Xi_{A\beta}$(33 yr) still increases 2.5- and 4.2-fold, with a heterotypic share of oligomer formation of 90% and 95%. The reduction of $R_{2,A\beta}$ grows with fibril length, to 21% (PHF) and 32% (CTE) at 30 μm and 74% and 84% without fragmentation (Fig. S13a). Shorter fibrils do not, however, make birth-generated sites more important, because fragmentation shortens fibrils without adding fibril-birth events. The departure of $\sigma_{D,A\beta}$ from $\eta_{grow}$ in Eq. (24) is then controlled by the fibril mass per birth event, $M_{A\beta}/N_{birth}$, where $N_{birth} = \int k_{conv,A\beta}m_{A\beta}S_{A\beta}dt$ is the cumulative number of birth events, rather than by $\bar{n}_F$. At 33 years, $M_{A\beta}/N_{birth} = 3.5\times10^6$ monomers for 1-μm fibrils, larger than $\bar{n}_F = 6.4\times10^5$ of the reference case, because the lower free-monomer concentration also slows nucleation. Accordingly, the matched-mass change in $D_{A\beta,eff}$ for $\eta_{birth}/(2\eta_{grow}) = 10^3$ is 0.03% for 1-µm fibrils (PHF coupling) compared with 0.3% without fragmentation, and it remains below 3.3% at $10^4$ for every length (Fig. S13b). The conclusions on site density therefore do not rely on the unrealistically long fibrils of the reference case, whereas the reduction of $R_{2,A\beta}$ at equal time in the absence of clearance is specific to long fibrils: with realistic lengths and without clearance, tau coupling leaves homotypic secondary nucleation nearly unchanged and raises $\Xi_{A\beta}$.

**Multi-parameter robustness.** Finally, nine parameters were varied simultaneously by Latin-hypercube sampling (400 samples, log-uniform distributions): the Aβ source $q_{A,0,A\beta}$ (1/3 to 3 times its reference value); the coefficients $k_{+,A\beta}$, $k_{conv,A\beta}$, $k_{o1,A\beta}$, and $k_{o2,A\beta,D,p}$ (each 0.1 to 10 times its reference value); the coupling coefficient $\chi_\tau$ ($1.86\times10^{-11}$ to $4.45\times10^{-9}$ µM$^{-2}$ s$^{-1}$, that is, from PHF coupling at $f_{acc} = 0.01$ to CTE coupling at $f_{acc} = 1$); the shell thickness $\delta_{A\beta}$ (0.0025 to 0.04 µm); a common factor ($10^{-3}$ to 1) multiplying the Aβ dissociation coefficients $k_{d1,A\beta}$ and $k_{d2,A\beta}$ of the dissociation scenario; and the birth-site ratio $\eta_{birth}/(2\eta_{grow})$ (0.1 to $10^3$). Soluble-protein clearance was switched off in this study (study A); it is added in study B (Section 4.6). Each sample was simulated with and without tau coupling. In all 400 samples, tau coupling lowered the homotypic secondary-nucleation rate at 33 years, both at equal time and at matched fibril mass, and raised $\Xi_{A\beta}$(33 yr); the largest change in $D_{A\beta,eff}$ at matched mass was 0.16% (median $3\times10^{-5}$%). The magnitudes, however, varied widely (Fig. S12a): $R_{2,A\beta}/R_{2,A\beta}^{(0)}$ ranged from 0.088 to 0.999 (median 0.77; 5th–95th percentiles 0.25–0.99) and $\Xi_{A\beta}/\Xi_{A\beta}^{(0)}$ from 1.001 to 11.6 (median 1.34; 5th–95th percentiles 1.01–4.3), and the heterotypic share of oligomer formation at 33 years had a 5th–95th percentile range of 3–98%. Partial rank correlation coefficients (PRCCs; Fig. S12b) indicate that both ratios are associated most strongly with $\chi_\tau$ (|PRCC| = 0.96), the per-site coefficient $k_{o2,A\beta,D,p}$ (0.91–0.92), and $\delta_{A\beta}$ (0.80–0.81), followed by $q_{A,0,A\beta}$ (0.57–0.59) and $k_{o1,A\beta}$ (0.44–0.46); $k_{+,A\beta}$, $k_{conv,A\beta}$, the dissociation factor, and $\eta_{birth}/(2\eta_{grow})$ have |PRCC| < 0.09. The magnitude of the tau effect is thus set by the relative strengths of the heterotypic channel and of the accessible-site homotypic channel.

Study B (Section 4.6) repeated this analysis with a tenth parameter, a factor $c_{clr}$ ($10^{-4}$ to 1, log-uniform) multiplying the Aβ monomer and oligomer clearance rates of the clearance scenario (Table S1; tau clearance remained switched off), and 400 new samples. $\Xi_{A\beta}$(33 yr) increased with tau coupling in all 400 samples (ratios 1.002 to $3.5\times10^3$; median 2.0). The equal-time $R_{2,A\beta}$ ratio was below unity in all 236 samples with $Da > 1$ and above unity in 134 of the 164 samples with $Da < 1$ (Fig. S17a); the 30 exceptions all had $0.17 < Da < 1$, whereas in all 124 samples with $Da < 0.1$ the ratio exceeded unity. At matched fibril mass, the $R_{2,A\beta}$ ratio was at most 0.998 in every sample with $Da > 1$; ratios above unity (up to 1.45) occurred only when almost no fibrillar Aβ formed in 33 years ($M_{A\beta} \leq 0.022$ µM) and the birth-site ratio was large (median $\eta_{birth}/(2\eta_{grow}) = 113$), as anticipated in Section 3.2. With clearance included, the PRCCs are largest for $c_{clr}$ (0.81 for the $\Xi_{A\beta}$ ratio and 0.74 for the $R_{2,A\beta}$ ratio), followed by $\chi_\tau$ (0.72 and −0.43) and $k_{o2,A\beta,D,p}$ (−0.55 and 0.30): the balance between elongation and clearance matters more for the tau effect than any active-site parameter.

Interpretation. The broad sensitivity study separates robust directions from uncertain magnitudes. Without soluble-protein clearance, tau coupling consistently lowers or leaves essentially unchanged homotypic secondary nucleation while increasing oligomer-equivalent exposure, but the size of both effects depends strongly on heterotypic coupling, the per-site secondary-nucleation coefficient, and plaque accessibility.

### S2.5. Can fibril morphology matter directly through accessibility or fragmentation?

Every accessibility closure of Section S2.4 makes $f_{surf}$ a function of $M_{A\beta}$ alone, and the fragmentation rate of Eq. (7) depends on $M_{A\beta}$ alone, so that neither can distinguish more, shorter fibrils from fewer, longer ones at matched mass. Two deliberately morphology-dependent laws remove this restriction. In the first, the exposed fraction of Eq. (16) is replaced by

$$f_{eff} = min[1, f_{surf}(M_{A\beta})g], \quad g = (\bar{n}_{F,ref}/\bar{n}_F)^{\gamma} \qquad \text{(S30)}$$

with $\bar{n}_{F,ref} = 6.4\times10^5$, the uncoupled mean fibril size of the reference case at 33 years, and $\gamma = 0$–$3$; $\gamma = 0$ recovers Eq. (16), and $\gamma > 0$ makes shorter fibrils expose more sites per unit mass, as might result from looser packing of short fibrils in a deposit. In the second, the fragmentation rate of Eq. (7) is replaced by

$$J_{\text{frag}} = k_{-,A\beta}M_{A\beta}(\bar{n}_F/\bar{n}_{F,ref})^{\beta_{frag}} \qquad \text{(S31)}$$

with $\bar{n}_{F,ref} = 4255$ monomers (a 1-μm fibril), $k_{-,A\beta} = 4.5\times10^{-13}$ s$^{-1}$ (mean length 1 μm for $\beta_{frag} = 0$; Section S2.4), $\beta_{frag} = -1$ to 2, and $\eta_{frag} = 10$ sites per fragmentation event; $\beta_{frag} > 0$ makes long fibrils break more often per unit mass. $J_{frag}$ replaces $k_{-,A\beta}M_{A\beta}$ in the fibril-number balance and in the fragmentation-site term of Eq. (13). Both factors are limited to the range $10^{-3}$–$10^3$, which matters only while $P_{A\beta}$ is still negligible. With $\gamma = \beta_{frag} = 0$, the code reproduces the reference implementation to within a relative difference of $3\times10^{-10}$.

**Accessibility law.** In the reference case ($Da = \infty$), the matched-mass ratio $R_{2,A\beta}/R_{2,A\beta}^{(0)}$ equals $(P_{A\beta}^{(tau)}/P_{A\beta}^{(0)})^{\gamma-2}$, as Eq. (30) predicts when $g$ replaces the unit accessible-site ratio, and reaches unity at $\gamma = 2$ for both PHF and CTE coupling at $f_{acc} = 1$ (Fig. S14a). $D_{A\beta,eff}$ at matched mass increases by up to 49% (PHF) and 85% (CTE) at $\gamma = 1$, and by less at larger $\gamma$, as the exposed fraction approaches unity (Fig. S14b). In the combined scenario ($Da = 0.023$), the monomer penalty is absent: the matched-mass $R_{2,A\beta}$ ratio is 1.8 (PHF) and 2.0 (CTE) at $\gamma = 0.5$ and rises to 41 and 71 at $\gamma = 2$, because the higher accessible-site concentration also accelerates further fibril formation. In the composite scenario, the ratio remains 1.000 for all $\gamma$, because fragmentation makes $P_{A\beta}$, and hence $\bar{n}_F$, a function of $M_{A\beta}$.

**Fragmentation law.** For $\beta_{frag} = -1$ to 2, $D_{A\beta,eff}$ at matched mass differed from its uncoupled value by at most 0.7% (Fig. S15a). For $\beta_{frag} \geq 0$, fragmentation dominates fibril formation and drives coupled and uncoupled systems to the same mean length (at 33 years, the mean fibril size $\bar{n}_F$, and hence the mean fibril length, differs between the tau-coupled and uncoupled systems by only about 0.1%), so that the fragmentation rate and the fragmentation-generated sites are again equal at matched mass. For $\beta_{frag} < 0$, fragmentation of long fibrils is suppressed, fragmentation-generated sites account for less than 15% of all sites, and the system approaches the reference case, in which tau coupling lowers $R_{2,A\beta}$ at equal time (to 0.20–0.33 for $\beta_{frag} = -1$; Fig. S15b). $\Xi_{A\beta}$(33 yr) increased 1.5- to 6.5-fold for every $\beta_{frag}$.

Interpretation. Morphology matters only when the closure explicitly allows it to matter. A length-dependent exposed fraction can reverse the matched-mass prediction because shorter fibrils can expose more catalytic sites, whereas a length-dependent fragmentation law mainly drives the coupled and uncoupled systems toward a common fibril length. Accessibility, rather than fragmentation itself, is therefore the more direct morphological route by which tau-induced shortening could enhance homotypic secondary nucleation.

### S2.6. What if tau-nucleated fibrils are intrinsically different?

All preceding calculations assume that fibrils nucleated on tau and fibrils nucleated homotypically form active sites with the same yields. Tau-seeded Aβ42 fibrils are, however, shorter and may differ structurally (Section 1). This assumption was therefore relaxed by splitting $S_{A\beta}$, $P_{A\beta}$, $M_{A\beta}$, and $D_{A\beta}$ into a homotypic lineage, fed by intrinsic primary and homotypic secondary nucleation, and a tau-derived lineage, fed by heterotypic nucleation. A lineage is thus the set of fibrils, together with their oligomer precursors and active sites, that descend from one type of nucleation event; a fibril keeps its lineage as it elongates, so monomer added to a tau-derived fibril becomes tau-derived fibril mass. Both lineages obey Eqs. (10)–(13) with a common free-monomer pool and a common plaque [$f_{surf}$ is evaluated with the total $M_{A\beta}$ in Eq. (16)], but the tau-derived lineage has the site yields $r\eta_{birth}$ and $r\eta_{grow}$. Oligomers formed by secondary nucleation on sites of the tau-derived lineage join the homotypic lineage in one variant and the tau-derived lineage in the other, in which the tau-derived structure is propagated by secondary nucleation. With $r = 1$, the lineage sums reproduce the single-population model to within a relative difference of $10^{-8}$. Site abundance and activity per site must, however, be distinguished. Let $r$ denote the factor that multiplies the site yields of the tau-derived lineage (site abundance) and $r_k$ the factor that multiplies its per-site secondary-nucleation coefficient (catalytic activity per site). Only the product $rr_k$ enters $R_{2,A\beta}$: for PHF coupling at $f_{acc}$ = 1, ($r$, $r_k$) = (10, 1), (1, 10), and ($\sqrt{10}$, $\sqrt{10}$) give the same $R_{2,A\beta}/R_{2,A\beta}^{(0)}$ = 1.03 and $\Xi_{A\beta}$ ratio 4.84 (agreement to $10^{-9}$), but total site densities $\sigma_{D,A\beta}/\sigma_0$ of 5.0, 1.0, and 2.0 at 33 years. Likewise, ($r$, $r_k$) = (0.3, 1) and (1, 0.3) give the same rate ratio, 0.127, but total site densities of 0.47$\sigma_0$ and 1.0$\sigma_0$. Bulk aggregation kinetics therefore constrain $rr_k$, whereas site counting, for example by BRICHOS binding, measures $r$ alone; together they separate site abundance from activity per site (Section S4.3). Unless stated otherwise, $r_k$ = 1 below, so that $r$ stands for $rr_k$ in the rate.

At matched fibril mass, the accessible-site ratio equals 1 + ($r$ − 1)$\varphi_\tau$ to four significant figures in the variant without propagation, where $\varphi_\tau$ is the fraction of fibril mass in the tau-derived lineage (Fig. S19a); for $r$ = 1, $\varphi_\tau$ = 0.71 for PHF coupling and 0.84 for CTE coupling at $f_{acc}$ = 1. Lineage-specific site formation therefore changes $D_{A\beta,eff}$ at matched mass in proportion to the tau-derived mass fraction, without the $1/\bar{n}_F$ suppression that limits birth-generated sites. In the reference case, $R_{2,A\beta}$ at 33 years nevertheless remains below its uncoupled value unless $r$ is about 10 or larger without propagation ($R_{2,A\beta}/R_{2,A\beta}^{(0)}$ = 0.53 for $r$ = 3 and 1.03 for $r$ = 10, PHF coupling at $f_{acc}$ = 1), or between 3 and 10 with propagation (0.63 and 1.54), because the squared monomer penalty must be overcome (Fig. S19b); $r$ < 1 strengthens the reduction (0.082 for $r$ = 0.1). $\Xi_{A\beta}$(33 yr) increases with $r$, from 3.6- to 5.9-fold for PHF coupling at $f_{acc}$ = 1 as r rises from 0.1 to 30. In the combined scenario of Section 4.6, in which tau coupling already raises the fibril mass, $r$ > 1 amplifies the equal-time increase of $R_{2,A\beta}$ (to 12-fold for $r$ = 10, PHF coupling at $f_{acc}$ = 1), and $r$ = 0.3 reverses it (0.76).

Whether shorter, tau-nucleated fibrils are more catalytically active is therefore determined by their site yield, not by their length. A product $rr_k$ several-fold above unity—a site density, a per-site activity, or both several-fold higher in tau-seeded Aβ42 fibrils than in control fibrils formed without tau—would be needed to raise homotypic secondary nucleation at matched mass, whereas equal or lower site densities would confirm the conclusions of Sections 4.3–4.5.

Synthesis of Section S2. Across the robustness analyses, four qualitatively distinct routes can make tau increase homotypic secondary nucleation: tau can increase total fibril mass while free monomer remains clearance-controlled; fibril birth can be an exceptionally strong source of sites; shorter fibrils can expose more sites per unit mass; or the sites of tau-nucleated fibrils can be intrinsically more abundant or more active. Mere shortening of otherwise equivalent fibrils is not sufficient. These are the experimentally distinguishable possibilities carried forward into the Discussion and Table 3.

## S3. Tau subsystem: the upstream driver of Aβ coupling

Because the coupling is one-way, the tau subsystem acts as an upstream driver: it determines the time-dependent tau fibrillar material available to Aβ, but Aβ does not feed back on tau. The equations are reproduced here so that the complete coupled model can be reconstructed without consulting the preceding paper [4]. They retain the same processes for tau supply, primary and secondary nucleation, oligomer dissociation and conversion, elongation, fragmentation, and monomer/oligomer turnover.

The free-tau-monomer balance reads

$$\frac{dm_{tau}}{dt} = -k_{o1,tau}m_{tau}(t) - k_{o2,tau}m_{tau}^2(t)M_{tau}(t) + k_{d1,tau}S_{tau}(t) + k_{d2,tau}S_{tau}(t)M_{tau}(t) \\ -2k_{+,tau}m_{tau}(t)P_{tau}(t) - k_{conv,tau}m_{tau}(t)S_{tau}(t) - \frac{m_{tau}(t)\ln(2)}{T_{1/2,m,tau}} + \frac{q_{tau}a_{21}}{V_{soma}} \tag{S32}$$

In Eq. (S32), free tau monomer is lost first to effective first-order primary nucleation and second to secondary nucleation catalyzed by fibril surfaces, a process whose rate scales as $m_{tau}^2(t)M_{tau}(t)$. Oligomer dissociation, whether spontaneous or catalyzed by fibrils, replenishes the monomer pool (third and fourth terms). Monomers are also consumed by growth at both fibril tips, hence the factor 2 in the fifth term, and by the monomer-addition step that turns an oligomer into an elongation-competent fibril (sixth term). The seventh term represents proteolytic degradation of monomers with rate constant $\ln(2)/T_{1/2,m,tau}$. Tau monomers are supplied at the volume-averaged rate $q_{tau}a_{21}/V_{soma}$ (last term).

For soluble on-pathway tau oligomers, the corresponding balance is

$$\frac{dS_{tau}}{dt} = k_{o1,tau}m_{tau}(t) + k_{o2,tau}m_{tau}^2(t)M_{tau}(t) - k_{d1,tau}S_{tau}(t) - k_{d2,tau}S_{tau}(t)M_{tau}(t) \\ -k_{conv,tau}m_{tau}(t)S_{tau}(t) - \frac{S_{tau}(t)\ln(2)}{T_{1/2,S,tau}} \tag{S33}$$

Eq. (S33) mirrors Eq. (S32): every oligomer produced by primary or fibril-surface-catalyzed secondary nucleation (first two terms) corresponds to a monomer lost in Eq. (S32), and every oligomer that dissociates, spontaneously or with fibril assistance (third and fourth terms), returns a monomer there. Oligomers also leave the pool when monomer addition converts them into elongation-competent fibrils (fifth term) and when they are degraded proteolytically with rate constant $\ln(2)/T_{1/2,S,tau}$.

Counting tau fibrils irrespective of their length gives the fibril-number equation

$$\frac{dP_{tau}}{dt} = k_{conv,tau} m_{tau}(t) S_{tau}(t) + k_{-,tau} max\left(M_{tau}(t) - n_{min,tau} P_{tau}(t), 0\right) \quad \text{(S34)}$$

New tau fibrils arise in Eq. (S34) in two ways: an oligomer that gains a monomer becomes an elongation-competent fibril (first term), and existing fibrils break into fragments (second term). The fragmentation term contains the coefficient $k_{-,tau}$, multiplied by $max\left(M_{tau}(t) - n_{min,tau} P_{tau}(t), 0\right)$, the fibril mass in excess of $n_{min,tau} P_{tau}(t)$, which is the least mass needed to sustain the present fibril population; taking the maximum with zero keeps the fragmentation rate non-negative. In this expression, $n_{min,tau}$ denotes the smallest fragment, in monomer units, that can still elongate.

Tau fibril mass, expressed in incorporated monomer units, evolves as

$$\frac{dM_{tau}}{dt} = 2k_{+,tau} m_{tau}(t) P_{tau}(t) + 2k_{conv,tau} m_{tau}(t) S_{tau}(t) \quad \text{(S35)}$$

Fibril mass in Eq. (S35) grows by elongation (first term), with elongation coefficient $k_{+,tau}$, free-monomer concentration $m_{tau}$, and fibril-number concentration $P_{tau}$; because monomers attach at both tips, the term carries a factor 2. The second term adds mass each time an oligomer and a monomer join to create a new fibril. Its rate is set by the conversion coefficient $k_{conv,tau}$ and the product $m_{tau}(t) S_{tau}(t)$ of the reactant concentrations; the factor 2 reflects the fact that each event incorporates both the free monomer and the oligomer, which counts as one activated-monomer equivalent.

Taken together, Eqs. (S32)–(S35) describe tau supply, primary and secondary nucleation, oligomer dissociation, oligomer-to-fibril conversion, elongation, fragmentation, and the turnover of monomers and oligomers. Since none of these processes involves Aβ, the tau equations can be integrated on their own, and tau influences Aβ only through $T_{tau}$ in Eq. (1). With the simulation started at the onset of AD pathology (Section S1.29), the initial conditions are

$$m_{tau}(0) = S_{tau}(0) = P_{tau}(0) = M_{tau}(0) = 0. \quad \text{(S36)}$$

This modularity is useful both computationally and conceptually. Every Aβ sensitivity study in Section S2 is driven by the same tau trajectory, so differences among the Aβ outcomes can be attributed to the altered Aβ assumptions rather than to a changing tau solution.

## S4. Analytical checks: what controls oligomers, mass, and parameter identifiability

Three compact analytical results make the numerical behavior easier to interpret. Section S4.1 shows why a larger active-site population does not automatically imply a larger soluble-oligomer burden; Section S4.2 gives the protein-mass identity used to verify the numerical solutions; and Section S4.3 explains why bulk aggregation kinetics cannot, by themselves, separate active-site abundance from catalytic activity per site.

### S4.1. Why active-site abundance alone does not determine oligomer burden

For a given instantaneous monomer concentration, fibrillar mass, total active-site concentration, and plaque geometry, the oligomer balance can be written in the form

$$\frac{dS_{A\beta}}{dt} = R_{1,A\beta} + R_{tau,A\beta} + R_{2,A\beta} - \Lambda_{A\beta} S_{A\beta} \tag{S37}$$

$$\Lambda_{A\beta} = k_{d1,A\beta} + k_{d2,A\beta} M_{A\beta} + k_{\mathrm{conv},A\beta} m_{A\beta} + \frac{\ln 2}{T_{1/2,S,A\beta}} \tag{S38}$$

The first three terms on the right-hand side of the oligomer balance in Eq. (S37) form soluble Aβ oligomers through intrinsic primary nucleation, tau-mediated heterotypic nucleation, and homotypic secondary nucleation, respectively. The final term, $-\Lambda_{A\beta} S_{A\beta}$, collects all first-order oligomer losses. In Eq. (S38), which defines $\Lambda_{A\beta}$, $k_{d1,A\beta}$ is spontaneous dissociation, $k_{d2,A\beta} M_{A\beta}$ is fibril-catalyzed dissociation, $k_{conv,A\beta} m_{A\beta}$ is oligomer-to-fibril conversion, and $\ln(2)/T_{1/2,S,A\beta}$ is the effective first-order oligomer clearance/turnover term.

If oligomer dynamics are fast relative to the evolution of $m_{A\beta}$, $M_{A\beta}$, $D_{A\beta}$, and $f_{surf}$, the corresponding quasi-steady approximation is

$$S_{A\beta,qs} = \frac{k_{o1,A\beta} m_{A\beta}^2 + \alpha_p T_{tau} m_{A\beta}^2 + k_{o2,A\beta,D,p} m_{A\beta}^2 D_{A\beta,\mathrm{eff}}}{k_{d1,A\beta} + k_{d2,A\beta} M_{A\beta} + k_{\mathrm{conv},A\beta} m_{A\beta} + \ln 2 / T_{1/2,S,A\beta}} \tag{S39}$$

The numerator on the right-hand side of Eq. (S39) is the total oligomer-formation rate. Its three terms represent intrinsic primary nucleation, tau-mediated heterotypic nucleation, and accessible-site-mediated homotypic secondary nucleation, respectively. The denominator is the total effective first-order oligomer-loss rate; its four terms represent spontaneous dissociation, fibril-catalyzed dissociation, oligomer-to-fibril conversion, and effective soluble-oligomer clearance/turnover. Their ratio gives the quasi-steady soluble Aβ oligomer-equivalent concentration.

In particular, increasing $D_{A\beta}$ does not necessarily increase accumulated oligomer-equivalent exposure if the same process substantially depletes free monomers, buries a growing fraction of sites within a compact plaque, or alters oligomer-removal rates. The full time-dependent system must therefore be solved to establish the net effect.

The analytical message is therefore the same as the numerical one: catalytic capacity is only one factor. Free monomer, plaque exposure, oligomer conversion, dissociation, and soluble-protein clearance jointly determine the oligomer-equivalent pool.

### S4.2. Why active sites do not enter the protein-mass balance

Summing Eqs. (9), (10), and (12) gives

$$\frac{d}{dt}\left(m_{A\beta} + S_{A\beta} + M_{A\beta}\right) = \frac{q_{A,0,A\beta}}{L} - m_{A\beta}\frac{\ln(2)}{T_{1/2,m,A\beta}} - S_{A\beta}\frac{\ln(2)}{T_{1/2,S,A\beta}} \quad \text{(S40)}$$

The first term on the right-hand side of Eq. (S40), $q_{A,0,A\beta}/L$, is the volume-averaged Aβ monomer source. The second term removes free monomer through effective first-order clearance/turnover at rate $\ln(2)/T_{1/2,m,A\beta}$. The third term removes soluble oligomer through effective first-order clearance/turnover at rate $\ln(2)/T_{1/2,S,A\beta}$. Nucleation, dissociation, conversion, and elongation cancel when the monomer, oligomer, and fibril-mass balances are summed because they redistribute Aβ among pools without changing total protein mass.

In the absence of clearance, the total Aβ mass in the control volume increases at the prescribed source rate, independently of how the mass is partitioned among monomers, oligomers, and fibrils. With the zero-Aβ-aggregate initial conditions used here, integration of Eq. (S40) gives $m_{A\beta} + S_{A\beta} + M_{A\beta} = (q_{A,0,A\beta}/L)t$.

Eq. (S40) provides an important check on the numerical implementation. The newly introduced active-site concentration must not be added to this mass balance because $D_{A\beta}$ is an active-site concentration, not a separate pool of Aβ protein.

This identity is more than a bookkeeping convenience: any implementation in which the active-site state variable changes total Aβ protein mass is inconsistent with the definition of that variable.

### S4.3. What bulk kinetics can—and cannot—identify about active sites

Bulk aggregation measurements generally constrain the product $k_{o2,A\beta,D,p}D_{A\beta,eff}$ rather than its factors separately. A change in total active-site concentration, exposed fraction, or catalytic activity per site can therefore produce similar bulk kinetics.

This creates a parameter-identification problem that cannot be resolved reliably using fibril-mass kinetics alone. Independent measurements of total active-site abundance, such as the BRICHOS binding measurements used in a recent study [5], should be combined with measurements of plaque or aggregate accessibility to estimate $D_{A\beta,eff}$ separately from $k_{o2,A\beta,D,p}$.

For experiments performed under matching conditions, a mass-based secondary-nucleation constant $k_{o2,A\beta,M}^{ref}$ and a measured reference site density $\sigma_{ref}$ are related by

$$k_{o2,A\beta,D}^{ref} = \frac{k_{o2,A\beta,M}^{ref}}{\sigma_{ref}} \tag{S41}$$

On the right-hand side of Eq. (S41), $k_{o2,A\beta,M}{}^{ref}$ is the reference secondary-nucleation coefficient expressed per unit fibril mass and $\sigma_{ref}$ is the measured reference number of active sites per fibrillar Aβ monomer. Dividing the mass-based coefficient by the site density gives $k_{o2,A\beta,D}{}^{ref}$, the corresponding coefficient per active site, which is the conditional value of $k_{o2,A\beta,D,p}$ in Table S1. This conversion is valid only when both reference quantities describe the same fibril preparation under fully accessible conditions.

The measured Aβ42 site density provides a useful experimental reference for $\sigma_{ref}$. It should not, however, be combined indiscriminately with a kinetic rate constant measured at another temperature or pH, or for another fibril preparation. Eq. (S41) applies to matched, fully accessible material; tissue-scale use additionally requires the geometric factor in Eqs. (15)–(18).

The practical implication is direct: active-site counting and secondary-nucleation kinetics should be measured on the same fibril preparation. Only such a matched experiment can distinguish a change in the number of catalytic sites from a change in the activity of each site.

**S5. Computational workflow and reproducibility**

All numerical results are generated from a single master script, providing one entry point for reproducing every figure and table in the paper. Calculations are performed in MATLAB R2023a with the stiff solver ode15s, using relative and absolute tolerances of $10^{-10}$. The master script calls the component analyses in the order required by their dependencies. Each component saves three types of output in a dedicated folder: the figures it generates, a text summary of the numerical quantities quoted in the manuscript, and a MATLAB data file. A complete run requires about 5 minutes on a desktop computer.

The code is organized into reusable model and analysis functions. The main right-hand-side function implements the coupled tau and Aβ equations and activates, when requested, the additional terms for masking, fibril removal, fragmentation-generated sites, and morphology-dependent effects. A separate right-hand-side function implements the two-lineage formulation of Section S2.6. Supporting functions define the parameters in Table S1, evaluate the accessibility closures and growth-history law, and calculate the reported observables. The implementation also contains automatic consistency checks. With all optional terms disabled, the model variants must agree to within a relative difference of $10^{-8}$; otherwise, execution stops with an error. Every run also repeats the site-density, mass-balance, and output-grid checks described in Section 4.1.

To make the connection between the code and the scientific results explicit, the component analyses are divided into five modules. Module 1 generates the reference case and the growth-history, masking, dissociation, clearance, and fragmentation results (Figs. 2–5, S1–S5, S7–S9, and S20–S23; Tables S2–S3). Module 2 examines plaque accessibility, fibril turnover, fragmentation-generated sites, and Latin-hypercube study A (Table S4; Figs. S10–S12). Module 3 covers the quasi-steady analysis, the monomer-control regime map, the combined scenario, the two-lineage model, and Latin-hypercube study B (Figs. S16, S17, and S19). Module 4 generates the growth-law sweep, fibril-length study, and composite scenario (Figs. S6, S13, and S18). Module 5 tests the length-dependent accessibility and fragmentation laws (Figs. S14–S15) and separates changes in site abundance from changes in activity per site. Fixed random seeds are used in both Latin-hypercube studies so that the uncertainty analyses are exactly reproducible.

Because the coupling is one-way, the tau subsystem is identical in all simulations (Fig. S1); its fibril mass $M_{tau}$ follows the supply-limited line $(q_{tau}a_{21}/V_{soma})t$, so that $T_{tau}$ increases almost linearly in time. All three checks of Section 4.1 are satisfied. In every reference simulation, $|\sigma_{D,A\beta}/\sigma_0 - 1|$ remains below $3\times10^{-15}$, and $m_{A\beta} + S_{A\beta} + M_{A\beta}$ agrees with $(q_{A,0,A\beta}/L)t$ to within a relative error below $10^{-14}$ (round-off level; Fig. S2a). Eq. (24) is satisfied at the same round-off level in every simulation with constant site yields, including those far from mass proportionality. Refinement of the output grid shows that $2\times10^3$ output times already fix $\Xi_{A\beta}(33\text{ yr})$ to better than $10^{-6}$ in the reference case, whereas in the dissociation scenario 250 output times underestimate $\Xi_{A\beta}(33\text{ yr})$ by 12% (Fig. S2b; 21% in a development version of the code of the preceding model); the $2\times10^4$ output times used here therefore yield converged results in all scenarios.

These checks make the computational workflow part of the argument rather than a separate implementation detail: the analytical identities are recovered at round-off accuracy, and the output-grid study verifies that the accumulated-exposure integrals reported in the paper are numerically converged.

## S6. Supplementary figures: visual tests of the mechanism

The figures are grouped by scientific role. Figs. S1–S4 establish the common tau driver, baseline behavior, and numerical checks; Figs. S5–S15 test active-site, kinetic, geometric, and morphology assumptions; Figs. S16–S19 examine monomer-control regimes and lineage-specific behavior; and Figs. S20–S23 provide detailed time courses and analytical benchmarks used to interpret the main-text results.

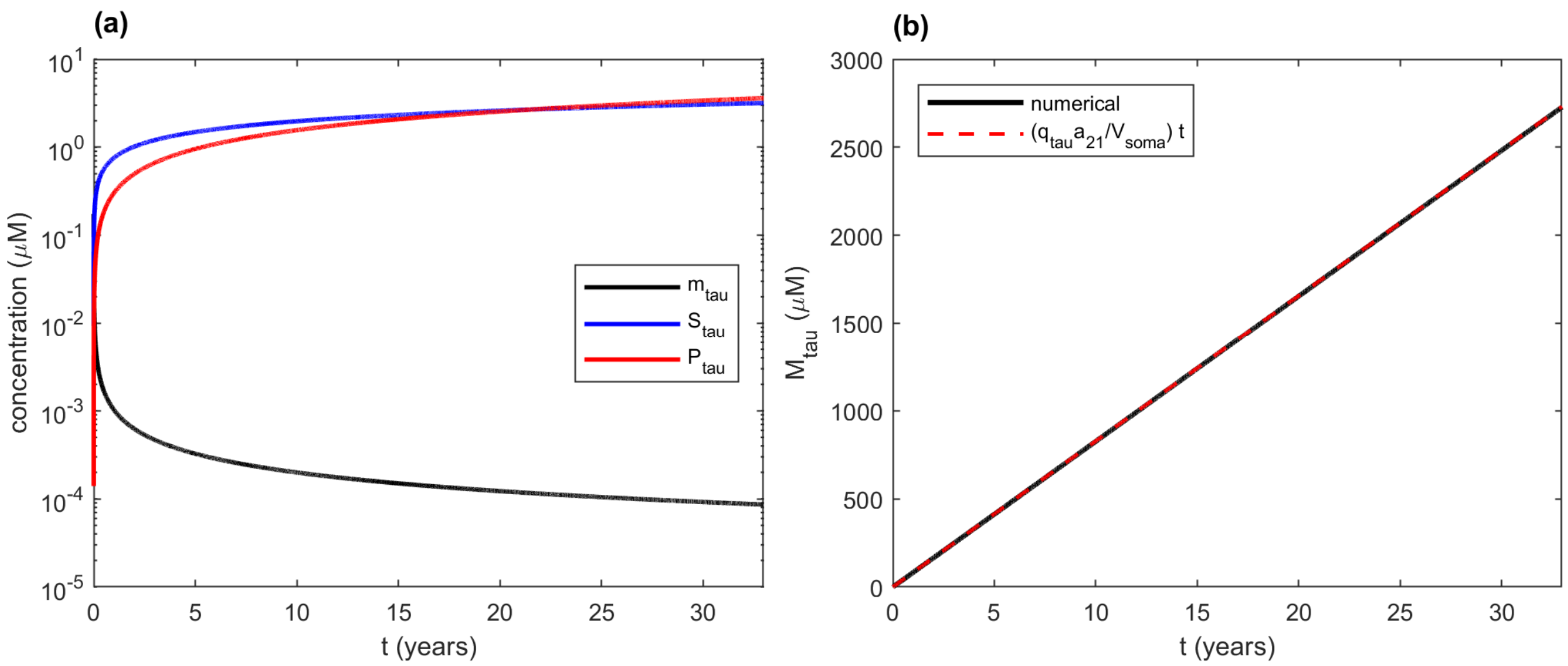

**Fig. S1.** Tau subsystem, identical in all simulations because the coupling is one-way. (a) Free monomer $m_{tau}$, oligomer $S_{tau}$, and fibril-number $P_{tau}$ concentrations. (b) Fibril mass $M_{tau}$ and the supply-limited line $(q_{tau}a_{21}/V_{soma})t$.

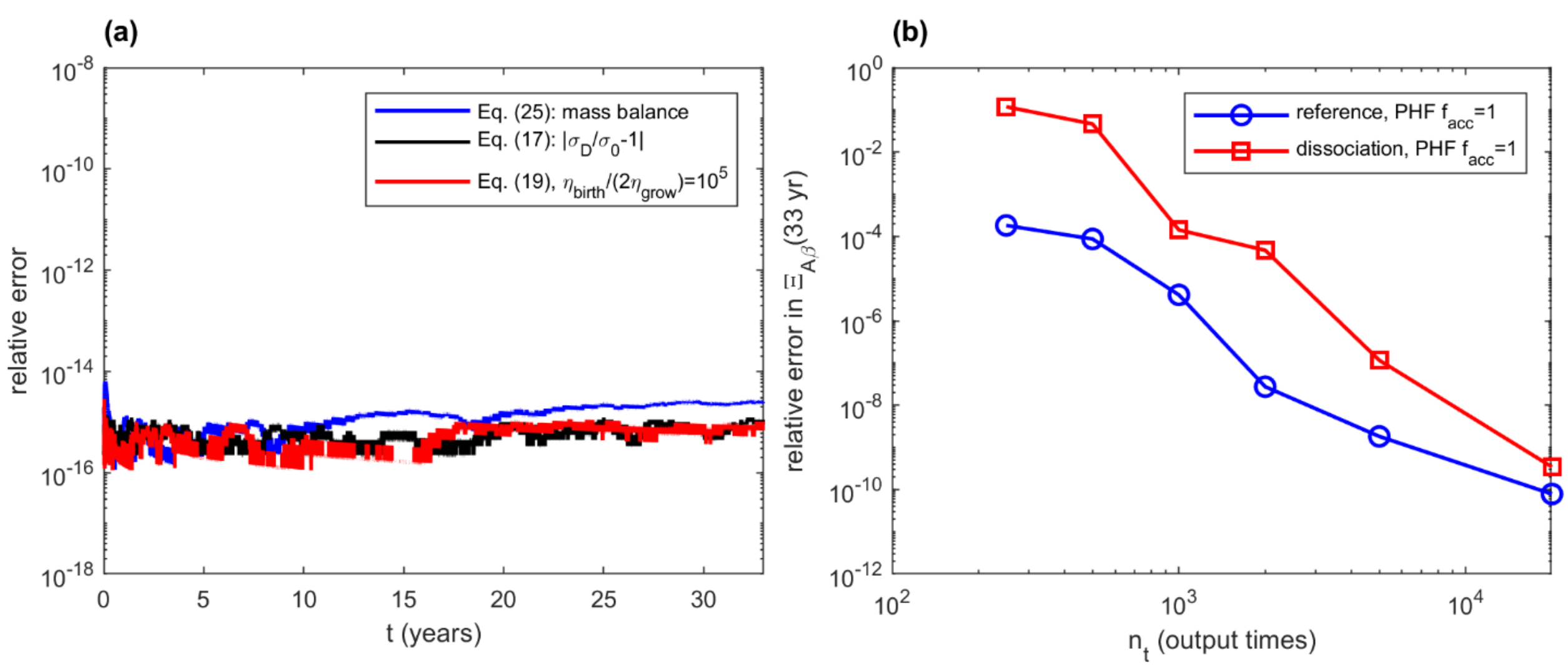

**Fig. S2.** Numerical verification. (a) Relative errors in the protein-mass balance [Eq. (S40)] and in the mass-proportional identity [Eq. (22)] for PHF coupling at $f_{acc} = 1$, and in Eq. (24) for $\eta_{birth}/(2\eta_{grow}) = 10^5$. (b) Relative error in $\Xi_{A\beta}(33\text{ yr})$ versus the number of output times $n_t$ used in the trapezoidal evaluation of Eq. (20), relative to $n_t = 5\times10^4$, for the reference and dissociation scenarios (PHF, $f_{acc} = 1$).

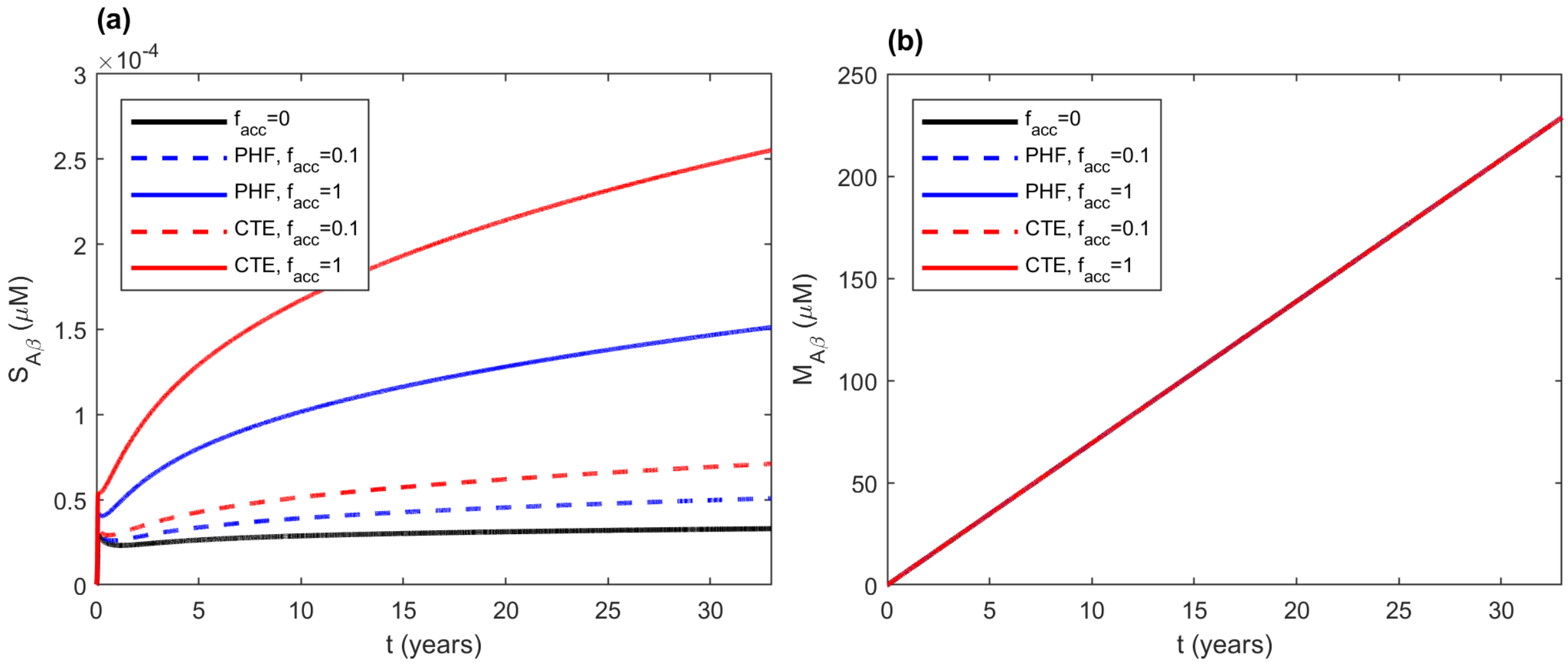

**Fig. S3.** Reference case; line styles as in Fig. 2. (a) Soluble Aβ oligomer-equivalent concentration $S_{A\beta}$. (b) Aβ fibril mass $M_{A\beta}$; the five curves coincide.

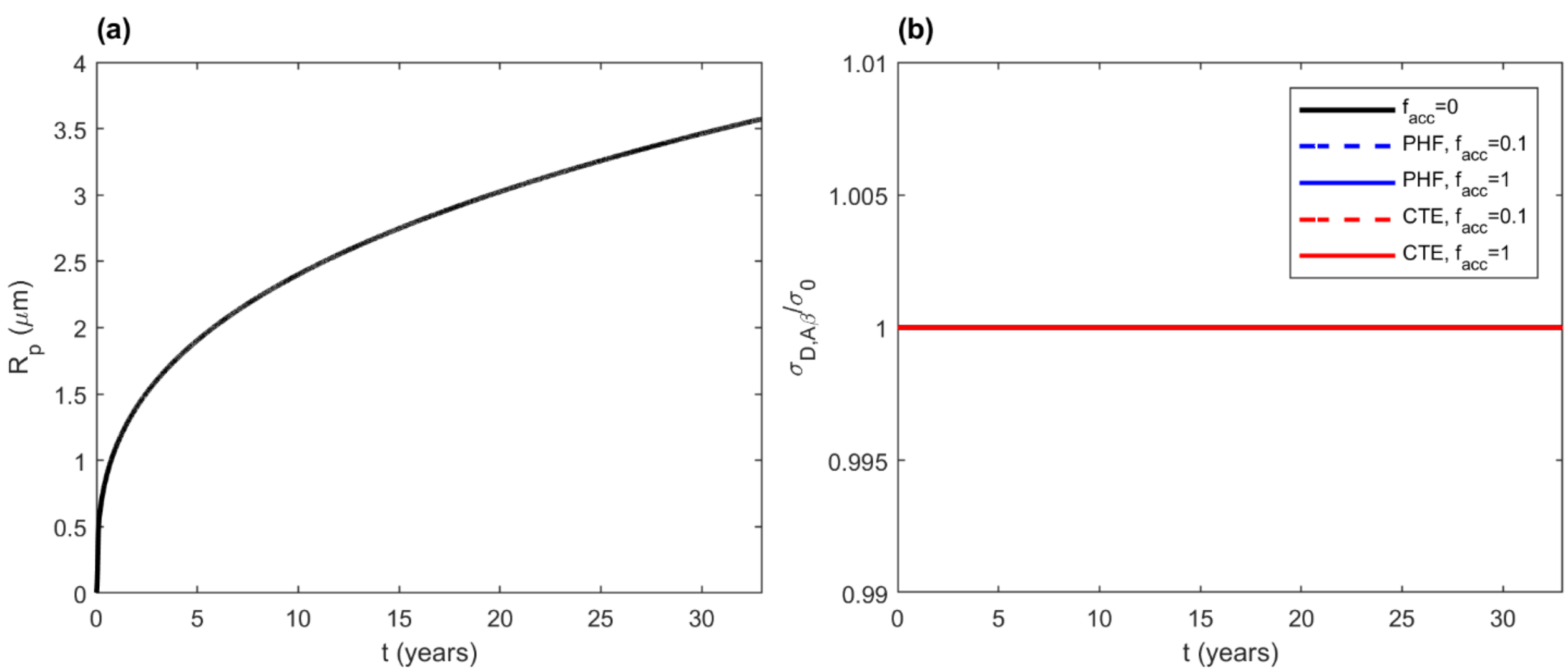

**Fig. S4.** (a) Equivalent plaque radius $R_p$ [Eq. (18)], identical in all reference simulations. (b) Active-site density $\sigma_{D,A\beta}/\sigma_0$ in the reference case, equal to unity as required by Eq. (22); line styles as in Fig. 2.

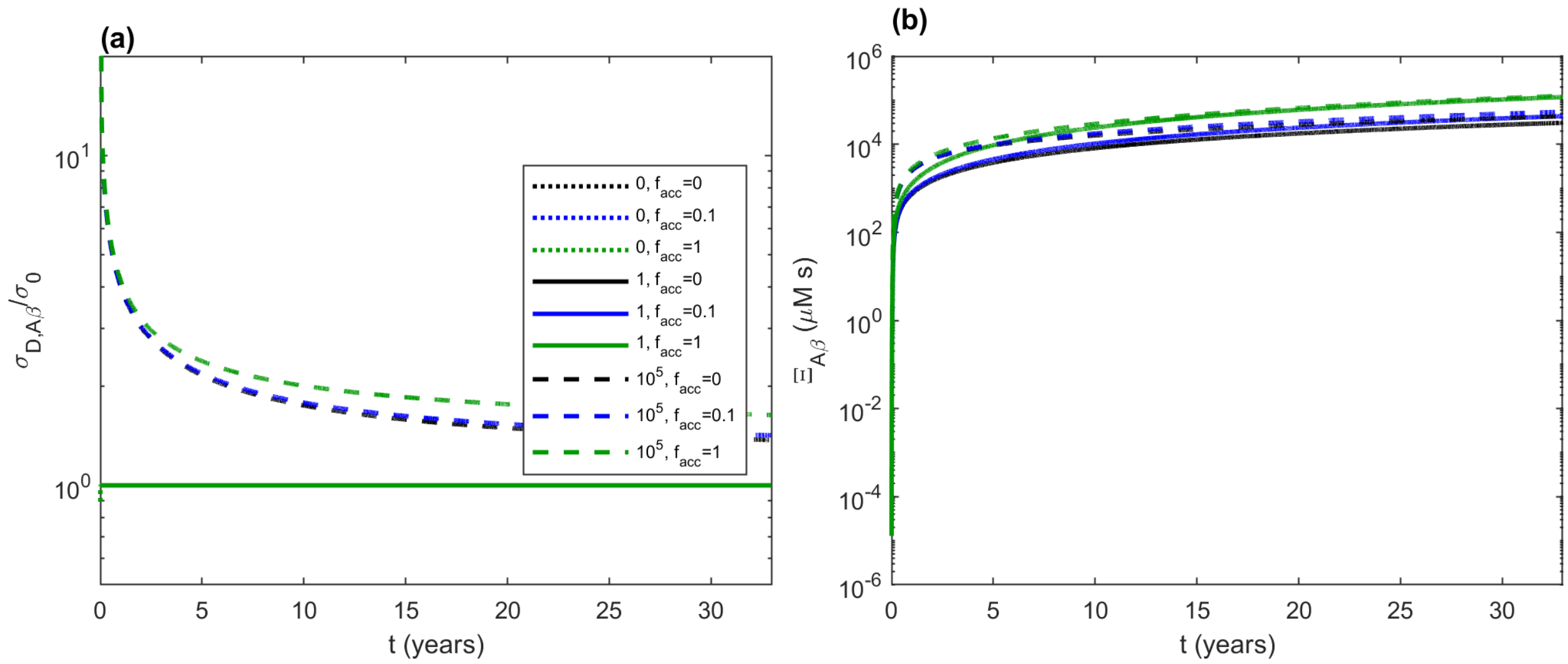

**Fig. S5.** Time histories for $\eta_{birth}/(2\eta_{grow}) = 0$ (dotted), 1 (solid), and $10^5$ (dashed) and $f_{acc} = 0$ (black), 0.1 (blue), and 1 (green), PHF coupling. (a) Active-site density $\sigma_{D,A\beta}/\sigma_0$; the curves for $\eta_{birth}/(2\eta_{grow}) = 0$ and 1 coincide at unity. (b) Accumulated oligomer-equivalent exposure $\Xi_{A\beta}$.

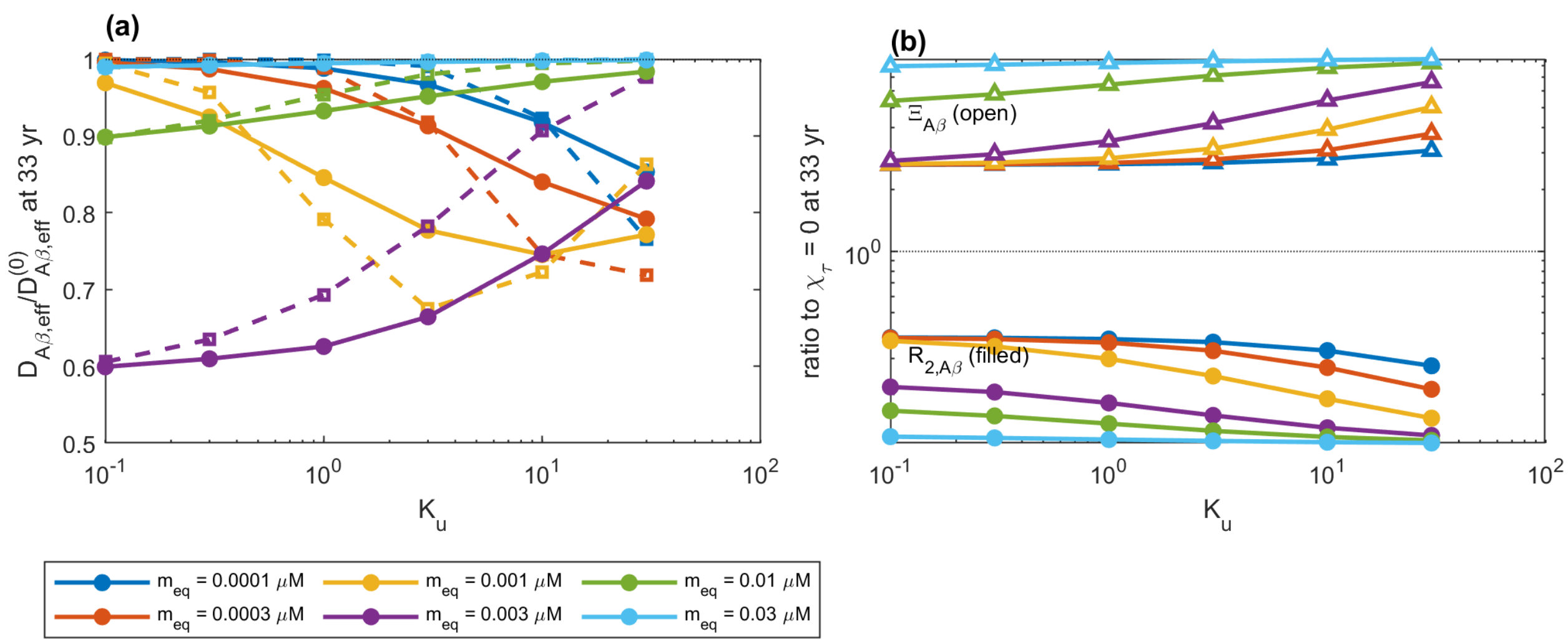

**Fig. S6.** Sensitivity of the growth-history study to the parameters of Eq. (14) (Section S2.1), PHF coupling at $f_{acc} = 1$, $\eta_{min} = \sigma_0/4$, $\eta_{max} = 2\sigma_0$. (a) $D_{A\beta,eff}/D_{A\beta,eff}^{(0)}$ at 33 years versus $K_u$ for $m_{eq,A\beta} = 10^{-4}$–$3\times10^{-2}$ μM (colors), with $h = 1$ (solid lines, filled circles) and $h = 2$ (dashed lines, open squares). (b) $R_{2,A\beta}/R_{2,A\beta}^{(0)}$ (filled circles) and $\Xi_{A\beta}(33\text{ yr})/\Xi_{A\beta}^{(0)}$ (open triangles) for $h = 1$; colors as in (a).

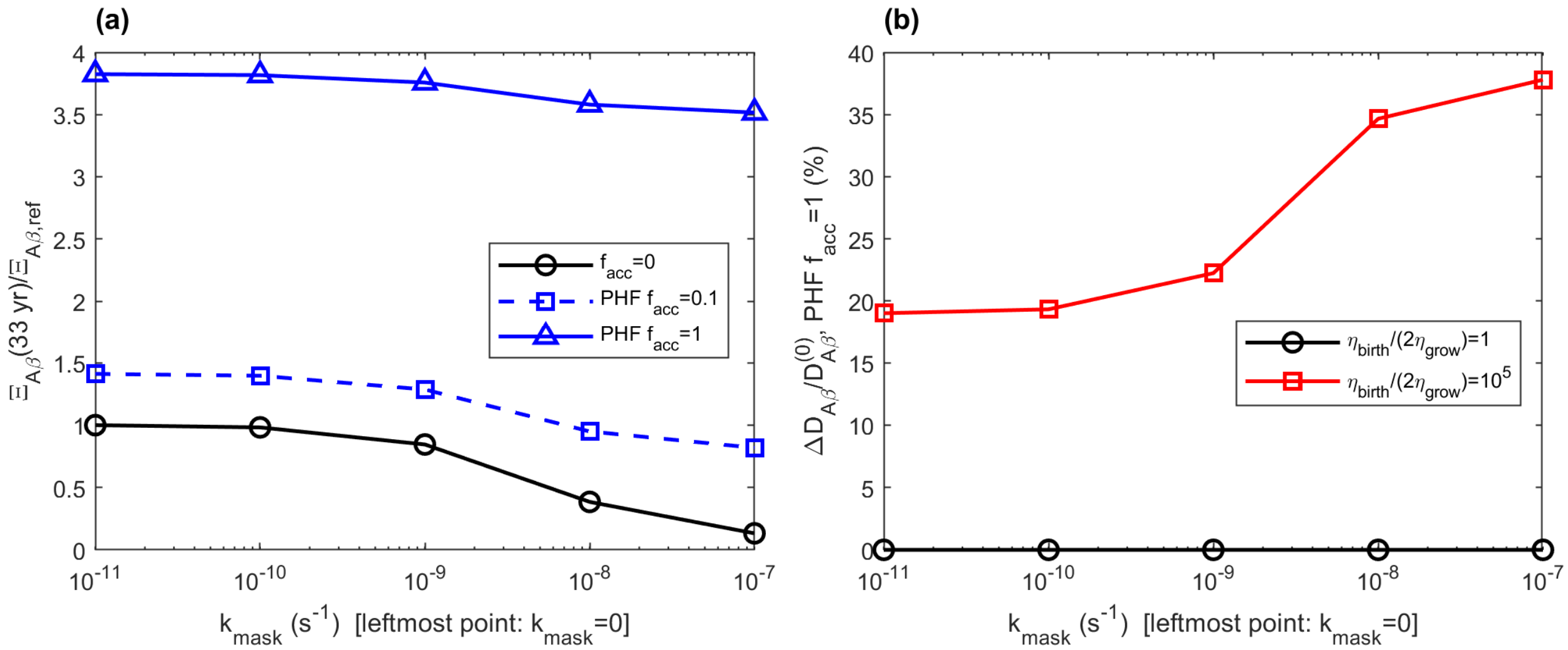

**Fig. S7.** Active-site masking at 33 years. (a) $\Xi_{A\beta}$(33 yr), normalized by the reference value for $f_{acc}$ = 0 and $k_{mask}$ = 0, for $f_{acc}$ = 0, 0.1, and 1 (PHF). (b) Relative difference in $D_{A\beta}$ between PHF coupling at $f_{acc}$ = 1 and $f_{acc}$ = 0 for $\eta_{birth}/(2\eta_{grow})$ = 1 and $10^5$. The leftmost points correspond to $k_{mask}$ = 0.

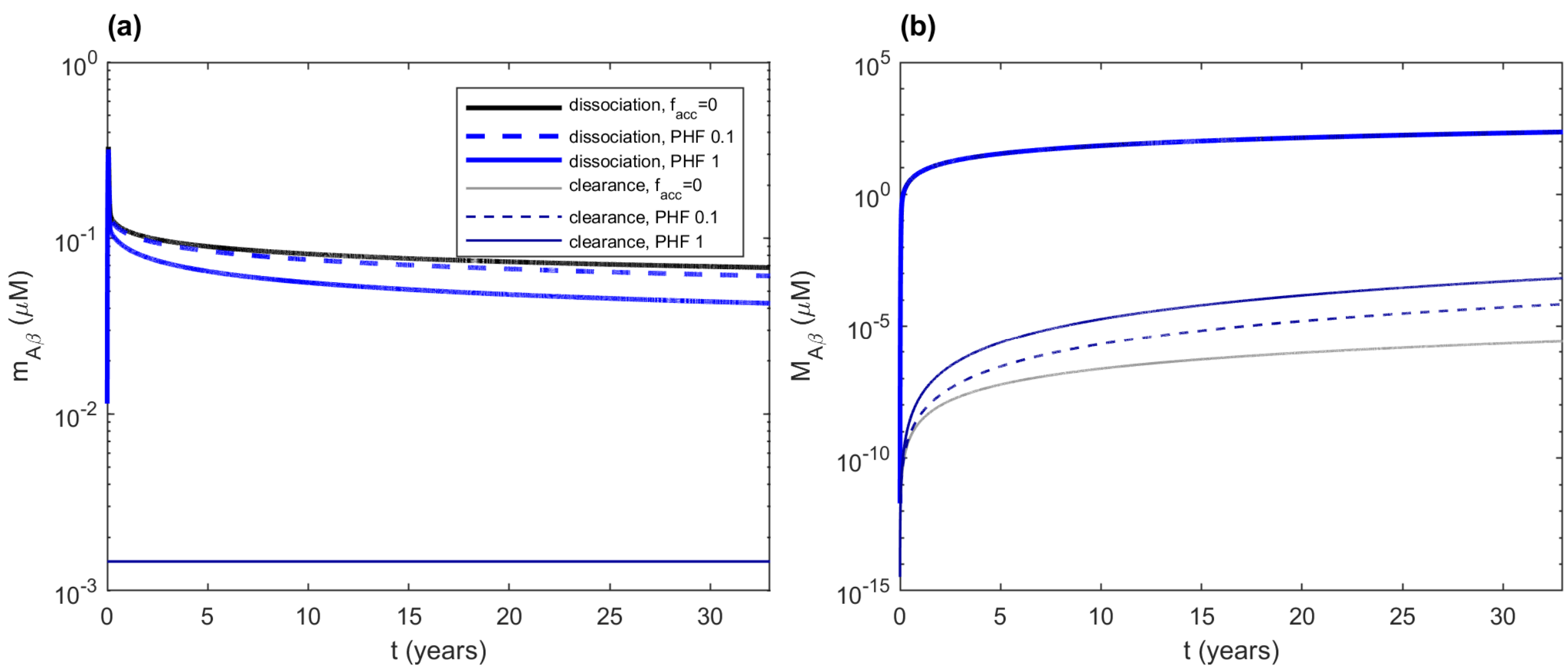

**Fig. S8.** Oligomer-dissociation scenario (thick lines) and soluble-protein-clearance scenario (thin lines) for $f_{acc}$ = 0, 0.1, and 1 (PHF). (a) Free Aβ monomer concentration; the three clearance curves coincide at $1.46\times10^{-3}$ μM. (b) Aβ fibril mass; the three dissociation curves coincide.

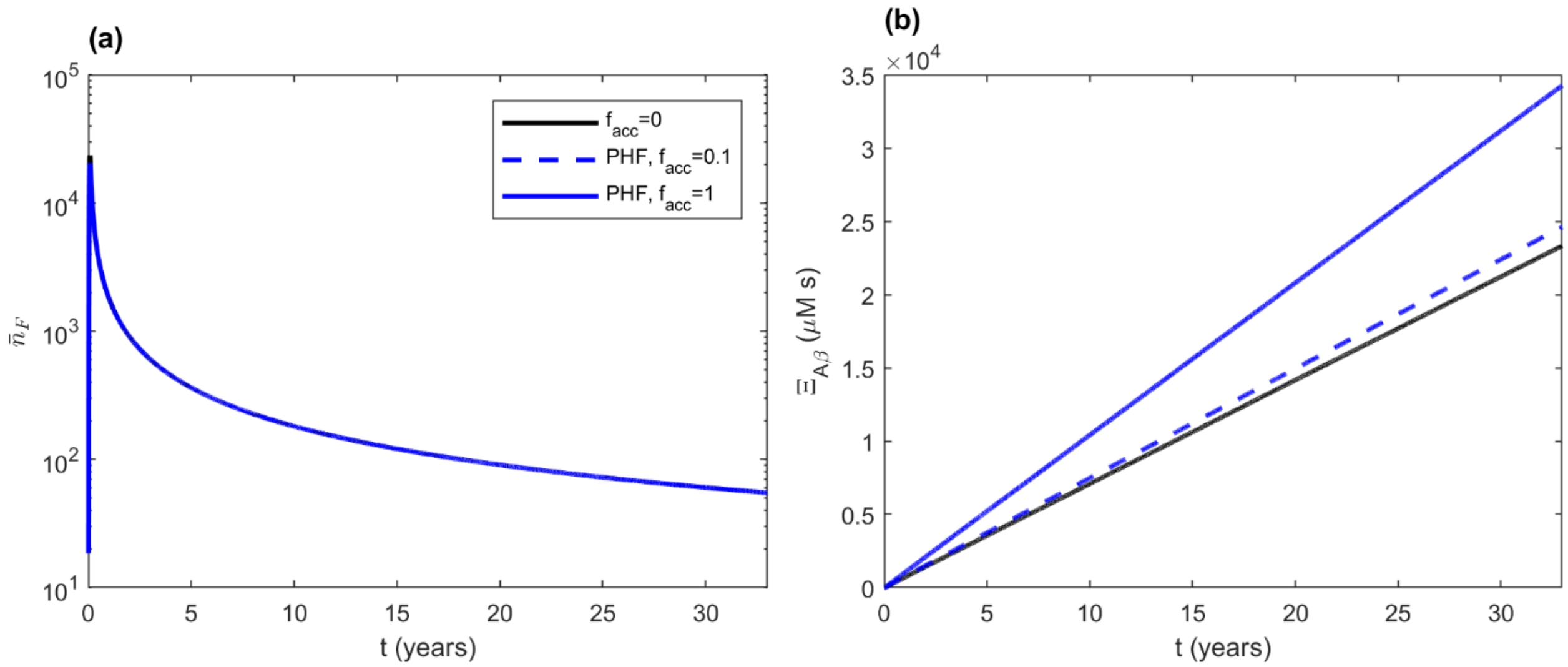

**Fig. S9.** Sensitivity to Aβ fibril fragmentation, $k_{-,A\beta} = 3.5\times10^{-11}$ s$^{-1}$, for $f_{acc}$ = 0, 0.1, and 1 (PHF); other parameters as in the reference case. (a) Mean fibril size $\bar{n}_F$. (b) Accumulated oligomer-equivalent exposure $\Xi_{A\beta}$.

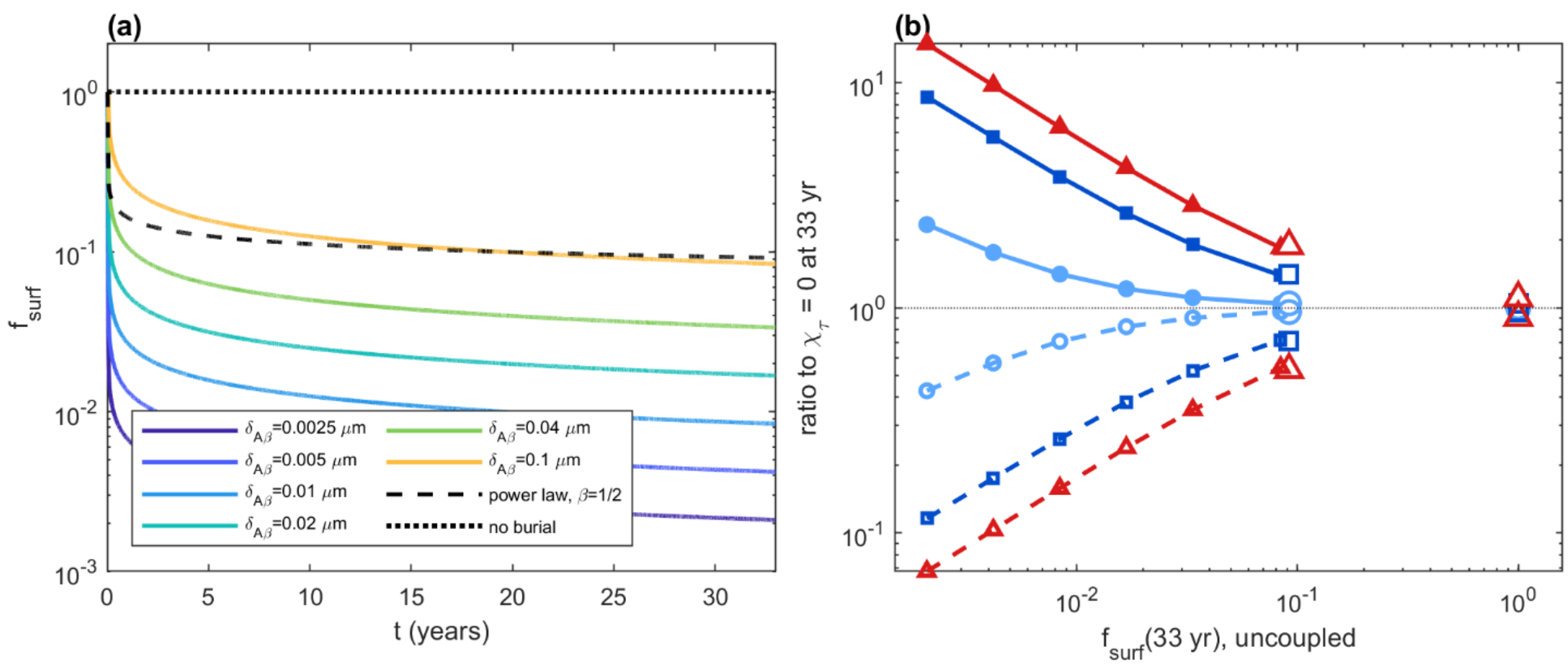

**Fig. S10.** Sensitivity to plaque accessibility (Sections 4.5 and S2.4). (a) Exposed fraction $f_{surf}$ without tau coupling for the shell closure [Eq. (16)] with $\delta_{A\beta}$ = 0.0025–0.1 µm (colored lines), the power-law closure $f_{surf} = min[1, (3\delta_{A\beta}/R_p)^{1/2}]$ with $\delta_{A\beta}$ = 0.010 µm (dashed), and no burial, $f_{surf}$ = 1 (dotted). (b) $\Xi_{A\beta}(33\text{ yr})/\Xi_{A\beta}^{(0)}$ (solid lines, filled symbols) and $R_{2,A\beta}/R_{2,A\beta}^{(0)}$ (dashed lines, open symbols) versus the uncoupled $f_{surf}$(33 yr), for $\chi_\tau = 1.86\times10^{-10}$ (PHF, $f_{acc}$ = 0.1; light blue circles), $1.86\times10^{-9}$ (PHF, $f_{acc}$ = 1; blue squares), and $4.45\times10^{-9}$ µM$^{-2}$ s$^{-1}$ (CTE, $f_{acc}$ = 1; red triangles). Small symbols: shell closure, $\delta_{A\beta}$ increasing from left to right; large symbols: power-law closure ($f_{surf} \approx 0.09$) and no burial ($f_{surf}$ = 1).

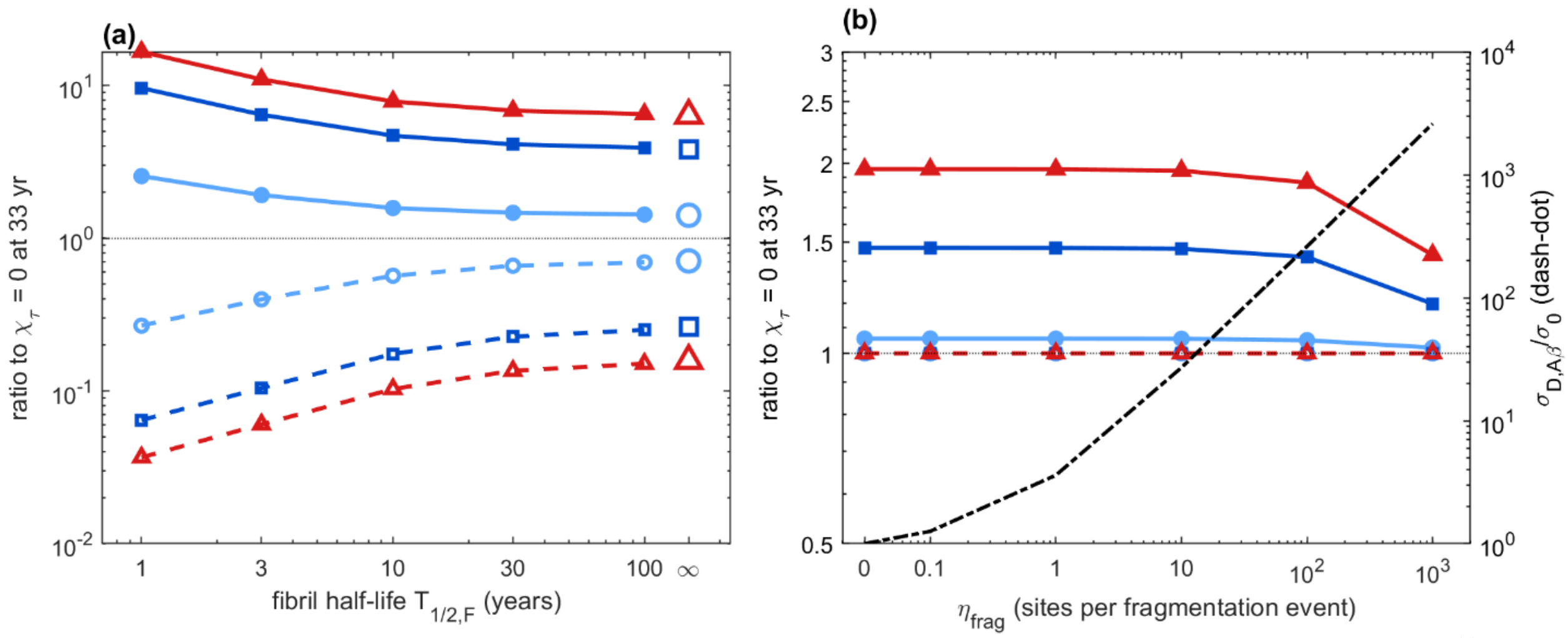

**Fig. S11.** (a) Slow removal of fibrillar Aβ with half-life $T_{1/2,F}$, applied to $P_{A\beta}$, $M_{A\beta}$, and $D_{A\beta}$: $\Xi_{A\beta}$(33 yr)/$\Xi_{A\beta}^{(0)}$ (solid lines, filled symbols) and $R_{2,A\beta}/R_{2,A\beta}^{(0)}$ (dashed lines, open symbols); the large symbols at $T_{1/2,F} = \infty$ show the reference case. (b) Fragmentation at $k_{-,A\beta} = 3.5\times10^{-11}$ s$^{-1}$ with $\eta_{frag}$ active sites created per fragmentation event: the same ratios (left axis; the $R_{2,A\beta}$ ratios coincide at unity) and the active-site density $\sigma_{D,A\beta}/\sigma_0$ without tau coupling (dash-dot line, right axis). Symbols and colors as in Fig. S10b.

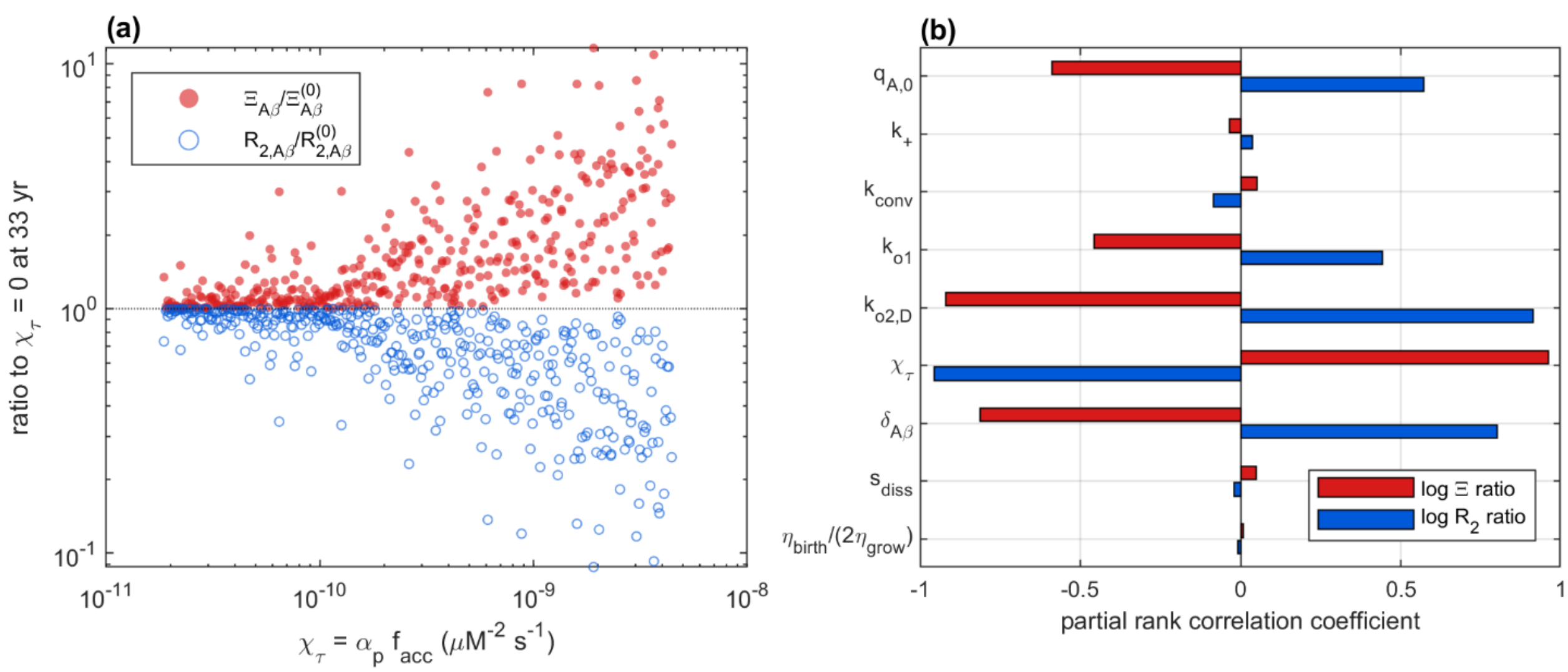

**Fig. S12.** Latin-hypercube robustness study A, without soluble-protein clearance (400 samples; Sections 4.5 and S2.4). (a) $\Xi_{A\beta}$(33 yr)/$\Xi_{A\beta}^{(0)}$ (filled red circles) and $R_{2,A\beta}/R_{2,A\beta}^{(0)}$ (open blue circles) versus the sampled coupling coefficient $\chi_\tau$; all $\Xi_{A\beta}$ ratios exceed unity and all $R_{2,A\beta}$ ratios are below unity. (b) Partial rank correlation coefficients between the logarithms of the sampled parameters and the logarithms of the two ratios. Sampled ranges (log-uniform): $q_{A,0,A\beta}$, 1/3–3 times the reference value; $k_{+,A\beta}$, $k_{conv,A\beta}$, $k_{o1,A\beta}$, and $k_{o2,A\beta,D,p}$, 0.1–10 times the reference values; $\chi_\tau$, $1.86\times10^{-11}$–$4.45\times10^{-9}$ μM$^{-2}$ s$^{-1}$; $\delta_{A\beta}$, 0.0025–0.04 μm; $s_{diss}$, $10^{-3}$–1 times the dissociation-scenario values of $k_{d1,A\beta}$ and $k_{d2,A\beta}$; $\eta_{birth}/(2\eta_{grow})$, 0.1–$10^3$.

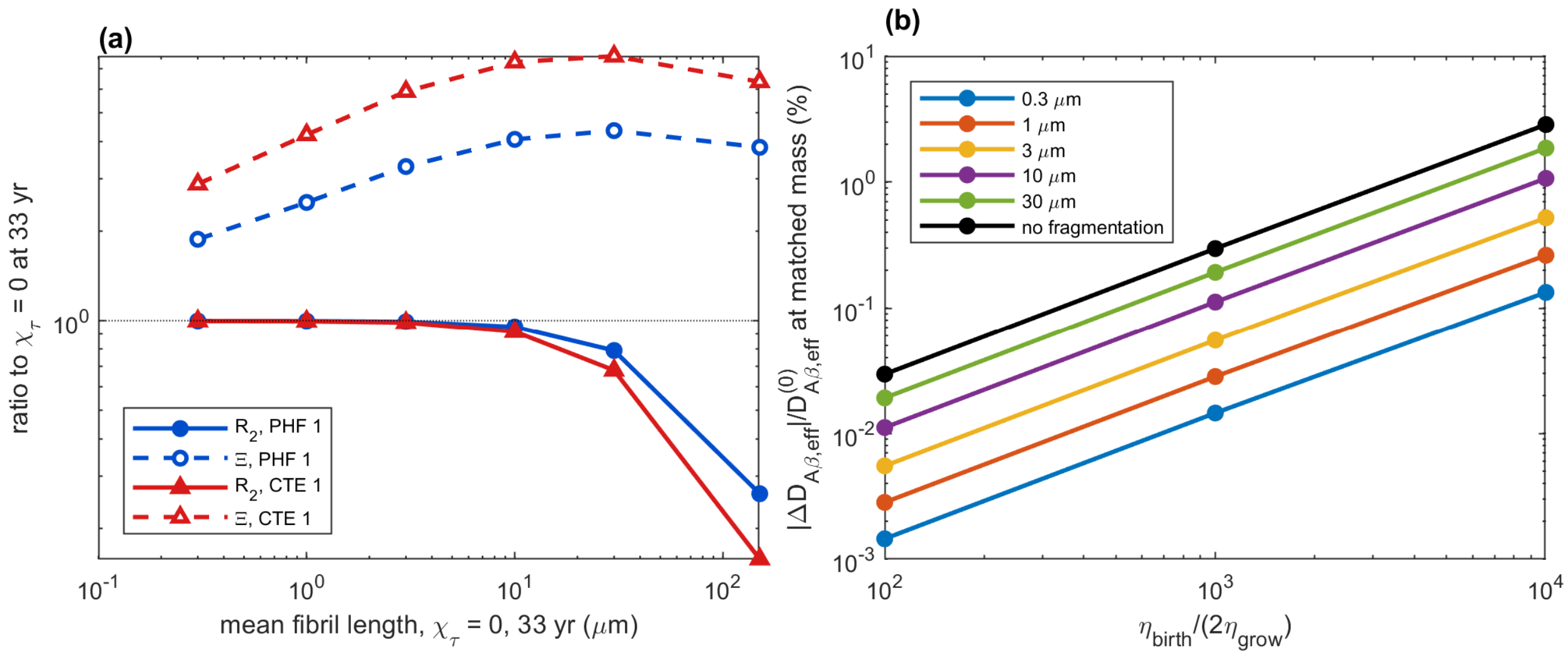

**Fig. S13.** Fibril length (Sections 4.5 and S2.4). The fragmentation coefficient $k_{-,A\beta}$ is chosen so that the uncoupled mean fibril length at 33 years is 0.3–30 μm; the rightmost points in (a), at about 150 μm, are the reference case without fragmentation. (a) $R_{2,A\beta}/R_{2,A\beta}^{(0)}$ (solid lines, filled symbols) and $\Xi_{A\beta}(33\text{ yr})/\Xi_{A\beta}^{(0)}$ (dashed lines, open symbols) for PHF (blue circles) and CTE (red triangles) coupling at $f_{acc}$ = 1, with mass-proportional sites. (b) Relative change of $D_{A\beta,eff}$ caused by PHF coupling at $f_{acc}$ = 1, at matched fibril mass, versus $\eta_{birth}/(2\eta_{grow})$, for the same fibril lengths.

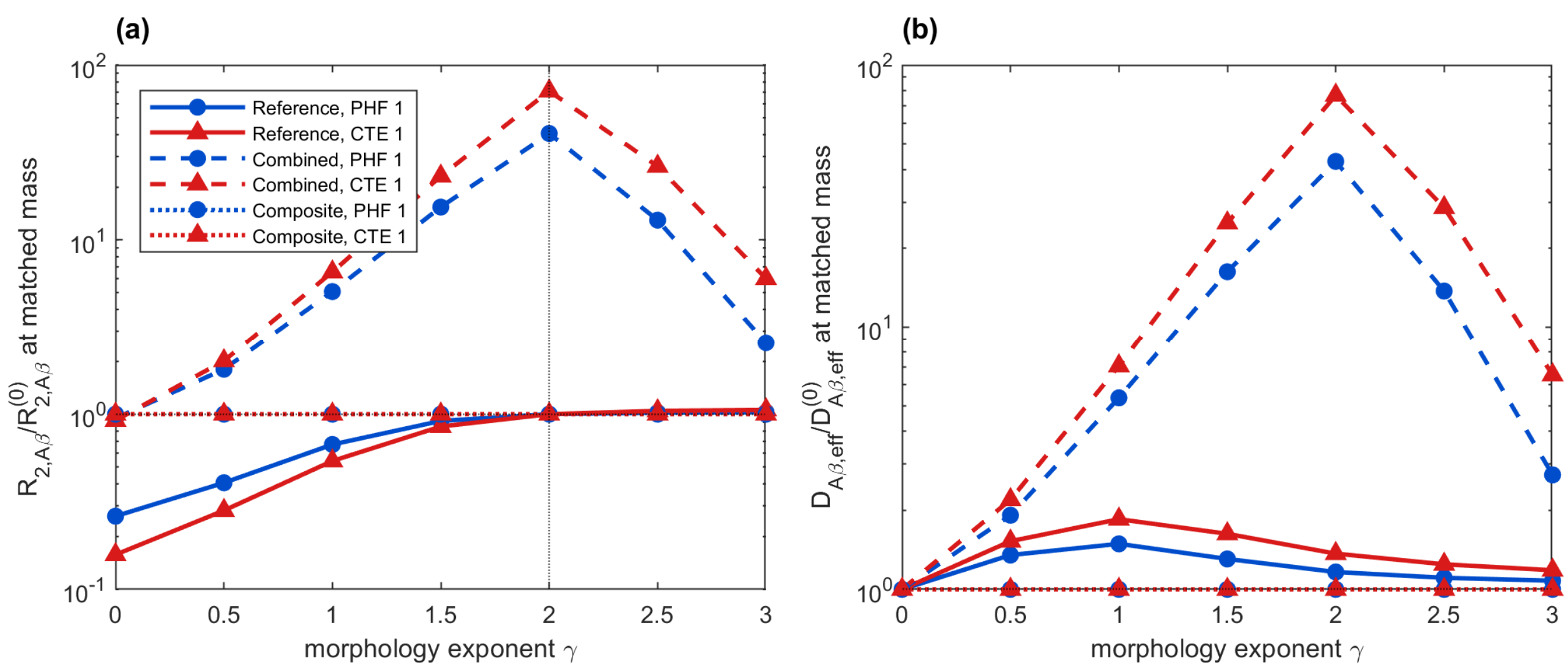

**Fig. S14.** Length-dependent accessibility, Eq. (S30) (Sections 4.5 and S2.5). (a) $R_{2,A\beta}/R_{2,A\beta}^{(0)}$ and (b) $D_{A\beta,eff}/D_{A\beta,eff}^{(0)}$ at matched fibril mass versus the exponent $\gamma$, for PHF (blue circles) and CTE (red triangles) coupling at $f_{acc}$ = 1, in the reference (solid lines), combined (dashed lines), and composite (dotted lines) scenarios. The vertical dotted line marks $\gamma$ = 2, at which the matched-mass rate ratio reaches unity in the reference case.

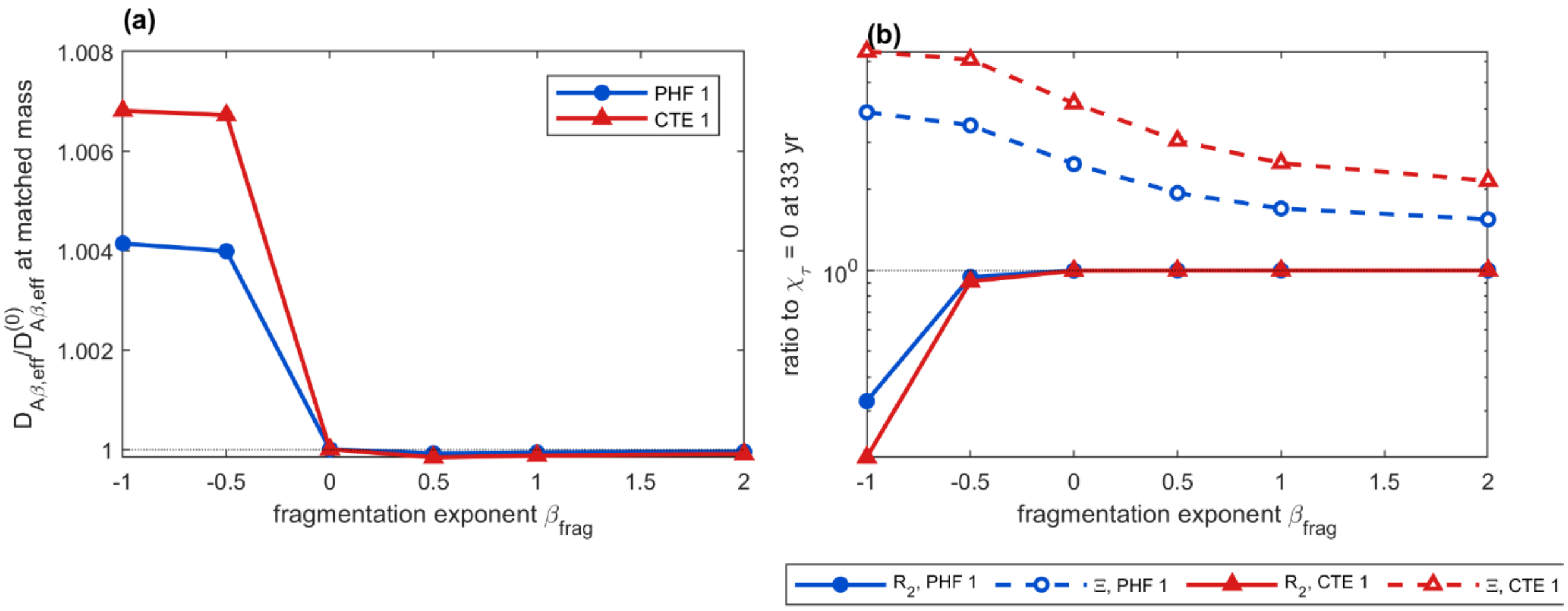

**Fig. S15.** Length-dependent fragmentation, Eq. (S31) (Sections 4.5 and S2.5), with $k_{-,A\beta} = 4.5\times10^{-13}$ s$^{-1}$ and $\eta_{frag}$ = 10. (a) $D_{A\beta,eff}/D_{A\beta,eff}^{(0)}$ at matched fibril mass and (b) $R_{2,A\beta}/R_{2,A\beta}^{(0)}$ (solid lines, filled symbols) and $\Xi_{A\beta}$(33 yr)/$\Xi_{A\beta}^{(0)}$ (dashed lines, open symbols) at 33 years versus the exponent $\beta_{frag}$, for PHF (blue circles) and CTE (red triangles) coupling at $f_{acc}$ = 1.

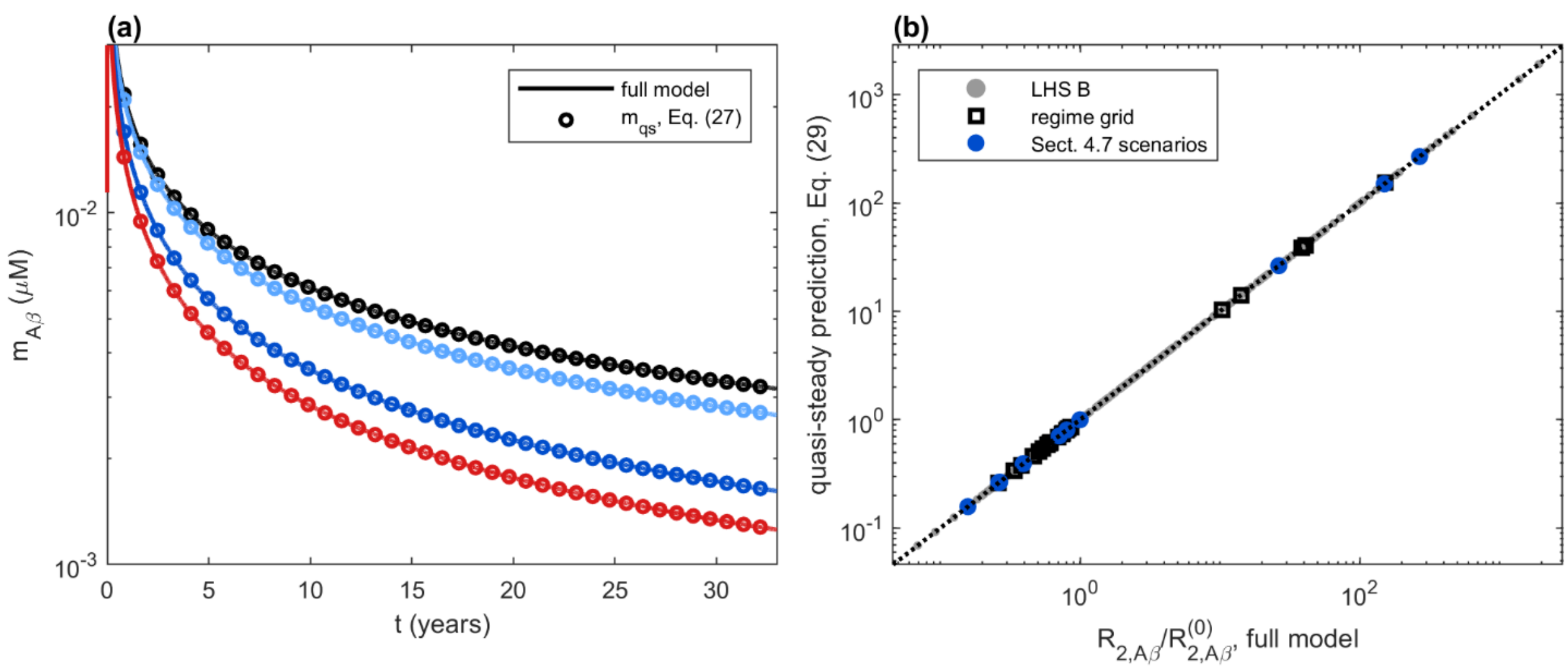

**Fig. S16.** Quasi-steady analysis (Sections 3.4 and 4.6). (a) Free Aβ monomer from the full model (lines) and from Eq. (28) (circles) in the reference case for $\chi_\tau = 0$ (black), $1.86\times10^{-10}$ (light blue), $1.86\times10^{-9}$ (blue), and $4.45\times10^{-9}$ μM$^{-2}$ s$^{-1}$ (red). (b) $R_{2,A\beta}/R_{2,A\beta}^{(0)}$ at 33 years predicted by Eq. (30), evaluated with the computed $P_{A\beta}$ and $D_{A\beta,eff}$, versus the full model, for the reference, dissociation, clearance, and fragmentation scenarios (blue circles), the regime grid of Fig. S17 (squares), and Latin-hypercube study B (gray dots); dotted line: equality.

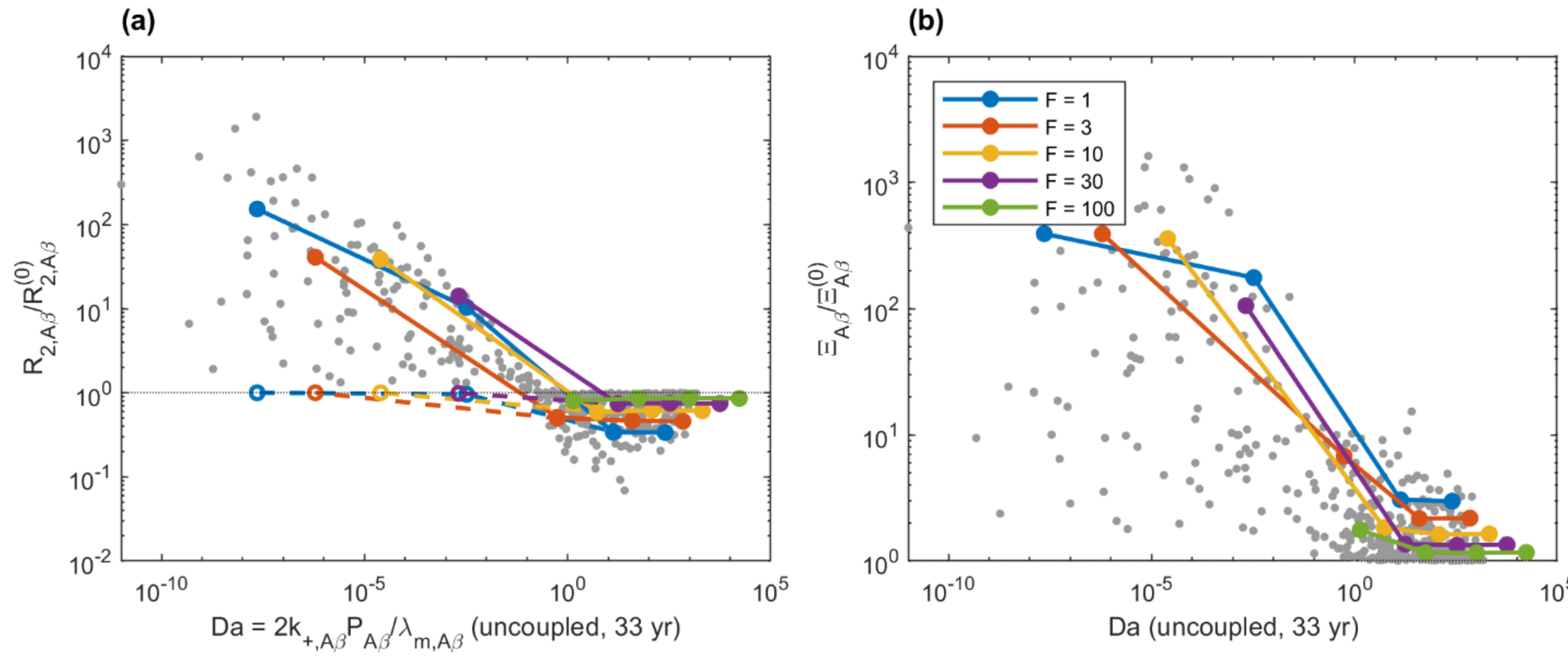

**Fig. S17.** Monomer-control regimes (Section 4.6). (a) $R_{2,A\beta}/R_{2,A\beta}^{(0)}$ at 33 years at equal time (solid lines, filled circles) and at matched fibril mass (dashed lines, open circles) for PHF coupling at $f_{acc} = 1$, versus the Damköhler number $Da = 2k_{+,A\beta}P_{A\beta}/\lambda_{m,A\beta}$ of the uncoupled system at 33 years, for Aβ source factors $F$ = 1–100 (colors) and Aβ clearance factors $c = 10^{-3}$–1 ($Da$ decreases with increasing $c$ along each line; $c = 0$, for which $Da \to \infty$, is not shown). Gray dots: equal-time ratios in Latin-hypercube study B (400 samples, $\chi_\tau$ sampled). (b) $\Xi_{A\beta}(33\text{ yr})/\Xi_{A\beta}^{(0)}$ for the same simulations.

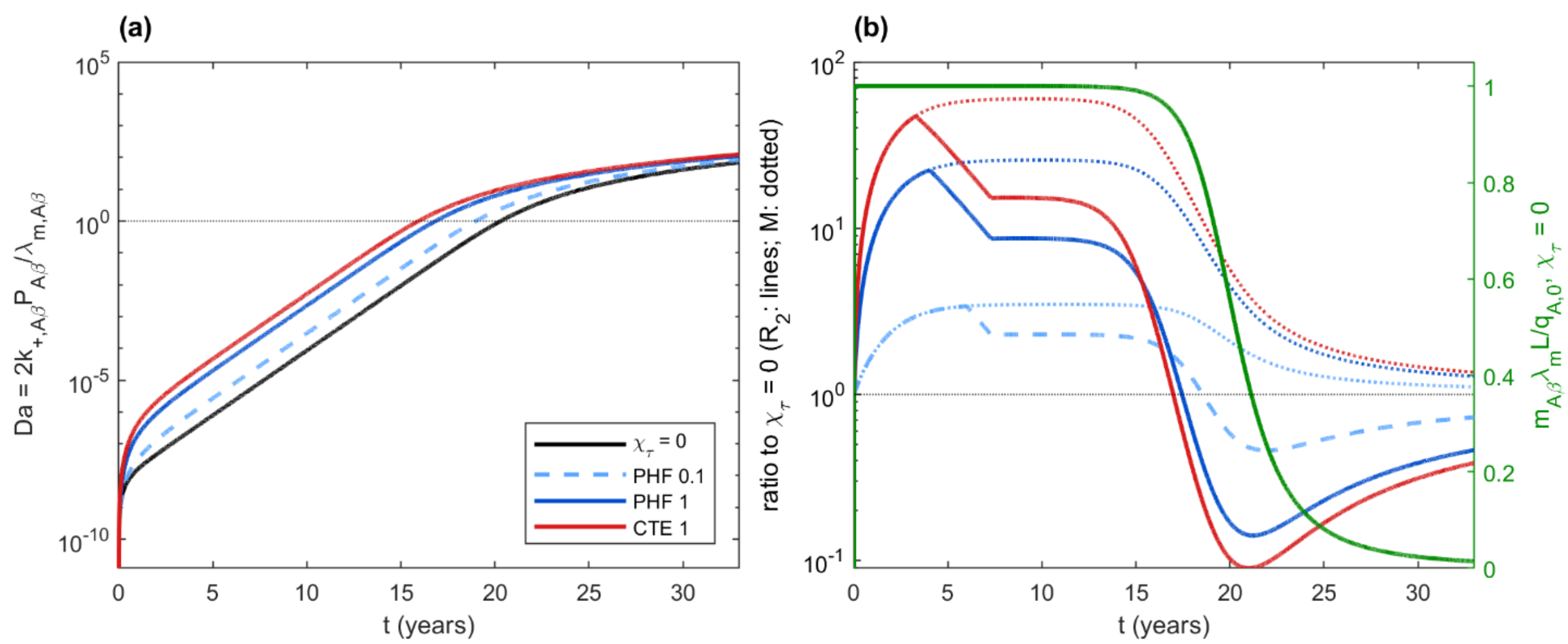

**Fig. S18.** Composite scenario (Section 4.6): oligomer dissociation, soluble-protein clearance, and fragmentation ($k_{-,A\beta} = 1.15\times10^{-12}$ s$^{-1}$), with the Aβ source raised 2.71-fold so that the uncoupled fibril mass at 33 years is 228.8 μM and the uncoupled mean fibril length is 1 μm. (a) Damköhler number $Da$ [Eq. (29)] for $\chi_\tau = 0$ (black), PHF coupling at $f_{acc} = 0.1$ (light blue, dashed) and 1 (blue), and CTE coupling at $f_{acc} = 1$ (red). (b) $R_{2,A\beta}/R_{2,A\beta}^{(0)}$ (lines) and $M_{A\beta}/M_{A\beta}^{(0)}$ (dotted lines), left axis, and free Aβ monomer of the uncoupled system normalized by its clearance-controlled value, $m_{A\beta}\lambda_{m,A\beta}L/q_{A,0,A\beta}$ (green, right axis). The kinks in the rate ratios at 3–7 years occur when $f_{surf}$ falls below unity, first in the coupled and later in the uncoupled system.

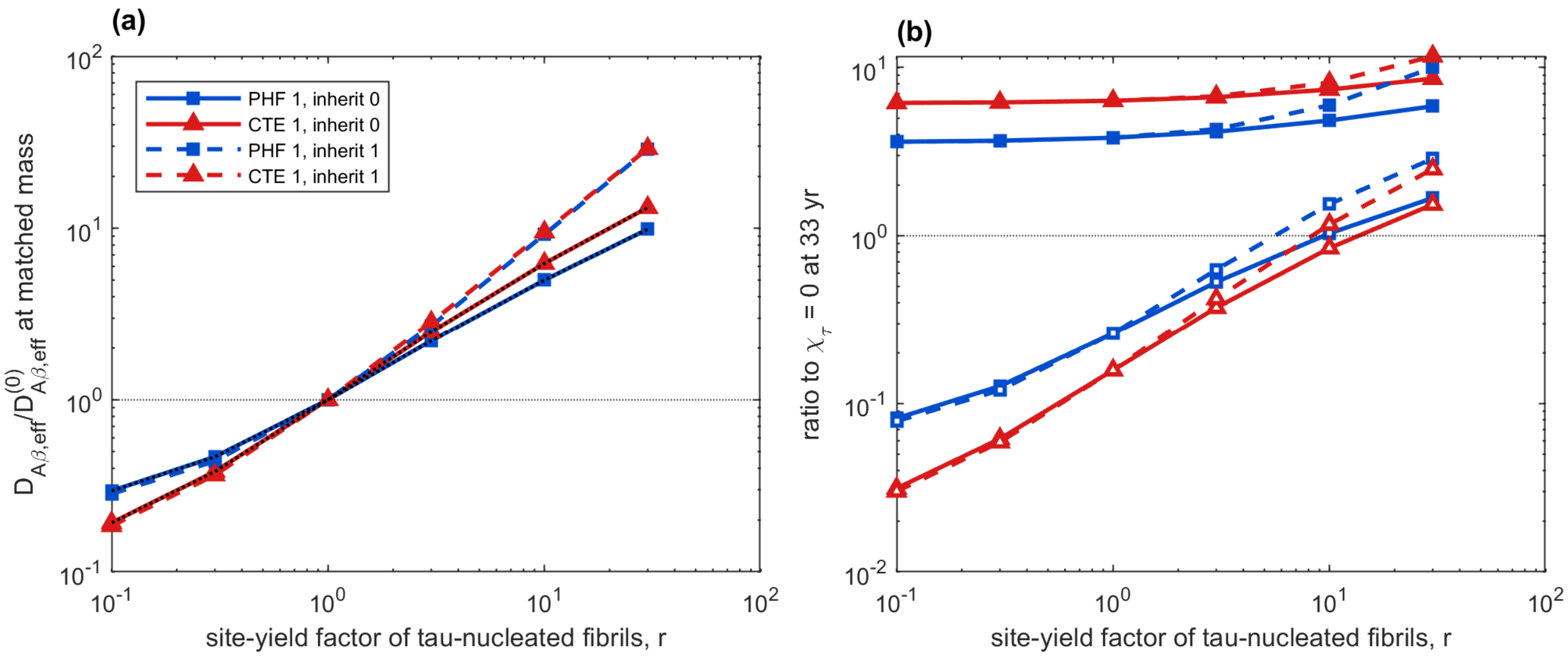

**Fig. S19.** Two-lineage model (Section S2.6), reference case. (a) Activity-weighted accessible-site concentration at matched fibril mass relative to $\chi_\tau = 0$, versus the site-abundance factor r of tau-nucleated fibrils, for PHF (blue squares) and CTE (red triangles) coupling at $f_{acc} = 1$, without (solid lines) and with (dashed lines) propagation of the tau-derived lineage by secondary nucleation; dotted lines: $1 + (r - 1)\varphi_\tau$. (b) $\Xi_{A\beta}(33\text{ yr})/\Xi_{A\beta}^{(0)}$ (filled symbols) and $R_{2,A\beta}/R_{2,A\beta}^{(0)}$ (open symbols) at 33 years; line styles as in (a).

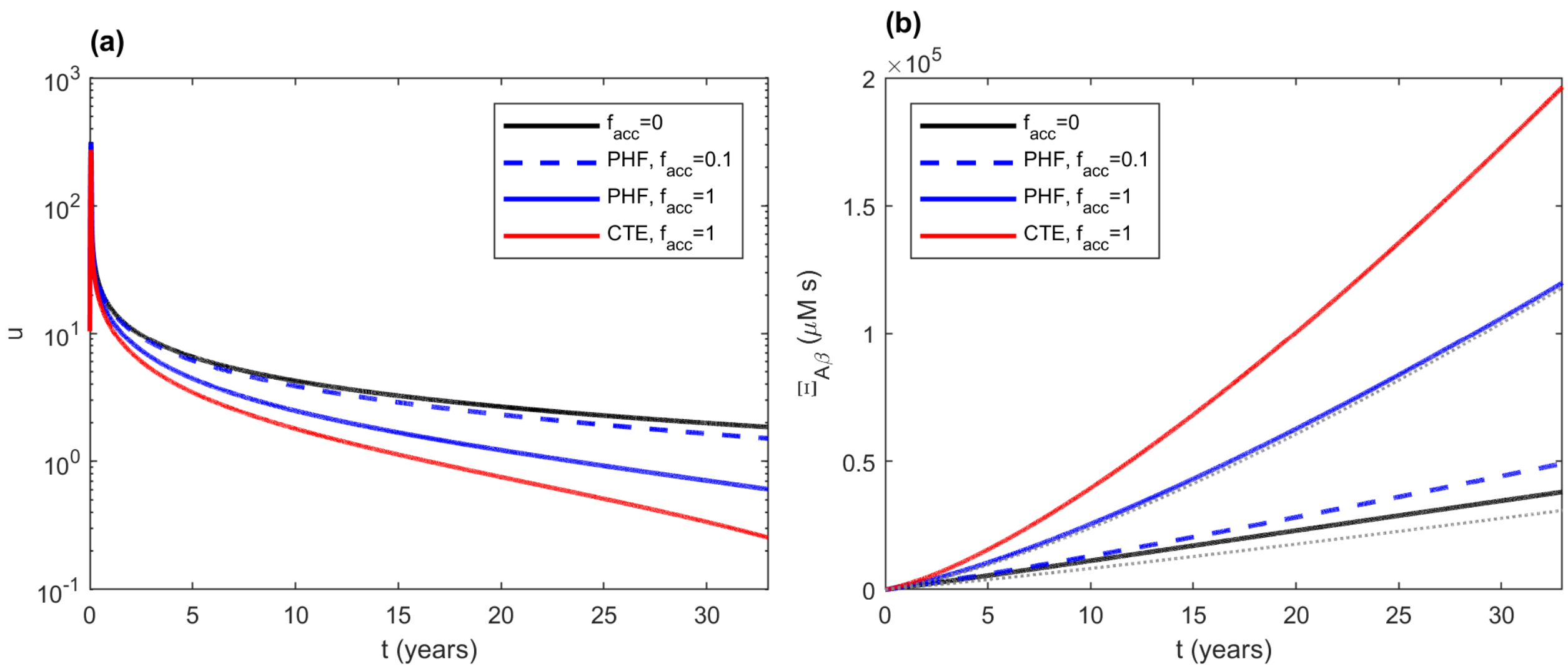

**Fig. S20.** Growth-history-dependent site formation [Eq. (14)] with the parameters of Fig. S21a. (a) Supersaturation $u$. (b) Accumulated oligomer-equivalent exposure $\Xi_{A\beta}$; gray dotted lines show the constant-yield reference case for $f_{acc} = 0$ and for PHF coupling at $f_{acc} = 1$.

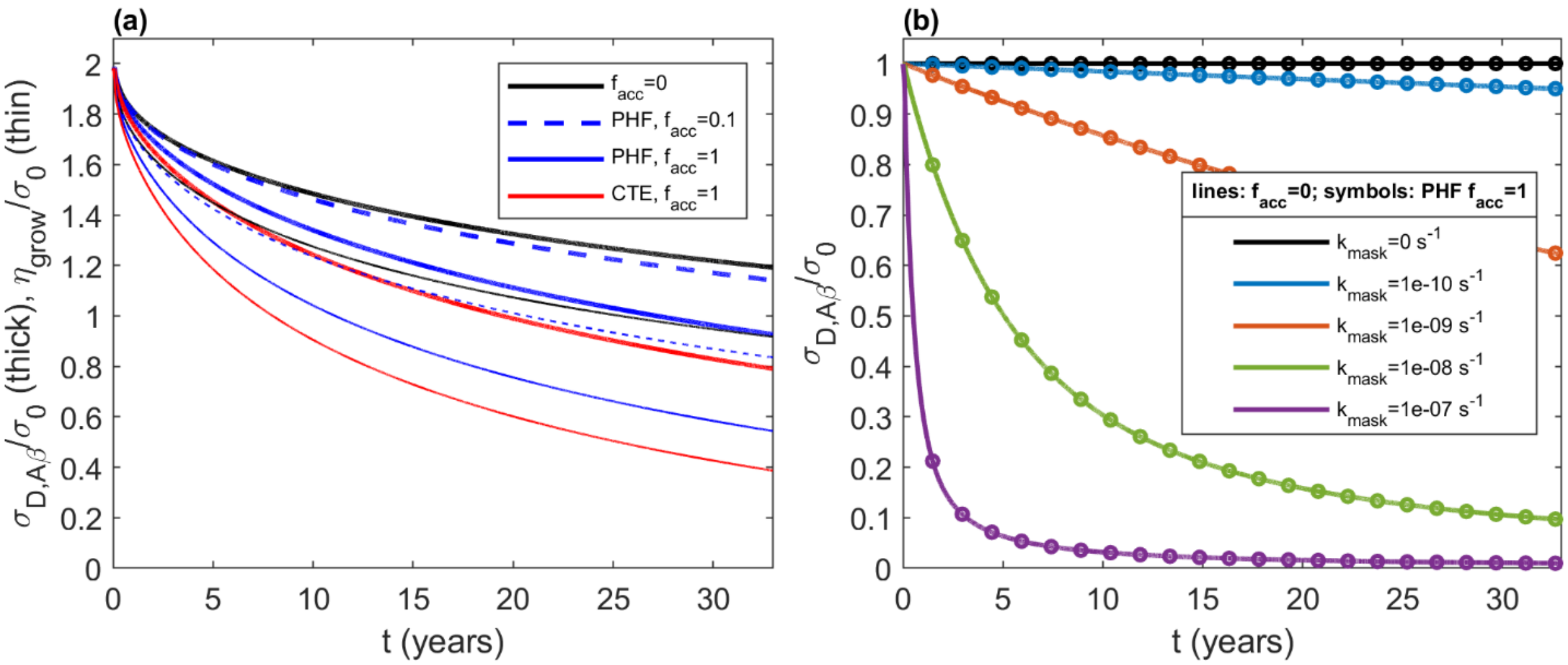

**Fig. S21.** Robustness of the active-site density. (a) Growth-history-dependent site formation [Eq. (14)] with $m_{eq,A\beta} = 10^{-3}$ μM, $\eta_{min} = \sigma_0/4$, $\eta_{max} = 2\sigma_0$, $K_u = 3$, and $h = 1$: active-site density $\sigma_{D,A\beta}/\sigma_0$ (thick lines) and instantaneous elongation yield $\eta_{grow}/\sigma_0$ (thin lines). (b) Active-site masking: $\sigma_{D,A\beta}/\sigma_0$ for $k_{mask} = 0$–$10^{-7}$ s$^{-1}$, for $f_{acc} = 0$ (lines) and PHF coupling at $f_{acc} = 1$ (symbols); $\eta_{birth} = 2\sigma_0$, $\eta_{grow} = \sigma_0$.

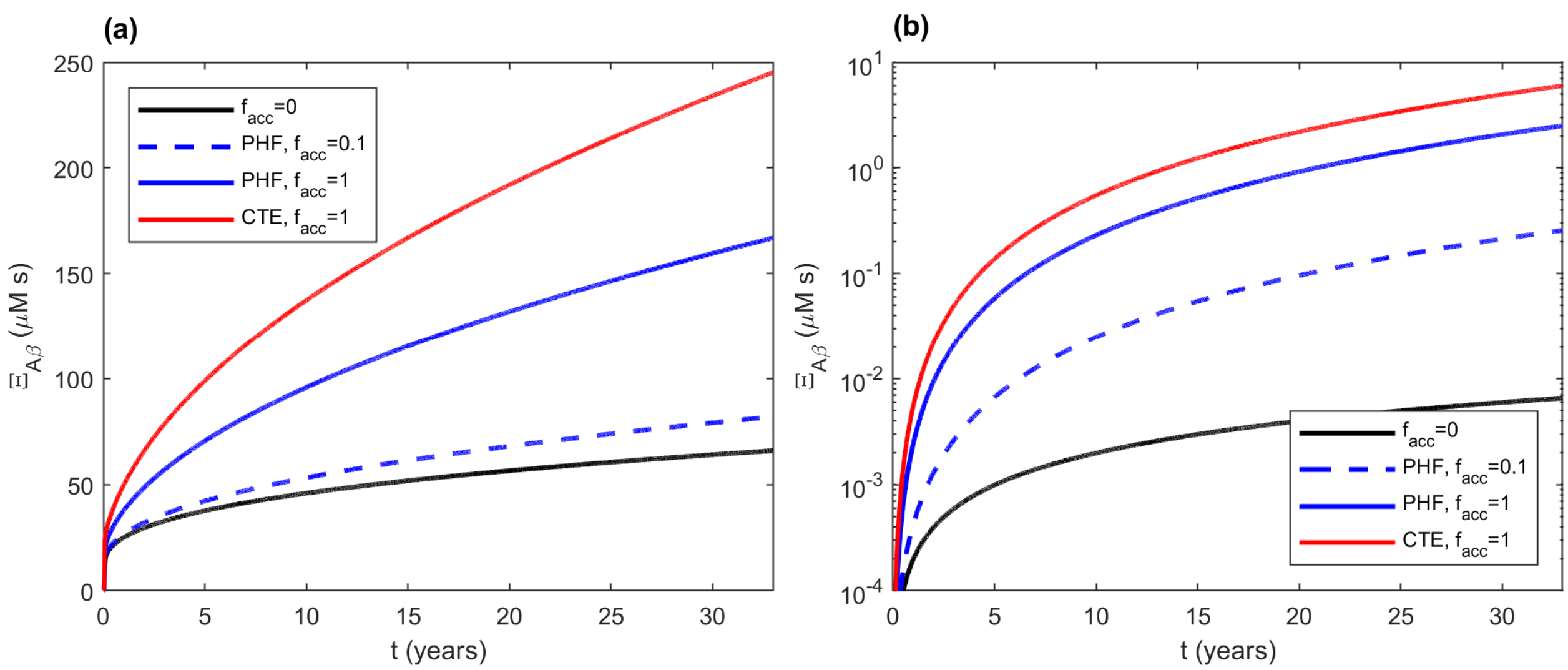

**Fig. S22.** Accumulated soluble Aβ oligomer-equivalent exposure $\Xi_{A\beta}$ (a) with oligomer dissociation ($k_{d1,A\beta} = 6.47\times10^{-7}$ s$^{-1}$, $k_{d2,A\beta} = 1.99\times10^{-5}$ μM$^{-1}$ s$^{-1}$, and the tau dissociation constants of Table S1) and (b) with soluble-protein clearance (half-lives of Table S1), for $f_{acc} = 0$ (black), PHF coupling at $f_{acc} = 0.1$ (blue dashed) and 1 (blue solid), and CTE coupling at $f_{acc} = 1$ (red). Note the logarithmic scale in (b).

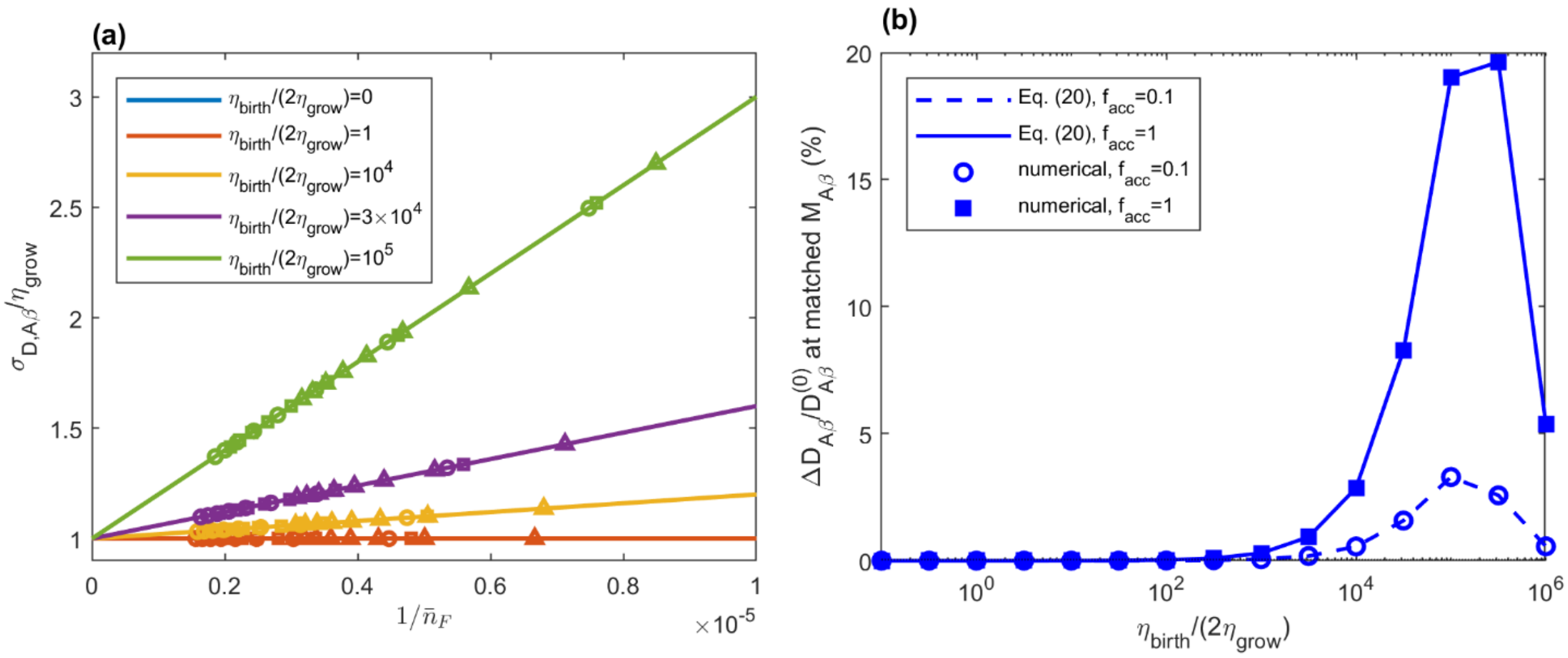

**Fig. S23.** Analytical benchmark for departures from mass-proportional site formation [Eqs. (24) and (25)], with $\eta_{grow} = \sigma_0$ and $k_{mask} = k_{-,A\beta} = 0$. (a) Active-site density normalized by the elongation yield versus the inverse mean fibril size. Lines: $\sigma_{D,A\beta}/\eta_{grow} = 1 + (\eta_{birth}/\eta_{grow} - 2)/\bar{n}_F$; symbols: full simulations for $f_{acc} = 0$ (circles), 0.1 (squares), and 1 (triangles; PHF), sampled between 3.3 and 33 years. The lines for $\eta_{birth}/(2\eta_{grow}) = 0$ and 1 are indistinguishable at this scale. (b) Relative difference in total active-site concentration between PHF-coupled ($f_{acc}$ = 0.1 and 1) and uncoupled systems at matched fibril mass ($M_{A\beta}$ = 228.8 μM): Eq. (25) evaluated with the computed $\Delta P_{A\beta}$ (lines) and full simulations (symbols).

## S7. Supplementary tables: parameters and numerical outcomes

The tables provide a compact lookup layer for the model. Table S1 gives the complete parameter set and its evidence status; Tables S2 and S3 report the principal numerical scenarios; Table S4 summarizes the plaque-accessibility, fibril-removal, and fragmentation-site sensitivity cases; and Table S5 defines every dependent and derived quantity. Together they allow any main-text conclusion to be traced back to its assumptions and numerical values.

**Table S1. Model parameters, inherited kinetic estimates, and new active-site coefficients.**

| Symbol | Definition | Units | Value/range | Source or basis | Value used / status |
|---|---|---|---|---|---|
| **Compartment geometry, source terms, and tau–Aβ coupling** | | | | | |
| $V_{tau}$ (= $V_{soma}$) | Volume of neuronal soma / intracellular tau compartment | μm$^3$ | 4.19×10$^3$ | Section S1.1 | 4.19×10$^3$ |
| $L$ | Side length of effective Aβ control volume | μm | 50 | [10], inter-plaque spacing of ≈100 μm estimated from Fig. 1(a), $L$ ≈ half the spacing; Section S1.2 | 50 |
| $V_{A\beta}$ | Volume of extracellular/interstitial Aβ compartment | μm$^3$ | $L^3$ = 1.25×10$^5$ | Geometric definition | $L^3$ |
| $a_{21}$ | Conversion from mol μm$^{-3}$ to μM | μM μm$^3$ mol$^{-1}$ | 10$^{21}$ | Unit conversion | 10$^{21}$ (tau source; plaque volume) |
| $q_{tau}$ | Effective tau monomer supply to soma | mol s$^{-1}$ | 1.10×10$^{-23}$ | Section S1.3 | Inherited; recheck if tau subsystem is refitted |
| $q_{A,0,A\beta}/L$ | Volume-averaged Aβ monomer production source | μM s$^{-1}$ | 2.2×10$^{-7}$ | [11]; Section S1.4 | 2.2×10$^{-7}$ |
| $f_{acc}$ | Effective fraction of tau fibril mass available to extracellular Aβ (phenomenological) | — | 0–1 | Coupling sensitivity parameter | 0, 0.1, 1; extend as needed |
| $\alpha_p$ | Tau-fold-dependent Aβ heterotypic nucleation coefficient (PHF/CTE) | μM$^{-2}$ s$^{-1}$ | PHF 1.86×10$^{-9}$; CTE 4.45×10$^{-9}$ | [3], seed series; Section S1.5 | Fold-specific; inherited |
| $\chi_\tau$ | Effective heterotypic coupling coefficient, $\alpha_p f_{acc}$ | μM$^{-2}$ s$^{-1}$ | 0–4.45×10$^{-9}$ | Derived (Section 2.2.1) | 0; 1.86×10$^{-10}$; 4.45×10$^{-10}$; 1.86×10$^{-9}$; 4.45×10$^{-9}$ |
| $\delta_{A\beta}$ | Thickness of the catalytically accessible surface layer of an Aβ plaque | μm | 0.010 | Approximately one fibril diameter; Section S1.6 | 0.010; 0.0025–0.1 (Section 4.5) |

| Symbol | Definition | Units | Value/range | Source or basis | Value used / status |
|---|---|---|---|---|---|
| $MW_{A\beta}$ | Aβ42 monomer molecular weight; converts fibril mass to protein-equivalent deposit volume | g mol$^{-1}$ | $4.51\times10^{3}$ | [11]; Section S1.7 | $4.51\times10^{3}$ |
| $\rho_{A\beta}$ | Effective protein-material density used to convert fibrillar Aβ mass to a protein-equivalent compact-deposit volume; not a measured bulk plaque density | g μm$^{-3}$ | $1.35\times10^{-12}$ (1.35 g cm$^{-3}$) | [14], conventional pure-protein density; Section S1.8 | $1.35\times10^{-12}$ |
| **Tau aggregation: inherited kinetic parameters** | | | | | |
| $n_{c,tau}$ | Effective tau primary-nucleation order | — | 1 | [24]; Section S1.9 | 1 |
| $n_{2,tau}$ | Effective tau fibril-surface secondary-nucleation order | — | 2 | [24] | 2 |
| $k_{o1,tau}$ | Effective first-order tau primary-nucleation coefficient | s$^{-1}$ | $2.78\times10^{-6}$ | [24], [23], upper-bound-based estimate; [4], note f (inherited value); [21], qualitative Finke–Watzky comparison only; Section S1.10 | Inherited estimate |
| $k_{o2,tau}$ | Effective tau surface-secondary-nucleation coefficient | μM$^{-2}$ s$^{-1}$ | $>1.67\times10^{-4}$ | [24], Table 1; [4], note g; derived using $k_{o1,tau}$; Section S1.11 | Lower-bound / sensitivity estimate; inherits $k_{o1,tau}$ uncertainty |
| $k_{d1,tau}$ | Spontaneous tau oligomer dissociation coefficient | s$^{-1}$ | $1.69\times10^{-4}$ | [22]; Section S1.12 | 0 (baseline); $1.69\times10^{-4}$ (dissociation scenario) |
| $k_{d2,tau}$ | Tau fibril-catalyzed oligomer dissociation coefficient | μM$^{-1}$ s$^{-1}$ | $\approx5.6\times10^{-5}$ | [7]; Section S1.13 | 0 (baseline); $5.6\times10^{-5}$ (dissociation scenario); cross-construct estimate |

| Symbol | Definition | Units | Value/range | Source or basis | Value used / status |
|---|---|---|---|---|---|
| $k_{+,tau}$ | Tau fibril elongation coefficient | $\mu M^{-1}$ $s^{-1}$ | $4.2\times10^{-3}$ | [23]; Section S1.14 | Inherited reference |
| $k_{-,tau}$ | Tau fibril fragmentation coefficient | $s^{-1}$ | $5.6\times10^{-11}$ | [23]; Section S1.15 | Inherited reference |
| $k_{conv,tau}$ | Tau oligomer-to-fibril conversion coefficient | $\mu M^{-1}$ $s^{-1}$ | $8.9\times10^{-6}$ | [22]; Section S1.16 | Inherited reference |
| $n_{min,tau}$ | Minimum stable tau fibril size in fragmentation law | monomers | $2\times10^{3}$ | Section S1.17 | Inherited assumption |
| $\lambda_{m,tau}$ | Effective clearance rate of soluble tau monomers | $s^{-1}$ | $\ln(2)/(2.16\times10^{5}$ s$) \approx 3.21\times10^{-6}$ | [17], Fig. 1A (approximately 60 h cellular-tau turnover); Section S1.18 | 0 (baseline, $T_{½} \to \infty$); $3.21\times10^{-6}$ (clearance scenario); cell-culture tau-turnover proxy, not monomer-specific |
| $\lambda_{S,tau}$ | Effective clearance rate of soluble tau oligomers | $s^{-1}$ | $\ln(2)/(8.38\times10^{5}$ s$) \approx 8.27\times10^{-7}$ | [18]; Section S1.19 | 0 (baseline, $T_{½} \to \infty$); $8.27\times10^{-7}$ (clearance scenario); soluble-tau turnover proxy, not oligomer-specific |
| **Aβ aggregation: inherited kinetic parameters** | | | | | |
| $n_{c,A\beta}$; $n_{2,A\beta}$ | Effective Aβ primary- and secondary-nucleation orders | — | 2; 2 | [3]; [29] | 2; 2 |
| $k_{o1,A\beta}$ | Intrinsic Aβ primary-nucleation coefficient in $R_{1,A\beta} = k_{o1,A\beta} m_{A\beta}^2$ | $\mu M^{-1}$ $s^{-1}$ | $(1.7–3.1)\times10^{-10}$ | [3]; [29]; Section S1.20 | $2.17\times10^{-10}$ |
| $k_{+,A\beta}$ | Aβ fibril elongation coefficient in the elongation term | $\mu M^{-1}$ $s^{-1}$ | 0.0972; 3.06 comparison | [3]; [29]; Section S1.21 | 0.0972 |

| Symbol | Definition | Units | Value/range | Source or basis | Value used / status |
|---|---|---|---|---|---|
| $k_{conv,A\beta}$ | Aβ oligomer-to-fibril conversion coefficient | $\mu M^{-1}\ s^{-1}$ | $1.83\times10^{-6}$ | [29]; Table S1 of that study's Supporting Information (pseudo-first-order rate at 5 μM); derived; Section S1.22 | Inherited derived estimate; first-order monomer dependence assumed (fitted order ≈ 2.7) |
| $k_{-,A\beta}$ | Aβ fibril fragmentation coefficient | $s^{-1}$ | $3.5\times10^{-11}$ (unliganded) | [30]; Section S1.23 | 0 reference; $3.5\times10^{-11}$ (Section S2.3); $1.3\times10^{-14}$–$1.5\times10^{-12}$ for mean fibril lengths of 0.3–30 μm (Sections 4.5, 4.6) |
| $k_{d1,A\beta}$ | Spontaneous Aβ oligomer dissociation coefficient | $s^{-1}$ | $6.47\times10^{-7}$ | [29], [7]; total dissociation rate + catalytic ratio; Section S1.24 | 0 (baseline); $6.47\times10^{-7}$ (derived split) |
| $k_{d2,A\beta}$ | Aβ fibril-associated oligomer dissociation coefficient | $\mu M^{-1}\ s^{-1}$ | $1.99\times10^{-5}$ | [29], [7]; total dissociation rate + catalytic ratio; Section S1.25 | 0 (baseline); $1.99\times10^{-5}$ (derived split); separate from active-site *D* |

| Symbol | Definition | Units | Value/range | Source or basis | Value used / status |
|---|---|---|---|---|---|
| $\ln(2)/T_{1/2,m,A\beta}$ | Effective clearance rate of soluble Aβ monomers | $s^{-1}$ | ln(2)/(1.28 h) ≈ $1.50\times10^{-4}$ | [27], turnover half-life assumed in slice modeling; adopted as an effective monomer-clearance rate; Section S1.26 | 0 (baseline, $T_{½}$ → ∞); $1.50\times10^{-4}$ (clearance scenario); turnover/clearance proxy, not a direct monomer-specific proteolysis measurement |
| $\ln(2)/T_{1/2,S,A\beta}$ | Effective clearance rate of soluble Aβ oligomers | $s^{-1}$ | ln(2)/($9.40\times10^{3}$ s) ≈ $7.37\times10^{-5}$; HMW-derived ≈$(1.0–3.5)\times10^{-5}$ (candidate sensitivity) | [28], size-exclusion-chromatography fractions: LMW 14–19 (LMW oligomers, including dimers), HMW 1–11; 8-h Aβ42 decrease 88.0% (LMW) vs 25.4% (HMW); [4], note t; first-order conversion, not a reported half-life; Section S1.27 | 0 (baseline, $T_{½}$ → ∞); $7.37\times10^{-5}$ (clearance scenario); LMW Aβ42 clearance proxy, not a directly reported half-life; candidate HMW sensitivity ≈$(1.0–3.5)\times10^{-5}$ $s^{-1}$ (not used in present simulations) |
| **New active-site-resolved secondary-nucleation parameters** | | | | | |
| $\sigma_{ref} = D_{A\beta,ref}/M_{A\beta,ref}$ | Measured reference density of active Aβ42 sites, $\sigma_{ref}$ | sites/monomer | ≈1/143 ≈ $6.99\times10^{-3}$ | [5], *in vitro* Aβ42 | Reference only; not tau-seeded value |
| $k_{o2,A\beta,M}^{ref}$ | Mass-based Aβ secondary-nucleation coefficient for comparison | $\mu M^{-2}\ s^{-1}$ | $1.00\times10^{-8}$ | [29]; Section S1.28 | Calibration comparison, not used in the new equations |

| Symbol | Definition | Units | Value/range | Source or basis | Value used / status |
|---|---|---|---|---|---|
| $k_{o2,A\beta,D,p}$ | Secondary-nucleation coefficient per accessible active site in $R_{2,A\beta} = k_{o2,A\beta,D,p} m_{A\beta}^2 D_{A\beta,eff}$ | $\mu M^{-2}\ s^{-1}$ | Unmeasured; conditional ≈$1.43\times10^{-6}$ | Eq. (S41), if $\sigma_{ref}$=1/143 and same mass basis | Fit to matched experiments; no universal value |
| $\sigma_0$ | Reference active-site density used in limiting case | sites/monomer | $\sigma_{ref}$ as illustrative choice | Eq. (22); [5] | 1/143 only for matched reference case |
| $\eta_{birth}$ | Mean number of active sites introduced per new fibril during conversion | sites/fibril | Unmeasured | New model, Eq. (13) | $2\sigma_0$ in mass-proportional limit; 0–$2\times10^6\sigma_0$ (Section 4.3) |
| $\eta_{grow}$ | Mean number of active sites introduced per monomer incorporated by elongation | sites/monomer | Unmeasured | New model, Eq. (13) | $\sigma_0$ in mass-proportional limit; varies with $u$ via Eq. (14) (Section S2.1) |
| $k_{mask}$ | Intrinsic loss or masking rate of catalytic-site activity without fibril-mass loss; separate from geometric shielding | $s^{-1}$ | Unmeasured | Phenomenological model coefficient | 0 reference; $10^{-10}$–$10^{-7}$ (Section S2.2) |
| $\eta_{frag}$ | Mean number of active sites created per fragmentation event; optional | sites/event | Not measured | New sensitivity parameter (Section 4.5) | 0 reference; 0.1–$10^3$ |
| $r$ | Site-abundance factor of tau-nucleated fibrils: $\eta_{birth}$ and $\eta_{grow}$ of the tau-derived lineage relative to those of the homotypic lineage | — | Not measured | Two-lineage model (Section S2.6) | 1 reference; 0.1–30 |
| $r_k$ | Activity factor of tau-nucleated fibrils: per-site secondary-nucleation coefficient of the tau-derived lineage relative to $k_{o2,A\beta,D,p}$ | — | Not measured | Two-lineage model (Section S2.6) | 1 reference; 0.3–10 |

| Symbol | Definition | Units | Value/range | Source or basis | Value used / status |
|---|---|---|---|---|---|
| $Da$ | Damköhler number $2k_{+,A\beta}P_{A\beta}/\lambda_{m,A\beta}$: monomer consumption by elongation relative to monomer clearance | — | Derived | Eq. (29), Section 3.4 | ∞ without clearance; 0.023 in combined scenario; rises from below $10^{-4}$ to 70 in composite scenario (Section 4.6) |
| $T_{1/2,F}$ | Half-life of fibrillar Aβ (fibril and plaque turnover); optional | s | Not established | New sensitivity parameter (Section 4.5) | ∞ reference; 1–100 yr |
| $\gamma$ | Exponent of the length-dependent accessibility factor $g = (\bar{n}_{F,ref}/\bar{n}_F)^\gamma$ [Eq. (S30)] | — | — | Sensitivity law (Sections 4.5, S2.5) | 0 reference; 0–3 |
| $\beta_{frag}$ | Exponent of the length factor of the fragmentation rate, $(\bar{n}_F/\bar{n}_{F,ref})^{\beta frag}$ [Eq. (S31)] | — | — | Sensitivity law (Sections 4.5, S2.5) | 0 reference; −1 to 2 |
| $\beta$ | Exponent of the alternative power-law accessibility closure $f_{surf} = min[1, (3\delta_{A\beta}/R_p)^\beta]$ | — | — | Sensitivity closure (Section 4.5) | 1/2 (shell closure of Eqs. (15)–(18): 1) |
| **Optional supersaturation-dependent active-site formation law [Eq. (14)]** | | | | | |
| $m_{eq,A\beta}$ | Equilibrium free Aβ monomer concentration | µM | Not established for tau-seeded preparation | Requires matched equilibrium measurements | $10^{-3}$ µM, illustrative; $10^{-4}$–$3\times10^{-2}$ µM (Section S2.1) |
| $\eta_{min}$ | Minimum elongation-related active-site yield | sites/monomer | Not measured | Proposed Eq. (14) | $\sigma_0/4$; $2\sigma_0/128$–$2\sigma_0$ (Section S2.1) |
| $\eta_{max}$ | Maximum elongation-related active-site yield | sites/monomer | Not measured | Proposed Eq. (14) | $2\sigma_0$ ($\eta_{max} \geq \eta_{min}$) |
| $K_u$ | Half-saturation constant in supersaturation variable $u$ | — | Not measured | Proposed Eq. (14) | 3; 0.1–30 (Section S2.1) |

| Symbol | Definition | Units | Value/range | Source or basis | Value used / status |
|---|---|---|---|---|---|
| $h$ | Hill-type exponent for active-site-yield dependence on $u$ | — | Not measured | Proposed Eq. (14) | 1; 1 and 2 (Section S2.1) |
| Numerical scenarios | | | | | |
| $t_{end}$ | Simulation interval for tissue-scale examples | years | 33 in preceding AD scenario | Section S1.29 | 33 |

Notes to Table S1. Values labeled "inherited" reproduce the kinetic conventions and estimates of the preceding coupled tau–Aβ model (derivation in Section S1); the new active-site variables require separate calibration. Concentration-based rate constants use μM and seconds unless otherwise indicated. The baseline mass-proportional limit is a mathematical reference scenario, not a measurement of the site yield in tau-seeded Aβ42 fibrils. In the baseline, oligomer dissociation and soluble-protein clearance are both switched off: all four dissociation constants are zero and all clearance rates are zero, with half-lives taken as infinite ($10^{20}$ s in the code). The dissociation scenario and the clearance scenario are each run separately with the values listed in the last column.

The estimate of one active site per approximately 143 Aβ42 monomers reported in a recent study [5] applies to that study's experimental fibril preparation. Accordingly, $\sigma_0 = 1/143$, $\eta_{birth} = 2\sigma_0$, and $\eta_{grow} = \sigma_0$ define an illustrative mass-proportional reference only. Values for tau-seeded Aβ fibrils require matched measurements.

The conditional value $k_{o2,A\beta,D,p} \approx 1.43 \times 10^{-6} \mu\text{M}^{-2}\text{s}^{-1}$ follows from $k_{o2,A\beta,M}^{ref} = 1.00 \times 10^{-8} \mu\text{M}^{-2}\text{s}^{-1}$ and $\sigma_{ref} = 1/143$ only when both measurements describe the same fibril preparation and catalytically accessible material. This calibration corresponds to $f_{surf} = 1$. In tissue-scale calculations, Eqs. (16) and (17) then reduce the number of exposed sites as the plaque consolidates without changing the intrinsic site density or per-site rate coefficient.

The optional growth-history law [Eq. (14)] is left uncalibrated until matched experiments provide equilibrium monomer concentration, active-site density, fibril mass, and fibril number. The values $m_{eq,A\beta} = 10^{-3}$ μM, $\eta_{min} = \sigma_0/4$, $\eta_{max} = 2\sigma_0$, $K_u = 3$, and $h = 1$ used in Section S2.1 are illustrative sensitivity choices, not calibrated values.

**Table S2.** Aβ quantities at $t$ = 33 years for the reference, dissociation, soluble-protein-clearance, and combined (Section 4.6) scenarios; Table S3 gives the remaining scenarios. Ratios with superscript (0) are relative to the uncoupled case ($\chi_\tau = 0$) of the same scenario. The coupling cases are $\chi_\tau = 1.86\times10^{-10}$ (PHF, $f_{acc}$ = 0.1), $1.86\times10^{-9}$ (PHF, $f_{acc}$ = 1), and $4.45\times10^{-9}$ μM$^{-2}$ s$^{-1}$ (CTE, $f_{acc}$ = 1); CTE coupling at $f_{acc}$ = 0.1 is omitted because it is equivalent to PHF coupling at $f_{acc}$ = 0.239 (Section 4.2). Combined: oligomer dissociation and soluble-protein clearance with the Aβ source raised 78.5-fold (Section 4.6). In every scenario, $D_{A\beta,eff}/D_{A\beta,eff}^{(0)}$ at matched fibril mass equals unity to within $10^{-4}$.

| Scenario | Case | $M_{A\beta}$ (μM) | $m_{A\beta}$ (μM) | $\bar{n}_F$ | $\sigma_{D,A\beta}/\sigma_0$ | $D_{A\beta,eff}/D_{A\beta,eff}^{(0)}$ | $R_{2,A\beta}/R_{2,A\beta}^{(0)}$ | $\Xi_{A\beta}$ (μM s) | $\Xi_{A\beta}/\Xi_{A\beta}^{(0)}$ |
|---|---|---|---|---|---|---|---|---|---|
| Reference | $\chi_\tau = 0$ | 228.8 | $3.16\times10^{-3}$ | $6.40\times10^{5}$ | 1.000 | 1 | 1 | $3.09\times10^{4}$ | 1 |
| | PHF 0.1 | 228.8 | $2.66\times10^{-3}$ | $5.39\times10^{5}$ | 1.000 | 1.000 | 0.709 | $4.36\times10^{4}$ | 1.41 |
| | PHF 1 | 228.8 | $1.62\times10^{-3}$ | $3.27\times10^{5}$ | 1.000 | 1.000 | 0.262 | $1.18\times10^{5}$ | 3.82 |
| | CTE 1 | 228.8 | $1.26\times10^{-3}$ | $2.54\times10^{5}$ | 1.000 | 1.000 | 0.158 | $1.96\times10^{5}$ | 6.34 |
| Dissociation | $\chi_\tau = 0$ | 228.7 | $6.81\times10^{-2}$ | $1.38\times10^{7}$ | 1.000 | 1 | 1 | 66.0 | 1 |
| | PHF 0.1 | 228.7 | $6.11\times10^{-2}$ | $1.23\times10^{7}$ | 1.000 | 1.000 | 0.804 | 82.1 | 1.24 |
| | PHF 1 | 228.8 | $4.27\times10^{-2}$ | $8.63\times10^{6}$ | 1.000 | 1.000 | 0.393 | 167 | 2.53 |
| | CTE 1 | 228.8 | $3.52\times10^{-2}$ | $7.11\times10^{6}$ | 1.000 | 1.000 | 0.266 | 245 | 3.71 |
| Soluble-protein clearance | $\chi_\tau = 0$ | $2.60\times10^{-6}$ | $1.46\times10^{-3}$ | $1.48\times10^{5}$ | 1.000 | 1 | 1 | $6.55\times10^{-3}$ | 1 |
| | PHF 0.1 | $6.85\times10^{-5}$ | $1.46\times10^{-3}$ | $9.96\times10^{4}$ | 1.000 | 26.4 | 26.4 | 0.257 | 39.2 |
| | PHF 1 | $6.62\times10^{-4}$ | $1.46\times10^{-3}$ | $9.85\times10^{4}$ | 1.000 | 150 | 150 | 2.51 | 383 |
| | CTE 1 | $1.58\times10^{-3}$ | $1.46\times10^{-3}$ | $9.84\times10^{4}$ | 1.000 | 268 | 268 | 5.99 | 915 |
| Combined ($F$ = 78.5) | $\chi_\tau = 0$ | 228.8 | 0.112 | $1.29\times10^{7}$ | 1.000 | 1 | 1 | 85.6 | 1 |
| | PHF 0.1 | 383.8 | 0.111 | $1.38\times10^{7}$ | 1.000 | 1.41 | 1.38 | 135 | 1.57 |
| | PHF 1 | 869.9 | 0.107 | $1.44\times10^{7}$ | 1.000 | 2.44 | 2.19 | 300 | 3.50 |
| | CTE 1 | 1236 | 0.103 | $1.43\times10^{7}$ | 1.000 | 3.08 | 2.61 | 435 | 5.08 |

**Table S3.** Aβ quantities at $t$ = 33 years for the growth-history, masking, fragmentation, fragmentation-site, and fibril-removal scenarios; columns and coupling cases as in Table S2. Growth history: Eq. (14) with the parameters of Fig. S21a. Masking: $k_{mask} = 10^{-8}$ s$^{-1}$. Fragmentation: $k_{-,A\beta} = 3.5\times10^{-11}$ s$^{-1}$; fragmentation sites: in addition $\eta_{frag}$ = 10; fibril removal: $T_{1/2,F}$ = 10 years. Except in the growth-history scenario, $D_{A\beta,eff}/D_{A\beta,eff}^{(0)}$ at matched fibril mass equals unity to within $10^{-4}$.

| Scenario | Case | $M_{A\beta}$ (μM) | $m_{A\beta}$ (μM) | $\bar{n}_F$ | $\sigma_{D,A\beta}/\sigma_0$ | $D_{A\beta,eff}/D_{A\beta,eff}^{(0)}$ | $R_{2,A\beta}/R_{2,A\beta}^{(0)}$ | $\Xi_{A\beta}$ (μM s) | $\Xi_{A\beta}/\Xi_{A\beta}^{(0)}$ |
|---|---|---|---|---|---|---|---|---|---|
| Growth history | $\chi_\tau = 0$ | 228.8 | $2.85\times10^{-3}$ | $5.76\times10^{5}$ | 1.192 | 1 | 1 | $3.80\times10^{4}$ | 1 |
| | PHF 0.1 | 228.8 | $2.51\times10^{-3}$ | $5.07\times10^{5}$ | 1.141 | 0.957 | 0.741 | $4.91\times10^{4}$ | 1.29 |
| | PHF 1 | 228.8 | $1.61\times10^{-3}$ | $3.25\times10^{5}$ | 0.927 | 0.777 | 0.247 | $1.20\times10^{5}$ | 3.15 |
| | CTE 1 | 228.8 | $1.25\times10^{-3}$ | $2.54\times10^{5}$ | 0.791 | 0.663 | 0.129 | $1.97\times10^{5}$ | 5.17 |
| Masking | $\chi_\tau = 0$ | 228.8 | $5.10\times10^{-3}$ | $1.03\times10^{6}$ | 0.096 | 1 | 1 | $1.19\times10^{4}$ | 1 |
| | PHF 0.1 | 228.8 | $3.25\times10^{-3}$ | $6.56\times10^{5}$ | 0.096 | 1.000 | 0.406 | $2.93\times10^{4}$ | 2.47 |
| | PHF 1 | 228.8 | $1.67\times10^{-3}$ | $3.38\times10^{5}$ | 0.096 | 1.000 | 0.108 | $1.11\times10^{5}$ | 9.31 |
| | CTE 1 | 228.8 | $1.28\times10^{-3}$ | $2.58\times10^{5}$ | 0.096 | 1.000 | 0.063 | $1.90\times10^{5}$ | 16.0 |
| Fragmentation | $\chi_\tau = 0$ | 228.8 | $2.72\times10^{-7}$ | 55.0 | 1.000 | 1 | 1 | $2.33\times10^{4}$ | 1 |
| | PHF 0.1 | 228.8 | $2.72\times10^{-7}$ | 55.0 | 1.000 | 1.000 | 1.000 | $2.46\times10^{4}$ | 1.06 |
| | PHF 1 | 228.8 | $2.72\times10^{-7}$ | 55.0 | 1.000 | 1.000 | 1.000 | $3.43\times10^{4}$ | 1.47 |
| | CTE 1 | 228.8 | $2.72\times10^{-7}$ | 54.9 | 1.000 | 1.000 | 1.000 | $4.56\times10^{4}$ | 1.96 |
| Fragmentation sites ($\eta_{frag}$ = 10) | $\chi_\tau = 0$ | 228.8 | $2.72\times10^{-7}$ | 55.0 | 27.0 | 1 | 1 | $2.35\times10^{4}$ | 1 |
| | PHF 0.1 | 228.8 | $2.72\times10^{-7}$ | 55.0 | 27.0 | 1.000 | 1.000 | $2.48\times10^{4}$ | 1.05 |
| | PHF 1 | 228.8 | $2.72\times10^{-7}$ | 55.0 | 27.0 | 1.000 | 1.000 | $3.44\times10^{4}$ | 1.46 |
| | CTE 1 | 228.8 | $2.72\times10^{-7}$ | 54.9 | 27.0 | 1.000 | 1.000 | $4.58\times10^{4}$ | 1.95 |
| Fibril removal ($T_{1/2,F}$ = 10 yr) | $\chi_\tau = 0$ | 89.9 | $6.29\times10^{-3}$ | $4.99\times10^{5}$ | 1.000 | 1 | 1 | $3.31\times10^{4}$ | 1 |
| | PHF 0.1 | 89.9 | $4.73\times10^{-3}$ | $3.76\times10^{5}$ | 1.000 | 1.000 | 0.567 | $5.21\times10^{4}$ | 1.57 |
| | PHF 1 | 89.9 | $2.63\times10^{-3}$ | $2.09\times10^{5}$ | 1.000 | 1.000 | 0.175 | $1.55\times10^{5}$ | 4.68 |
| | CTE 1 | 89.9 | $2.02\times10^{-3}$ | $1.60\times10^{5}$ | 1.000 | 1.000 | 0.103 | $2.60\times10^{5}$ | 7.84 |

**Table S4.** Sensitivity of the tau-coupling effects at $t$ = 33 years (Section 4.5). Ratios are relative to $\chi_\tau = 0$ for the same sensitivity case; columns give $\chi_\tau = 1.86\times10^{-10}$ (PHF 0.1), $1.86\times10^{-9}$ (PHF 1), and $4.45\times10^{-9}$ $\mu M^{-2}$ $s^{-1}$ (CTE 1). All other parameters are those of the reference case. The characteristic value refers to the uncoupled simulation. In every row, $D_{A\beta,eff}/D_{A\beta,eff}^{(0)}$ at matched fibril mass equals unity to within $10^{-4}$.

| Sensitivity case | Characteristic value | $R_{2,A\beta}$ ratio, PHF 0.1 | $R_{2,A\beta}$ ratio, PHF 1 | $R_{2,A\beta}$ ratio, CTE 1 | $\Xi_{A\beta}$ ratio, PHF 0.1 | $\Xi_{A\beta}$ ratio, PHF 1 | $\Xi_{A\beta}$ ratio, CTE 1 |
|---|---|---|---|---|---|---|---|
| Reference, $\delta_{A\beta}$ = 0.010 μm | $f_{surf} = 8.40\times10^{-3}$ | 0.709 | 0.262 | 0.158 | 1.41 | 3.82 | 6.34 |
| $\delta_{A\beta}$ = 0.0025 μm | $f_{surf} = 2.10\times10^{-3}$ | 0.427 | 0.116 | 0.067 | 2.34 | 8.66 | 14.9 |
| $\delta_{A\beta}$ = 0.005 μm | $f_{surf} = 4.20\times10^{-3}$ | 0.568 | 0.175 | 0.103 | 1.76 | 5.73 | 9.73 |
| $\delta_{A\beta}$ = 0.02 μm | $f_{surf} = 1.68\times10^{-2}$ | 0.824 | 0.381 | 0.239 | 1.21 | 2.63 | 4.19 |
| $\delta_{A\beta}$ = 0.04 μm | $f_{surf} = 3.36\times10^{-2}$ | 0.902 | 0.525 | 0.352 | 1.11 | 1.90 | 2.84 |
| $\delta_{A\beta}$ = 0.1 μm | $f_{surf} = 8.40\times10^{-2}$ | 0.959 | 0.719 | 0.542 | 1.04 | 1.39 | 1.85 |
| Power-law closure, $\beta$ = 1/2 | $f_{surf} = 9.16\times10^{-2}$ | 0.957 | 0.709 | 0.531 | 1.05 | 1.41 | 1.88 |
| No burial | $f_{surf} = 1$ | 0.996 | 0.957 | 0.905 | 1.005 | 1.045 | 1.105 |
| $T_{1/2,F}$ = 1 yr | $M_{A\beta}$ = 10.0 μM | 0.267 | 0.064 | 0.037 | 2.55 | 9.59 | 16.6 |
| $T_{1/2,F}$ = 10 yr | $M_{A\beta}$ = 89.9 μM | 0.567 | 0.175 | 0.103 | 1.57 | 4.68 | 7.84 |
| $T_{1/2,F}$ = 100 yr | $M_{A\beta}$ = 204.5 μM | 0.694 | 0.251 | 0.151 | 1.43 | 3.91 | 6.49 |
| Fragmentation, $\eta_{frag} = 0$ | $\sigma_{D,A\beta}/\sigma_0 = 1$ | 1.000 | 1.000 | 1.000 | 1.06 | 1.47 | 1.96 |
| Fragmentation, $\eta_{frag} = 1$ | $\sigma_{D,A\beta}/\sigma_0 = 3.60$ | 1.000 | 1.000 | 1.000 | 1.06 | 1.47 | 1.96 |
| Fragmentation, $\eta_{frag} = 10$ | $\sigma_{D,A\beta}/\sigma_0 = 27.0$ | 1.000 | 1.000 | 1.000 | 1.05 | 1.46 | 1.95 |
| Fragmentation, $\eta_{frag} = 10^3$ | $\sigma_{D,A\beta}/\sigma_0 = 2.60\times10^3$ | 1.000 | 1.000 | 1.000 | 1.02 | 1.20 | 1.43 |

**Table S5. Dependent variables and derived quantities used in the active-site-resolved model.**

| Symbol | Definition | Units |
|---|---|---|
| **Tau aggregation: intracellular state variables** | | |
| $m_{tau}$ | Free tau monomer concentration. | μM |
| $S_{tau}$ | Soluble on-pathway tau oligomer-equivalent concentration, in activated-monomer equivalents. | μM |
| $P_{tau}$ | Number concentration of tau fibrillar species of any length. | μM |
| $M_{tau}$ | Concentration of tau monomer units contained in fibrils (fibril mass). | μM |
| **Aβ aggregation: extracellular/interstitial state variables** | | |
| $m_{A\beta}$ | Free Aβ monomer concentration. | μM |
| $S_{A\beta}$ | Soluble on-pathway Aβ oligomer-equivalent concentration, in activated-monomer equivalents; not a particle concentration. | μM |
| $P_{A\beta}$ | Number concentration of Aβ fibrillar species of any length. | μM |
| $M_{A\beta}$ | Concentration of Aβ monomer units contained in fibrils (fibril mass). | μM |
| $D_{A\beta}$ | Molar concentration of catalytically active Aβ secondary-nucleation sites; not an additional protein-mass pool. | μM (sites) |
| **Derived coupling variables and observables** | | |
| $T_{tau}$ | Effective concentration of accessible tau fibrillar material in the Aβ compartment [Eq. (1)]. | μM |
| $D_{A\beta,eff}$ | Geometrically accessible concentration of Aβ secondary-nucleation sites [Eq. (17)]. | μM (sites) |
| $f_{surf}$ | Fraction of total Aβ active sites exposed at the plaque surface [Eq. (16)]. | — |
| $\sigma_{D,A\beta}$ | Total Aβ active-site density $D_{A\beta}/M_{A\beta}$, defined for $M_{A\beta} > 0$ [Eq. (15)]. | sites/monomer |
| $\overline{n}_F$ | Mean number of Aβ monomer units per fibril, $M_{A\beta}/P_{A\beta}$, defined for $P_{A\beta} > 0$ [Eq. (19)]. | monomers/fibril |
| $\Xi_{A\beta}$ | Time-integrated soluble Aβ oligomer-equivalent exposure, $\int S_{A\beta}\, dt$ [Eq. (20)]; model diagnostic, not directly measurable. | μM s |
| $u$ | Dimensionless positive monomer supersaturation used in the optional growth-history law [Eq. (14)]. | — |